\documentclass[twocolumn]{aastex63}

\usepackage{booktabs}
\usepackage{amsmath}
\newcommand{\Mnom} {\hbox{M$_\odot$}}
\newcommand{\si}{$\mathcal{SI}$}

\received{June 1, 2021}
\revised{January 10, 2021}
\accepted{\today}
\submitjournal{AJ}

\shorttitle{High Mass-ratio Binaries in Open Clusters}
\shortauthors{Jadhav et al.}

\graphicspath{{./}{figures/}}

\begin{document}

\title{High Mass-ratio Binary Population in Open Clusters:
Segregation of early type binaries and an increasing binary fraction with mass}
%

\correspondingauthor{Vikrant V. Jadhav}
\email{vikrant.jadhav@iiap.res.in}

\author[0000-0002-8672-3300]{Vikrant V. Jadhav}
\affiliation{Indian Institute of Astrophysics, Sarjapur Road, Koramangala, Bangalore, India.}
\affiliation{Joint Astronomy Programme and Department of Physics, Indian Institute of Science, Bangalore, India}

\author[0000-0003-1828-6476]{Kaustubh Roy}
\affiliation{Indian Institute of Science, Bangalore, India.}

\author[0000-0002-9938-7166]{Naman Joshi}
\affiliation{Indian Institute of Science Education and Research (IISER), Bhopal, India.}

\author{Annapurni Subramaniam}
\affiliation{Indian Institute of Astrophysics, Sarjapur Road, Koramangala, Bangalore, India.}

\begin{abstract}
Binary stars play a vital role in astrophysical research, as a good fraction of stars are in binaries. Binary fraction (BF) is known to change with stellar mass in the Galactic field, but such studies in clusters require binary identification and membership information. Here, we estimate the total and spectral-type-wise high mass-ratio (HMR) BF ($f^{0.6}$) in 23 open clusters using unresolved binaries in color-magnitude diagrams using \textit{Gaia} DR2 data. We introduce the segregation index (\si) parameter to trace mass segregation of HMR (total and mass-wise) binaries and the reference population. This study finds {that} in open clusters, (1) HMR BF for the mass range 0.4--3.6 \Mnom\ (early M to late B type) has a range of {0.12 to 0.38} with a peak at {0.12--0.20}, (2) older clusters have a relatively higher HMR BF, (3) the mass-ratio distribution is unlikely to be a flat distribution and BF(total) $\sim$ (1.5 to 2.5) $\times f^{0.6}$, (4) a decreasing BF(total) from late B-type to {K-type}, in agreement with the Galactic field stars, (5) older clusters show radial segregation of HMR binaries, (6) B and A/F type HMR binaries show radial segregation in some young clusters suggesting a primordial origin. This study will constrain the initial conditions and identify the major mechanisms that regulate binary formation in clusters. Primordial segregation of HMR binaries could result from massive clumps spatially segregated in the collapse phase of the molecular cloud.
\end{abstract}



\section{Introduction} \label{sec:intro}
Stars form as singles, binaries, and multiples as a result of star formation. Binary stars play a vital role in star clusters, and they are of importance in a wide variety of astrophysical research. The evolution of binary systems leads to the formation of exotic systems such as Supernovae (SN), cataclysmic variables (CVs), blue straggler stars (BSSs), sub-sub giants, chemically peculiar stars, etc. As a result of star formation, it is found that most of the stars are formed as part of a binary/multi-component system \citep{duquennoy1991multiplicity} {\citep{Raghavan2010ApJS..190....1R,Moe2017ApJS..230...15M}}. Star clusters are found to be home to binary systems. As they undergo dynamical effects, the binary fraction (BF) found in them is likely to be different from the primordial binaries \citep{Fregeau2009}. The presence of binaries also modifies the dynamic evolution of star clusters \citep{bailyn1995blue}, resulting in evaporation of single/low-mass systems and central segregation of massive stellar systems.

Binary systems with white dwarfs close to the Chandrasekhar mass limit are potential progenitors of various SN of type Ia. There is an increase in the variety of potential SN Ia progenitor binaries and the swelling number of different kinds of thermonuclear SN \citep{Jha2019NatAs...3..706J}. These systems imply that a detailed understanding of binaries with WDs and a greater understanding of binaries in the low mass range (1.0--3.0 \Mnom), which are the starting point of such systems, is necessary. 

There are several studies estimating the BF in the Galactic field, along with simulations to recreate these fractions.
\citet{DeRosa2014MNRAS.437.1216D} estimated the BF between 1.5--3.0 \Mnom\ as 32\% (up to 44\% for non-single systems), \citet{Raghavan2010ApJS..190....1R} found  33\% binaries between 0.8--1.2 \Mnom (and up to 46\% non-single systems). \citet{Bergfors2010A&A...520A..54B} estimated the multiplicity fraction of 34\% between 0.08--0.45 \Mnom, suggesting a decreasing fraction of binary stars with mass. 
A recent review by \citet{Lee2020SSRv..216...70L} presents our current understanding of binary and multiple systems in the Galactic field, along with a comprehensive review of estimations of BF across a range of mass. \citet{Parker2014MNRAS.442.3722P} suggested that the binary population in the field is indicative of the primordial binary population in star-forming regions, at least for systems with primary masses in the range 0.02-–3.0 \Mnom.

In order to study the BF in open clusters (OCs) and their radial segregation, binaries need to be identified first. Binaries can be identified spectroscopically or photometrically. A few relatively old star clusters are studied using spectroscopic monitoring of bright stars to study binarity. For example, \citet{Mathieu1986} {and \citet{Geller2021AJ....161..190G}} studied M67 and found the spectroscopic binaries and BSS to be segregated with respect to stars near the turn-off. \citet{Geller_2012} studied NGC 188 and the dynamical status of the spectroscopic binaries. Both the studies are based on magnitude limited stars, consisting of upper main-sequence (MS) stars and brighter stars. \citet{2012A&A...540A..16M} studied globular clusters using photometric binaries. Recently, \citet{Thompson2021} used photometric data to estimate BF in eight OCs. Such studies of BF in more OCs using photometric data are needed to cover a range of cluster ages and stellar mass. 
Such a study was not feasible earlier, as membership information was lacking for numerous OCs. The situation has changed with the \textit{Gaia} data releases \citep{2016A&A...595A...1G,2018A&A...616A...1G}, and the membership information for a large number of star clusters is now available.

On an average, binaries have a higher total mass than their single-star counterparts. Therefore, energy exchange between these two groups tends to cause binaries to sink toward the cluster's core. Locally, at different radii within a cluster, the binary frequency also could evolve due to two-body relaxation and dynamical friction. 
This dynamical mass segregation could result in the binary frequency rising in the cluster core and falling in the halo. Indeed, many old open and globular clusters show a rising binary frequency toward the cluster core, which is interpreted as the result of mass segregation (e.g., \citealt{Mathieu1986, Geller_2012, 2012A&A...540A..16M}).

To quantitatively estimate the central segregation of the binary population in these OCs, we decided to use an approach similar to that of \citet{Alessandrini_2016} by constructing radial profiles of the binary population and comparing it with a reference population. We introduced a new parameter to quantitatively measure the extent of this segregation, the \si\ parameter (as in Segregation Index), which can be calculated from the radial profiles. We also divided stars by magnitude and analyzed their radial profiles to see how binary segregation changes with stellar mass.

In this work, we calculated BF in 23 OCs. Qualitative measurements of binary segregation is done using a newly introduced \si\ parameter. The details of calculations are given in \S~\ref{sec:binary_selection} and \ref{sec:method_of_parameter}. Results for individual cluster are discussed in \S~\ref{sec:individual}. The overall discussion and conclusions are presented in \S~\ref{sec:discussion} and \ref{sec:conclusions}.


\section{Data and Method} \label{sec:data}
\subsection{Gaia data of OCs and isochrone fitting} \label{sec:gaia_data}

\begin{table*}
    \centering
    \begin{tabular}{lcccc ccccc r}
    \toprule
Name	&	log($age$)	&	DM	&	Metallicity	&	E(B$-$V)	&	\multicolumn{2}{c}{Core radius}			&	$N_{relax}$	&	Total	&	$f^{0.6}$			&	\si			\\ \cmidrule(lr){6-7}
	&		&		&		&		&	[\arcmin]	&	[pc]	&		&	Members	&				&				\\ \hline
IC4651	&	9.3	&	9.77	&	0.1	&	0.12	&	2.5	&	0.66	&	8.89	&	960	&	0.16$\pm$0.02	&	0.3$\pm$0.1	\\ 
IC4756	&	8.95	&	8.4	&	0	&	0.2	&	6	&	0.83	&	7.48	&	543	&	0.31$\pm$0.03	&	-0.24$\pm$0.08	\\ 
NGC0188	&	9.78	&	11.3	&	0.11	&	0.08	&	2.1	&	1.12	&	10.29	&	1181	&	0.25$\pm$0.02	&	0.11$\pm$0.08	\\ 
NGC0752	&	9.15	&	8.26	&	0.12	&	0.05	&	6.2	&	0.81	&	11.23	&	433	&	0.19$\pm$0.03	&	-0.05$\pm$0.16	\\ 
NGC1039	&	8.4	&	8.53	&	0.1	&	0.08	&	5.6	&	0.83	&	1.04	&	764	&	0.17$\pm$0.02	&	0.24$\pm$0.13	\\ 
NGC2168	&	8.13	&	9.6	&	-0.2	&	0.3	&	4.3	&	1.04	&	0.49	&	1794	&	0.23$\pm$0.01	&	0.13$\pm$0.06	\\ 
NGC2360	&	9	&	10.23	&	0.1	&	0.1	&	2.4	&	0.79	&	8.6	&	1037	&	0.19$\pm$0.02	&	0.46$\pm$0.1	\\ 
NGC2422	&	8.16	&	8.44	&	0.14	&	0.11	&	4.9	&	0.7	&	2.05	&	907	&	0.14$\pm$0.02	&	0.03$\pm$0.1	\\ 
NGC2423	&	9	&	9.8	&	0.15	&	0.1	&	3.1	&	0.82	&	6.49	&	694	&	0.17$\pm$0.02	&	0.11$\pm$0.12	\\ 
NGC2437	&	8.4	&	11.1	&	0.05	&	0.18	&	5.7	&	2.74	&	0.68	&	3032	&	0.3$\pm$0.01	&	0.02$\pm$0.04	\\ 
NGC2447	&	8.78	&	10	&	-0.05	&	0.04	&	3.2	&	0.94	&	3.83	&	926	&	0.19$\pm$0.02	&	0.24$\pm$0.1	\\ 
NGC2516	&	8.25	&	8.08	&	0.07	&	0.13	&	10.2	&	1.22	&	1.97	&	2518	&	0.16$\pm$0.01	&	-0.05$\pm$0.06	\\ 
NGC2547	&	7.7	&	7.68	&	-0.14	&	0.05	&	8.1	&	0.81	&	1.1	&	644	&	0.22$\pm$0.04	&	-0.23$\pm$0.21	\\ 
NGC2548	&	8.74	&	9.43	&	0.13	&	0.02	&	13.8	&	3.09	&	1.98	&	509	&	0.17$\pm$0.02	&	0.19$\pm$0.09	\\ 
NGC2682	&	9.6	&	9.62	&	0	&	0.05	&	3.8	&	0.92	&	48.68	&	1520	&	0.22$\pm$0.02	&	0.25$\pm$0.08	\\ 
NGC3532	&	8.6	&	8.35	&	0.1	&	0.02	&	9.4	&	1.28	&	1.49	&	1879	&	0.13$\pm$0.01	&	0.08$\pm$0.08	\\ 
NGC6025	&	8.18	&	9.51	&	0.2	&	0.17	&	2.5	&	0.57	&	0.94	&	452	&	0.21$\pm$0.02	&	-0.03$\pm$0.09	\\ 
NGC6281	&	8.54	&	8.62	&	0.15	&	0.15	&	3.3	&	0.52	&	7.83	&	573	&	0.27$\pm$0.02	&	0.06$\pm$0.11	\\ 
NGC6405	&	7.96	&	8.1	&	0.07	&	0.14	&	4.5	&	0.54	&	0.64	&	967	&	0.17$\pm$0.02	&	-0.06$\pm$0.11	\\ 
NGC6774	&	9.3	&	7.46	&	0.16	&	0.08	&	9	&	0.82	&	40.67	&	234	&	0.38$\pm$0.06	&	0.24$\pm$0.21	\\ 
NGC6793	&	8.78	&	8.65	&	0.1	&	0.27	&	4	&	0.63	&	6.05	&	465	&	0.17$\pm$0.03	&	-0.42$\pm$0.17	\\ 
Pleiades	&	8.04	&	5.67	&	0.09	&	0.04	&	25	&	0.99	&	1.12	&	1326	&	0.14$\pm$0.02	&	0.14$\pm$0.13	\\ 
Trump10	&	7.74	&	7.82	&	-0.12	&	0.04	&	24.7	&	2.64	&	0.14	&	947	&	0.12$\pm$0.02	&	0.28$\pm$0.18	\\ 

    \bottomrule
    \end{tabular}
    \caption{Parameters for the 23 OCs. log($age$), DM, metallicity and E(B$-$V) are derived from isochrone fitting. Core radii, $f^{0.6}$ and \si\ are estimated in this work. Total members are taken from \citet{Gaia2018}.}
    \label{tab:master_table}
\end{table*} 

The study of the binary population across various OCs requires a homogeneous catalog of clusters. We used the OC catalog given by \citet{Gaia2018} in this study.
We selected OCs within 130--1600 pc of various ages with well-defined MS, minimal differential reddening, and a fairly good number of stars. The list of selected clusters is given in Table \ref{tab:master_table}.

The isochrones used to fit the cluster data were obtained from PARSEC\footnote{http://stev.oapd.inaf.it/cgi-bin/cmd} stellar tracks \citep{2012MNRAS.427..127B}. {In order to select binary stars, a good isochrone fit to the MS is a must. Hence, we refitted the isochrones to the \textit{Gaia} CMDs using literature parameters as starting points \citep{2002A&A...389..871D, Netopil2016, Gaia2018}. Some isochrone fits were modified to achieve better match to the MS, MS turnoff (MSTO) and red giants (RGs) present in the CMDs visually. The $\chi^2$ values of our fits were compared with the literature \citep{ Gaia2018, Bossini_2019A&A...623A.108B} and were found to be similar or better.}
The parameters derived are tabulated in Table \ref{tab:master_table}. 

\subsection{Isochrones with various mass-ratios} \label{sec:iso_q_generation}

\begin{figure*}
    \centering
    \includegraphics[width=0.98\textwidth]{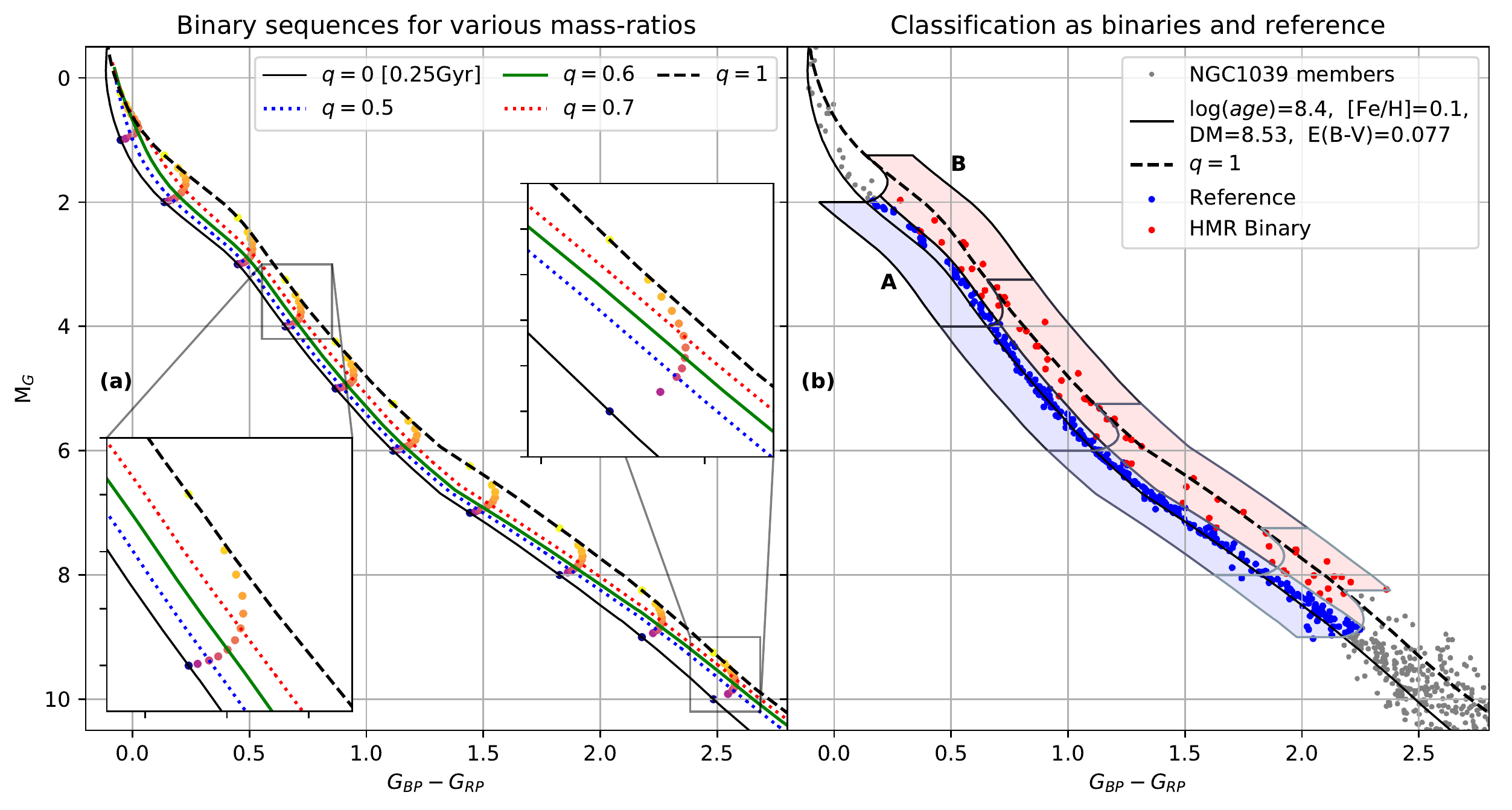}
    \caption{{(a) Binary sequences for $q=0$, $q=0.5$, $q=0.6$, $q=0.7$ and $q=1$. The {colored filled circles} show sequences with the same primary mass (or magnitude). The two insets near M$_G$ 4 and 10 illustrate the different shapes of binary sequences. (b) Schematic for selecting the HMR binary stars in OC NGC 1039. Isochrones for $q=0$ and $q=1$ are shown as solid and dashed back curves. The A (light-blue) and B (light-red) regions show the reference and HMR binary region, respectively. The panel also shows the division of stars into magnitude bins.}}
    \label{fig:schematic}
\end{figure*}

{The unresolved binaries are redder and brighter than the primary star, but the shifts in magnitude and colors depend on the magnitude of the primary star. We used interpolated PARSEC isochrones to calculate the magnitudes of the primary star, secondary star and the combined unresolved binary for various $q$ values. The insets in Fig.~\ref{fig:schematic} (a) show examples of binaries with different $q$ values for same primary magnitude. 
Fig.~\ref{fig:schematic} (a) also shows the binary sequences for $q=0$ to 1. The gap between these binary isochrones with the original isochrone depends on the shape of the isochrone and the mass-luminosity relation. For example, the $q=0.6$ sequence goes close to $q=1$ at M$_G\sim 0$ and at $\sim$ 10 mag, while it comes near $q=0$ at M$_G\sim 6$.} 
In general, there is not much separation between the $q=0$ and $q=0.5$ sequences resulting in the low mass-ratio binaries creating an over-density near $q=0$ sequence. 
The unresolved binaries congregate near either $q=0$ or $q=1$ isochrones due to the uneven gaps in the sequences (see Fig. 2 of \citealt{2020arXiv200804684L}.
Based on {the above factors}, we chose $q = 0.6$ as the cut-off to select HMR binaries. $q = 0.6$ isochrone stays away from both $q = 0$ and $q = 1$ for 1--10 mag range, which is the range of MS stars in our sample. $q=0.6$ cut-off also ensures that the single MS stars are 3$\sigma$ away from the binary sequence, even at the faint limit.

\begin{table*}
    \centering
    \begin{tabular}{rll}
    \toprule
Term	&	Formula	&	Remark	\\ \toprule
$N_A,\ N_B$	&	---	&	Number of stars in region A (reference) and region B (HMR binaries)	\\
$N_{B,4-6}$	&	---	&	Number of stars in region B within 4--6 mag	\\ \hline
$f^{0.6}$	&	$\displaystyle =\frac{N_B}{N_A+N_B}$	&	HMR BF of the whole cluster	\\
$f^{0.6}_{4-6}$	&	$\displaystyle = \frac{N_{B,4-6}}{N_{A,4-6} + N_{B,4-6}}$	&	HMR BF within 4--6 mag	\\ \hline
$r,\ r_{core}$	&	---	&	Radius of a star and King's core radius of the cluster	\\
$x$	&	$= \log(r/r_{core})$	&	Radius of stars normalized to King's radius	\\
$xmin,\ xmax$	&	---	&	Normalized radius of innermost and outermost star	\\
$\mathcal{SI} (sample,ref)$	&	$\displaystyle = \int_{xmin}^{xmax} \left(\phi_{sample}(x)-\phi_{ref}(x)\right) dx$	&	Segregation index	\\
\si	&	$=\mathcal{SI} (B,A)$	&	Segregation index for HMR binaries	\\
$\mathcal{SI}_{4-6}$	&	$=\mathcal{SI}(B_{4-6},A)$	&	Segregation index for HMR binaries within 4--6 mag	\\
$\mathcal{SI}_{4-6}^{ref}$	&	$=\mathcal{SI}(A_{4-6},A)$	&	Segregation index for reference stars within 4--6 mag	\\ \hline
$r_{eff}(sample)$	&	$\displaystyle =\frac{\Sigma\ r_{sample}} {N_{sample}}$	&	Effective radius of a sample	\\
$\mathcal{R}(sample)$	&	$\displaystyle = \frac{r_{eff}(sample)}{r_{eff}(cluster)}$	&	Normalized effective radius of a sample	\\
$\mathcal{R}$	&	$=\mathcal{R}(B)$	&	Normalized effective radius of HMR binaries	\\
$\mathcal{R}_{4-6}$	&	$=\mathcal{R}(B_{4-6})$	&	Normalized effective radius of HMR binaries within 4--6 mag	\\
$\mathcal{R}^{ref}$	&	$=\mathcal{R}(A)$	&	Normalized effective radius of reference stars	\\
$\mathcal{R}_{4-6}^{ref}$	&	$=\mathcal{R}(A_{4-6})$	&	Normalized effective radius of reference stars within 4--6 mag	\\ \hline
$T_{relax}$	&	$\displaystyle = \frac{8.9 \times 10^5 (N_{cluster}\ r^3)^{0.5} }{\langle m \rangle^{0.5} \textrm{log}(0.4N_{cluster})}$	&	{Dynamical relaxation time}	\\
$N_{relax}$	&	$= age/T_{relax}$	&	{Number of relaxation times passed}	\\ \bottomrule

    \end{tabular}
    \caption{Definitions of all the derived parameters from the HMR binary and reference stars are listed here.}
    \label{tab:definitions}
\end{table*} 

\subsection{Estimating cluster parameters}
\label{sec:method_of_parameter}

Fig.~\ref{fig:schematic} (b) shows the schematic for selecting binary stars. We classified the stars red-wards of $q=0.6$ isochrone as binaries (region B) and blue-wards as reference stars (region A). {More details on classifying binary stars are given in Appendix \S~\ref{sec:binary_selection}.
\S~\ref{sec:error} gives the details of error calculation {using bootstrap method}.
{\S~\ref{sec:stats} gives details of the statistical tests used in the analysis: i) Kolmogorov--Smirnov (KS) test is used to compare differences between two populations; ii) Spearman rank-order correlation test is used to test the monotonic nature of any trends. 
We have mainly used p-value of $<$0.003 ($\sim3\sigma$ significance) to judge the similarity in KS tests and monotonicity in Spearman tests. The trends with \textbar\textit{Spearman coefficient}\textbar $>$ 0.7 are considered monotonic for this study.} \S~\ref{sec:data_limitation} contains possible biases due to field contamination {(which is typically $<$ 5\% and increases with apparent magnitude)} and \textit{Gaia} data limitations {(which can increase BF)}.}

Table~\ref{tab:definitions} shows the definitions of all the derived parameters using  the HMR binary and reference stars. We also divided the stars into magnitude bins and calculated the same parameters for each magnitude bin for further analysis. The notable parameters are discussed below:
\begin{enumerate}
    \item \textit{HMR Binary Fraction ($f^{0.6}$):} The fraction of HMR binaries is calculated as the ratio of HMR binary and total stars within the same magnitude range. Since the OC catalog consists of cluster members only, a separate correction for field stars is not applied.
    \item \textit{Core radius ($r_{core}$):} For comparing radial stellar distributions between clusters of various sizes, we calculated the core radius by fitting King's surface density profile \citep{King_1962} to the cluster radial profile.
    \item \textit{Segregation index (\si):} Similar to the $A^{+}$ parameter used by \citet{Alessandrini_2016}, we formulated \si\ to measure the level of HMR binary segregation. \si\ is defined as the area between the normalized cumulative radial frequency of sample population such as HMR binaries ($\phi_{sample}$) and the reference population ($\phi_{ref}$; consisting of stars in region A), in the $\phi(r) - \log(r/r_{core})$ plane. A positive value for \si\ indicates relative segregation and vice-versa.
    \item \textit{Normalized effective radius ($\mathcal{R}$):} Another way to study radial segregation of stars is by estimating the effective radius of their radial distribution. The effective radius of a sample is defined as the average of the radii of all stars in the sample. We use the normalized effective radius for easy comparison of effective radii across clusters and their subsets. $\mathcal{R} < 1$ for a population indicates segregation and vice-versa. 
    \item \textit{Dynamical relaxation time ($T_{relax}$):} {To study the effect of dynamical evolution of the cluster on the binary segregation, we estimated the $T_{relax}$ using the number of stars in the cluster ($N_{cluster}$), radius containing half cluster members ($r$) and average mass of the cluster members ($\langle m \rangle$; \citealt{Spitzer1971ApJ...164..399S, Jadhav2021MNRAS.tmp.2079J}).}
\end{enumerate}

\subsubsection{Bin-wise analysis} \label{sec:bins}
To study the population of HMR binary stars across various masses, we divided the cluster MS according to the absolute magnitudes of the sources. The MS was divided in bins with absolute magnitudes in following ranges:
\begin{center}
\begin{tabular}{ccc}
    M$_G$ range & Mass range & Spectral type\\ \hline
    0--2 mag &  3.6--1.9 \Mnom & late B-type\\
    2--4 mag &  1.9--1.2 \Mnom & A\&F-type\\
    4--6 mag &  1.2--0.85 \Mnom & G-type\\
    6--8 mag &  0.85--0.6 \Mnom & K-type\\
    8--10 mag &  0.6--0.38 \Mnom & early M-type\\
\end{tabular}
\end{center}
In many clusters, the bin limits were adhered to. In some cases, the bins do not cover the full bin width, either due to the presence of a large scatter near the MS turn-off, or a large scatter near the limiting magnitude. In general, this study explores HMR binaries in the mass range 0.38--3.6 \Mnom\ (early M to late B spectral types), though the exact range studied vary from cluster to cluster. The mass range studied in most clusters is 0.5--2.0 \Mnom\ (K to A spectral types).


\section{Comments on individual clusters} \label{sec:individual}

\begin{figure*}
    \centering
    \includegraphics[width=0.97\textwidth]{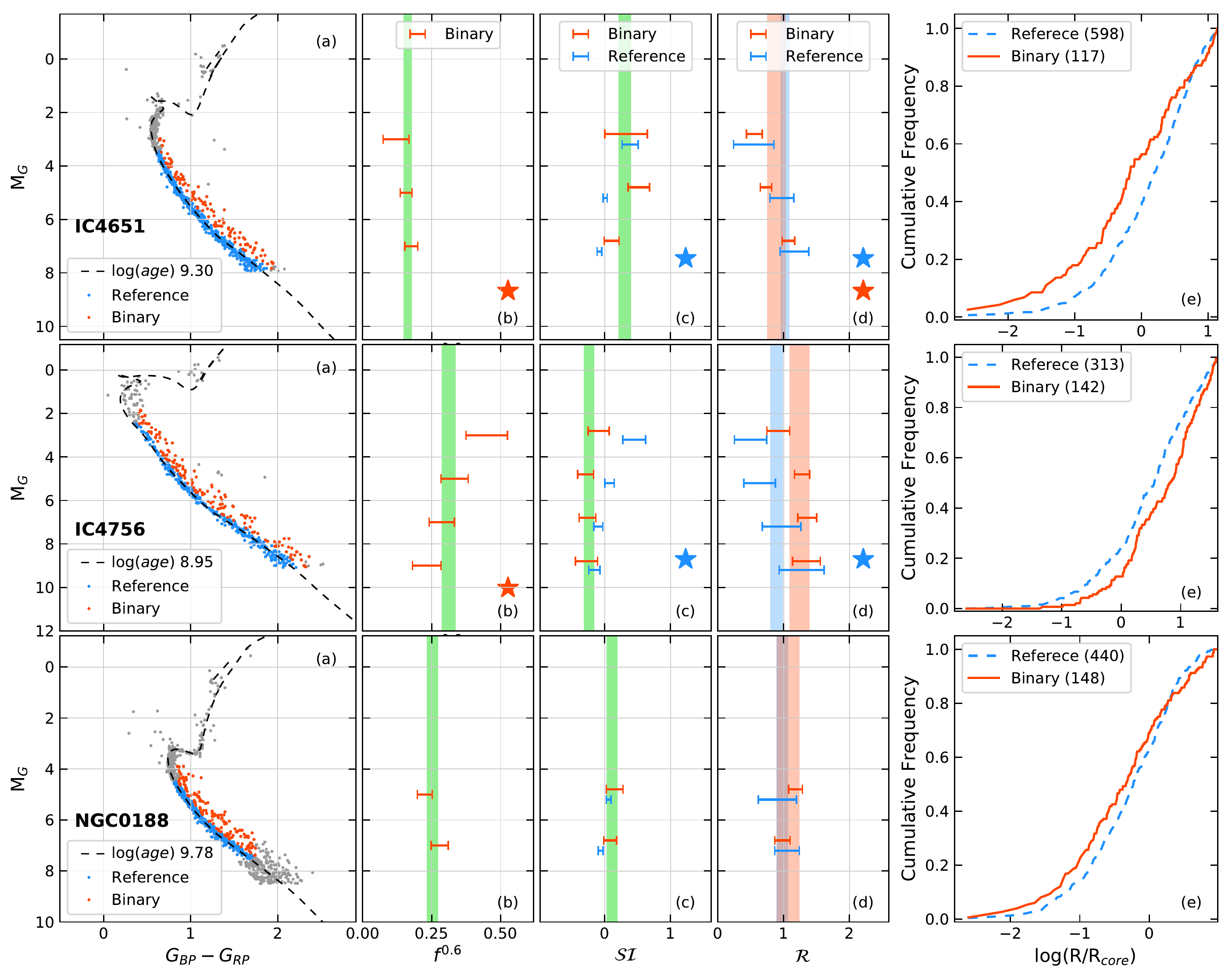}
    \caption{(a): CMDs of IC 4651, IC 4756 and NGC 188. The red dots are the HMR binaries corresponding to those in region B, and the blue dots are reference population from region A. (b): $f_{binned}^{0.6}$ variation across different magnitude bins. (c): \si$_{binned}$ variation across various magnitude bins. The red error-bars are \si$_{binned}$ for HMR binaries and the blue error-bars are for \si$^{ref}_{binned}$ (the segregation index of {the binned} reference population). The green vertical lines in (b) and (c) correspond to the mean $f^{0.6}$ and \si, while their thickness corresponds to the 1-sigma error. (d): Effective radii of reference and binary binned population are shown as blue and red error-bars. The blue and red vertical lines show effective radii of reference and HMR binary population, while their thickness corresponds to the 1-sigma error. (e): Cumulative radial profiles of reference and binary populations. {Statistically significant monotonic trends in binary or reference populations are noted by red or blue stars, respectively.}}
    \label{fig:my_label_1}
\end{figure*}

In this section, we discuss the individual clusters. The analysis for 23 clusters are presented in figures \ref{fig:my_label_1} to \ref{fig:my_label_8}. The CMD is shown in the first panel, with the fitted isochrone, {HMR binaries and reference population.}
{The cluster members that are outside the faint and bright limit mentioned in \S \ref{sec:binary_selection} are shown as gray dots.}
The second panel shows the HMR BF for various bins ($f_{binned}^{0.6}$). The third panel shows the \si\ for various bins (\si$_{binned}$ and \si$^{ref}_{binned}$), along with the mean value as a vertical line. The fourth panel shows the normalized effective radii for reference and HMR binaries of various bins, along with the mean values. The cumulative radial profiles of reference and HMR binaries are shown in the last panel. The results obtained, and the information conveyed by the plots for each cluster are discussed below.

\subsection{IC 4651} \label{sec:cluster_1}
IC 4651 is an intermediate-age OC with an age of 1.59 Gyr \citep{2019MNRAS.490.1383R}, whereas we estimate an age of $\sim$ 2 Gyr. 
The CMD (Fig. \ref{fig:my_label_1}) shows the presence of a red clump, a few BSS and two sub sub-giants.
The cluster has an average $f^{0.6}$ value of $0.16\pm0.02$; {and it shows statistically significant monotonic increase with increasing magnitude.}

From the bin-wise analysis, {we see that the $\mathcal{R}_{binned}$ is increasing as we go fainter}. Overall the cluster shows HMR binary segregation, with an average value of \si\ {$= 0.3\pm0.1$}.
The reference population is also found to be radially segregated, similar to the HMR binaries, with more segregation for the more massive population.

\subsection{IC 4756} \label{sec:cluster_2}
IC 4756 is an intermediate-age OC with an age of 955 Myr \citep{Gaia2018}.
We fitted the isochrone using lower MS and red clump/RGs in the CMD. The MSTO of the cluster is relatively broad, and the cluster is a candidate for having an extended MSTO (eMSTO). 
The CMD (Fig. \ref{fig:my_label_1}) shows a varying number of binaries along the binary sequence. The cluster CMD shows that stars redder than $q = 1$ isochrone are present across all magnitudes. Differential reddening and differential rotation \citep{2015A&A...580A..66S} could be the causes for the observed spread towards redder colors.

IC 4756 has one of the highest $f^{0.6}$ value {($0.31\pm 0.03$)} among the clusters studied here, suggesting that it is very rich in HMR binaries. However, the cluster shows declining HMR BF at fainter magnitudes.
The cluster has an overall negative \si\ value {($-0.24\pm0.08$)}, and the HMR binaries in the most massive bin show relative segregation. The effective radii suggest that the reference population is significantly segregated with respect to the HMR binaries in the three most massive bins. On an average, the HMR binaries are located relatively outward when compared to the reference population.

\subsection{NGC 188} \label{sec:cluster_3}
NGC 188 is an old OC with an age of about 5.5 Gyr \citep{Gaia2018}.
This cluster has a prominent binary sequence.  \citet{mathieu2009binary} reported BF of 76\% among BSSs in this cluster. Hard-binary population in MS, giants, and BSSs in this cluster have been studied by \citet{Geller_2012}, who reported a BF of 29\% among the MS stars and 76\% among the BSSs. {\citet{Cohen2020AJ....159...11C} reported $f^{0.5}$ of $42\pm4$\%.} 

The $f^{0.6}$ {($0.25\pm0.02$)} of this cluster is consistent throughout the MS, covered by two bins (Fig. \ref{fig:my_label_1}).
The \si\ value is positive {($0.11\pm08$)}, indicating a segregation of HMR binaries in the cluster.
Despite NGC 188 being one of the oldest  clusters analyzed, the cluster shows an average \si, with respect to the rest of the clusters. The higher mass HMR binaries $(M_{primary} > 0.8\ \Mnom)$ are tentatively more segregated than the lower mass HMR binaries (which are actually not segregated). We also note that the reference population and the HMR binaries show similar radial segregation.

\begin{figure*}
    \centering
    \includegraphics[width=0.97\textwidth]{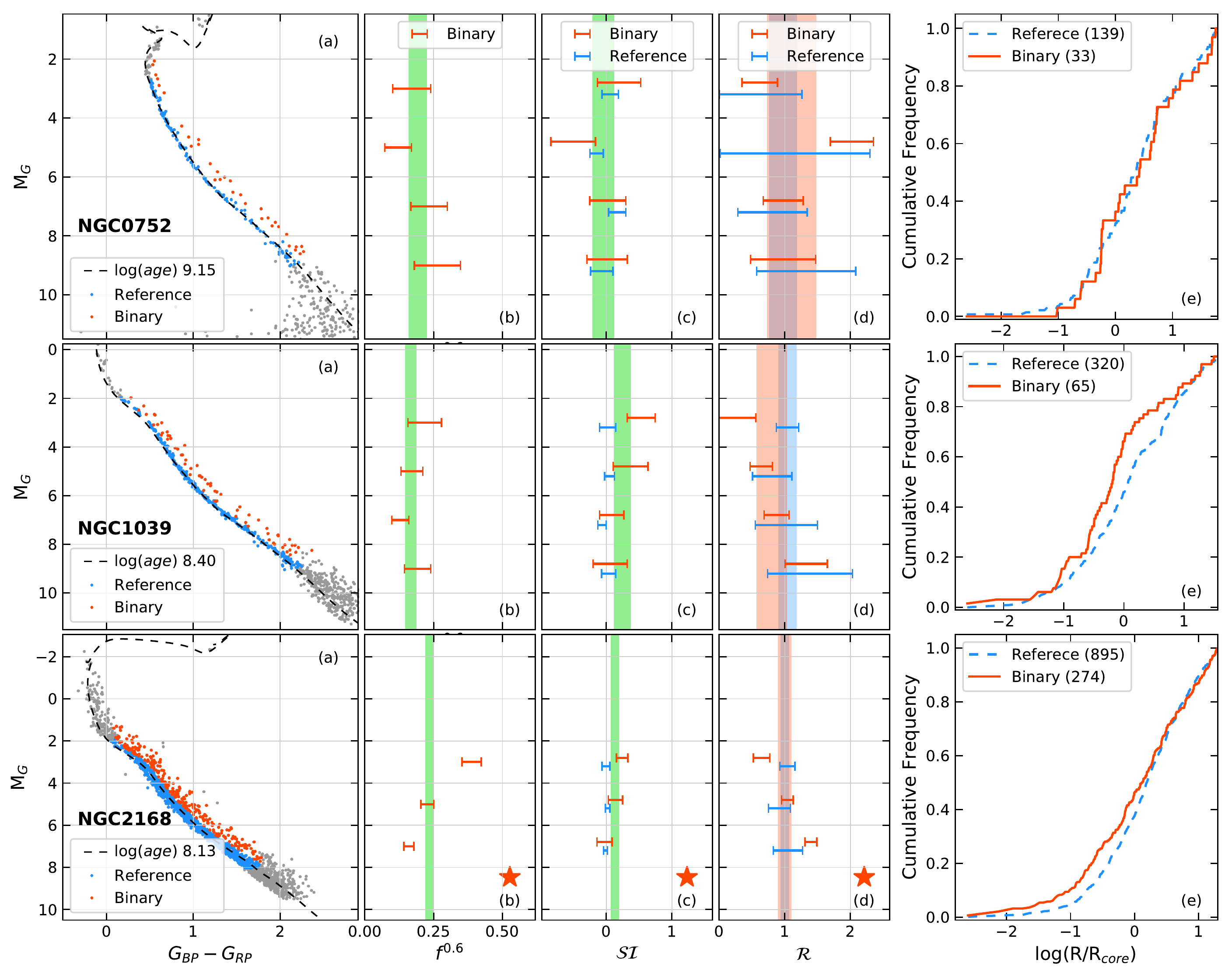}
    \caption{Plots of NGC 752, NGC 1039 and NGC 2168. All subplots are similar to Fig.~\ref{fig:my_label_1}.
    }
    \label{fig:my_label_2}
\end{figure*}

\subsection{NGC 752} \label{sec:cluster_4}
NGC 752 is an intermediate-age OC with an age of 2.0 Gyr \citep{Dinescu_1995}. \citet{Twarog_1983} investigated the bimodal distribution of stars on the MS, hypothesizing it to be a result of the presence of binaries as well as differing rotational velocities.

This cluster has a $f^{0.6}$ value of $0.19\pm0.03$ (Fig. \ref{fig:my_label_2}). The \si\ value for this cluster is {close to zero indicating no segregation}, and the \si$_{binned}$ values are statistically insignificant since each bin has very few binary members ($\sim$ 10--20). Any trends or variations cannot be inferred due to poor statistics. 

\subsection{NGC 1039} \label{sec:cluster_5}
NGC 1039 is a young OC with an age of 250 Myr \citep{2019MNRAS.490.1383R}.
The binary sequence is sparsely populated with an increasing number of members at fainter magnitudes (Fig. \ref{fig:my_label_2}). The HMR BF is found to have an average value of $0.17\pm0.02$, with the brightest and the faintest bins having {larger} fraction. $f_{binned}^{0.6}$ is found to be decreasing for fainter bins till 8 mag, with a {slight} increase in the faintest bin, $f_{8-10}^{0.6}$. 
Overall, the cluster shows binary segregation with a positive value for \si\ ($0.24\pm0.13$).
The \si$_{binned}$ parameter is observed to increase with increasing stellar mass, but the trend is not statistically significant due to large errors. The brighter bins of the HMR binaries are also radially segregated with respect to the reference population. It is to be noted that the data suffer from poor statistics in certain bins and the results are to be considered with caution. 

\begin{figure*}
    \centering
    \includegraphics[width=0.97\textwidth]{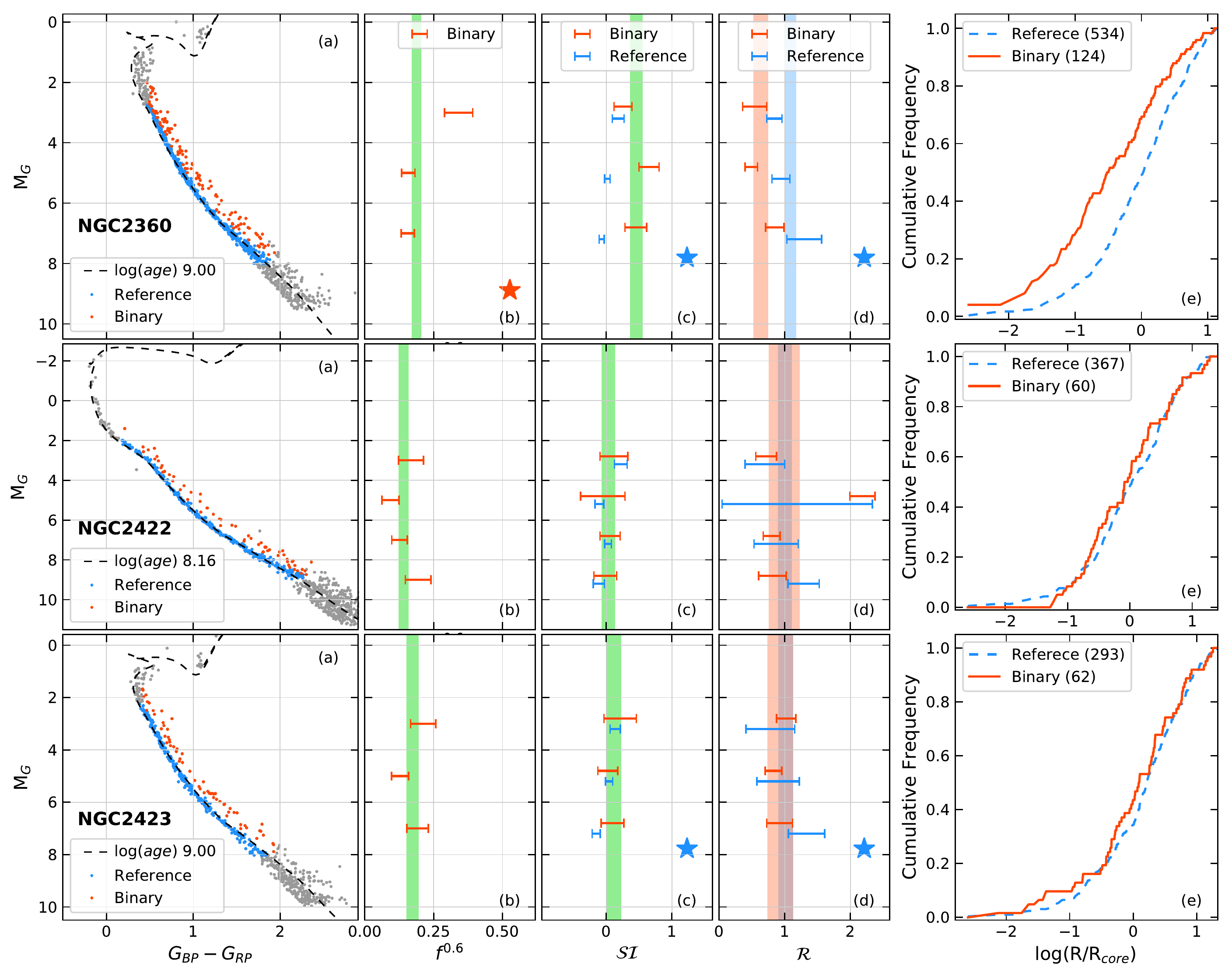}
    \caption{Plots of NGC 2360, NGC 2422 and NGC 2423. All subplots are similar to Fig.~\ref{fig:my_label_1}.}
    \label{fig:my_label_3}
\end{figure*}

\subsection{NGC 2168} \label{sec:cluster_6}
NGC 2168 (Messier 35) is a young rich OC with an age of 180 Myr \citep{Kalirai_2003}.
\citet{Sung_1999} predicted a minimum BF of $35\% \pm 5\%$. This cluster is rich in stars, and the CMD (Fig. \ref{fig:my_label_2}) shows the presence of two RGs, which were used for fitting the isochrone.
This cluster has a relatively high value of $f^{0.6}$ ($0.23\pm0.01$), and a decreasing trend of $f_{binned}^{0.6}$ was observed with increasing magnitude. 
On an average, the binaries {are segregated} (\si\ $= 0.13\pm0.06$) in this young OC. Massive HMR binaries are more centrally located compared to low mass HMR binaries. We also note that the reference population in this cluster does not show the presence of radial segregation.

\subsection{NGC 2360} \label{sec:cluster_7}
\citet{Sales_Silva_2014} and \citet{Mermilliod1989} studied the spectroscopic binaries in the cluster.
This intermediate-age OC (1.8 Gyr; \citealt{Gunes2012}) is a known eMSTO cluster \citep{Cordoni2018ApJ...869..139C}, which makes the age determination and selection of binary stars near turnoff difficult. 
This cluster has an average $f^{0.6}$ value of ($0.19\pm0.02$; Fig. \ref{fig:my_label_3}). However, the higher $f_{2-4}^{0.6}$ value ($>$ 1.3 \Mnom) could be affected by the MS spread near the turnoff.
The cluster shows a positive \si\ value across all mass ranges (with \si\ $=0.46\pm0.10$). The bin-wise analysis shows the binary segregation peaking in the middle bin (\si$_{4-6}=0.65\pm0.16$).
This is one of the rich clusters with high segregation of HMR binaries.

\subsection{NGC 2422} \label{sec:cluster_8}
NGC 2422 is a young OC with an age of about 130 Myr \citep{2019MNRAS.490.1383R,2002A&A...389..871D}. \citet{1975AJ.....80..131D} did a spectroscopic study of this cluster and predicted the presence of a large number of spectroscopic binaries in this cluster.
We observed an average $f^{0.6}$ value of $0.14\pm0.02$ and a nearly consistent $f_{binned}^{0.6}$ along the MS (Fig. \ref{fig:my_label_3}).
The radial profile shows that the HMR binary population is not segregated, with \si\ $0.03\pm0.10$. The radial distribution and radial distribution of the reference and the HMR binaries are found to be very similar.

\begin{figure*}
    \centering
    \includegraphics[width=0.97\textwidth]{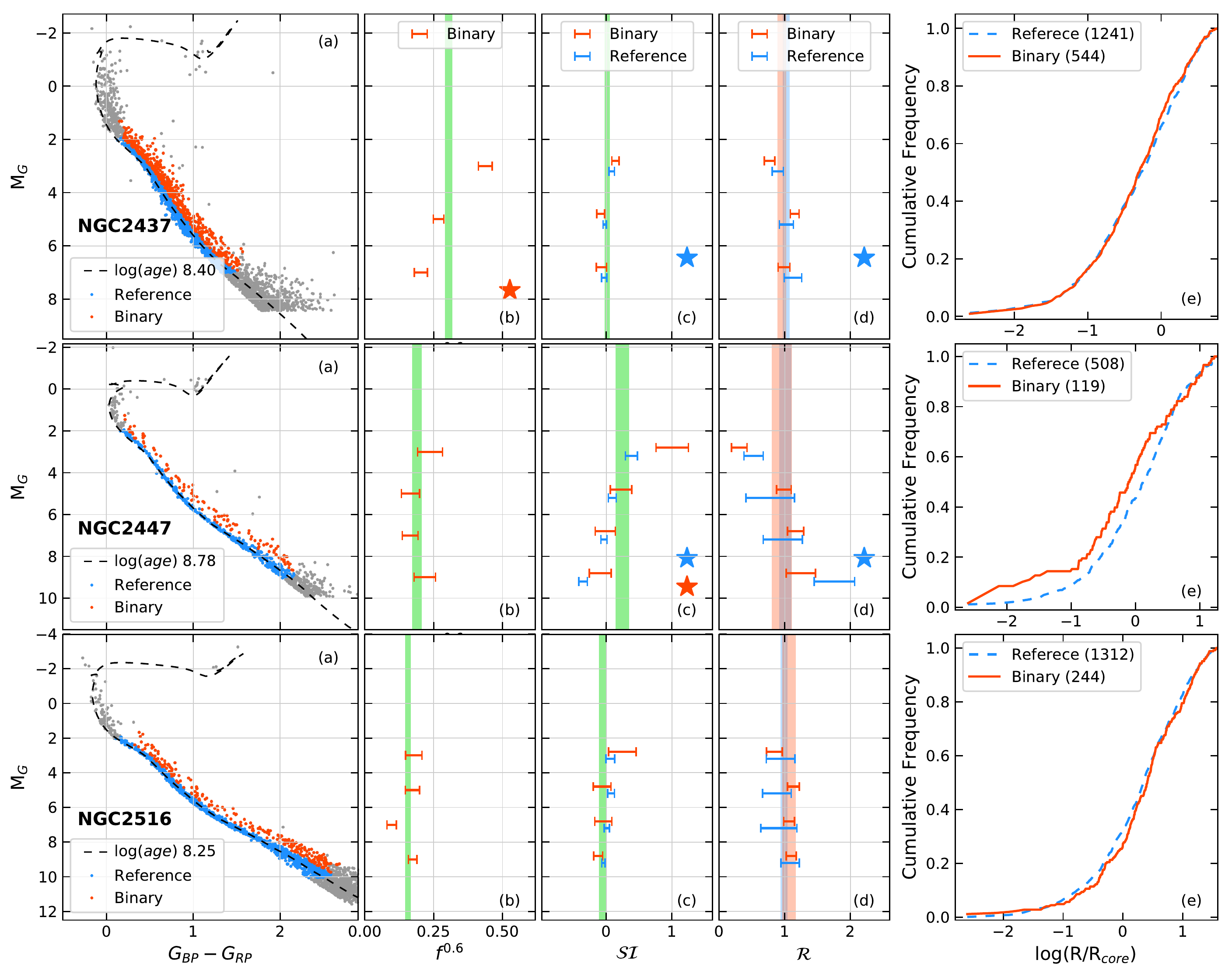}
    \caption{Plots of NGC 2437, NGC 2447 and NGC 2516. All subplots are similar to Fig.~\ref{fig:my_label_1}.}
    \label{fig:my_label_4}
\end{figure*}

\subsection{NGC 2423} \label{sec:cluster_9}
NGC 2423 is an intermediate-age OC with an age of 750 Myr \citep{Lovis_2007}.
The red clump and lower MS were used for fitting the isochrone, as we detect a spread near the MSTO {it is} a candidate eMSTO cluster (Fig. \ref{fig:my_label_3}). The $f^{0.6}$ value is found to be $0.17\pm0.02$ and is consistent throughout the MS.
The binary population seems to show negligible segregation (\si\ $= 0.11\pm0.12$). The effective radii are found to be similar between the HMR and the reference population. {The reference population shows increasing segregation with stellar mass.}

\subsection{NGC 2437} \label{sec:cluster_10}
NGC 2437 (Messier 46) is a young rich cluster, with an age of about 220 Myr \citep{Davidge_2013}.
The cluster may be physically associated with nearby planetary nebula (NGC 2438; \citealt{Bonatto_2008}). The upper MS shows a large spread, suggesting it to be a candidate eMSTO cluster (Fig. \ref{fig:my_label_4}). We observed a fairly high $f^{0.6}$ value of $0.30\pm0.01$ for this cluster and a decreasing trend of $f_{binned}^{0.6}$ along the MS. The brightest bin has a large HMR BF. 
The radial profiles indicate no binary segregation in the cluster (\si\ $=0.02\pm0.04$). Bin-wise comparison of \si$_{binned}$ also yields similar values. 

The cluster has a high BF and shows a strong correlation between primary mass and $f_{binned}^{0.6}$. This cluster is the most massive cluster in our sample. Despite it's large $f^{0.6}$, there is no binary segregation in this young cluster. The reference and the HMR binary population are not segregated.

\begin{figure*}
    \centering
    \includegraphics[width=0.97\textwidth]{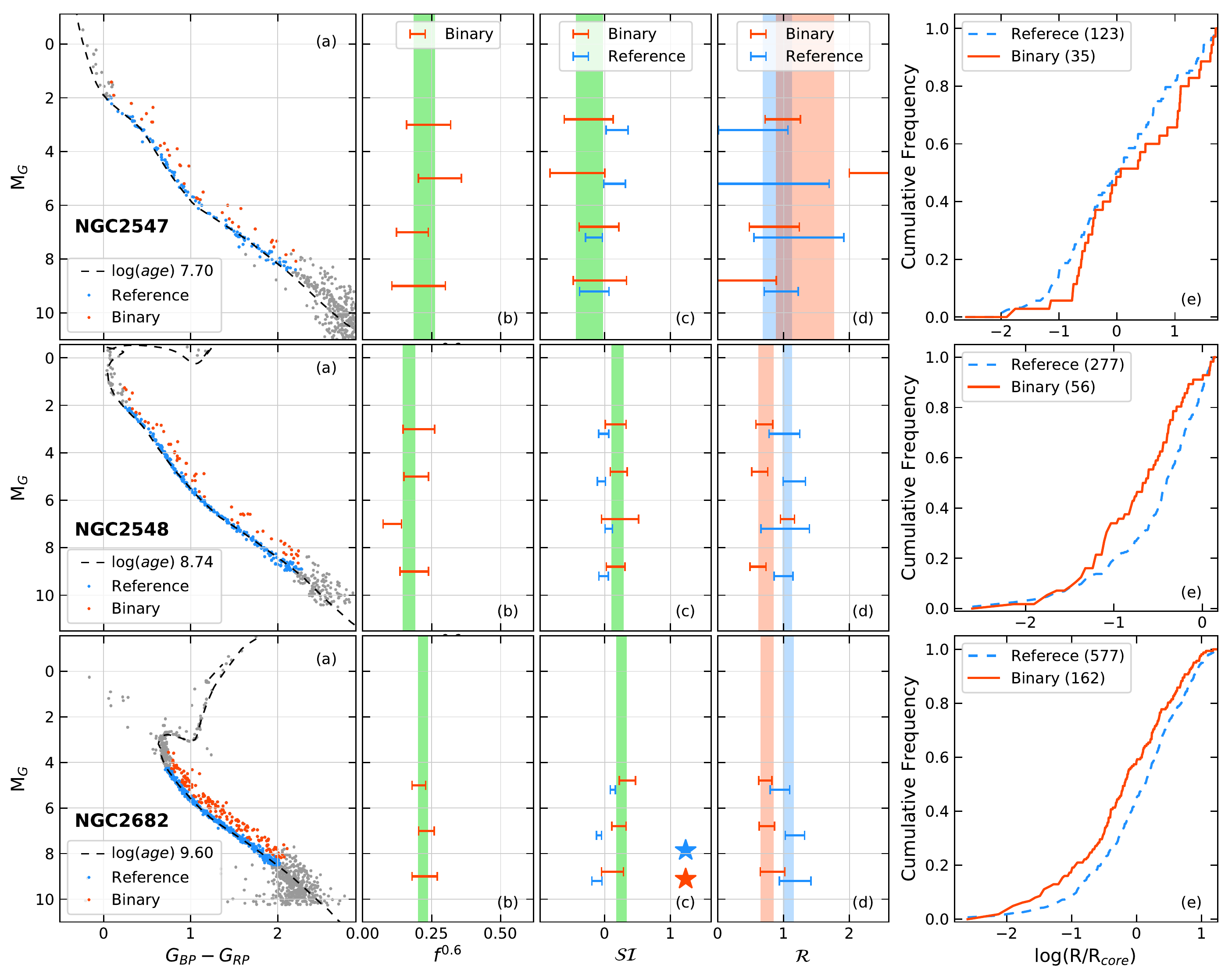}
    \caption{Plots of NGC 2547, NGC 2548 and NGC 2682. All subplots are similar to Fig.~\ref{fig:my_label_1}.}
    \label{fig:my_label_5}
\end{figure*}

\subsection{NGC 2447} \label{sec:cluster_11}
NGC 2447 (Messier 93) is an intermediate-age OC, with an age of about 450 Myr \citep{hamdani2000chemical}.
\citet{Mermilliod1989} found three spectroscopic binaries, with an RG primary and an MS secondary A-type star (yellow straggler). \citet{Eyer_2010} found 54 variable stars in this cluster, including an ellipsoidal binary. The cluster has a BSS, red clump, and prominent binary sequence (Fig. \ref{fig:my_label_4}).

This cluster has an average $f^{0.6}$ value of $0.19\pm0.02$.
From the radial profiles, the HMR binary population of the cluster seems to show central segregation (\si\ $= 0.24\pm0.10$), however \si$_{2-4}$ in the brightest bin has an anomalously high value, though it is not due to outliers. We note that both the reference and the HMR binaries in the brightest bin are significantly segregated. Both reference and binary population show increasing segregation with stellar mass.

\subsection{NGC 2516} \label{sec:cluster_12}
NGC 2516 is a young OC, with an age of 150 Myr \citep{Jeffries_2001}.
\citet{Jeffries_2001} estimated a BF of 65--85\% in the cluster, and also observed mass segregation by comparing the radial distribution of stars above and below 0.8 \Mnom. \citet{Gonz_lez_2000} and \citet{Abt_1972} studied the spectroscopic binaries in this cluster.
The CMD shows that the binary sequence is unevenly populated, with large gaps near 1.5, 4.5 and 6.5 mag (Fig. \ref{fig:my_label_4}). The sporadically populated binary sequence is typically found only in less populated clusters, but is an interesting feature in this cluster.
This cluster has an average $f^{0.6}$ value of $0.16\pm0.01$. The \si\ parameter for this cluster is $-0.05\pm0.06$, indicating no segregation of the HMR binaries. There is no mass dependent segregation of either HMR binaries or reference population. The HMR binaries are found to be distributed outward when compared to the reference population. 

\begin{figure*}
    \centering
    \includegraphics[width=0.97\textwidth]{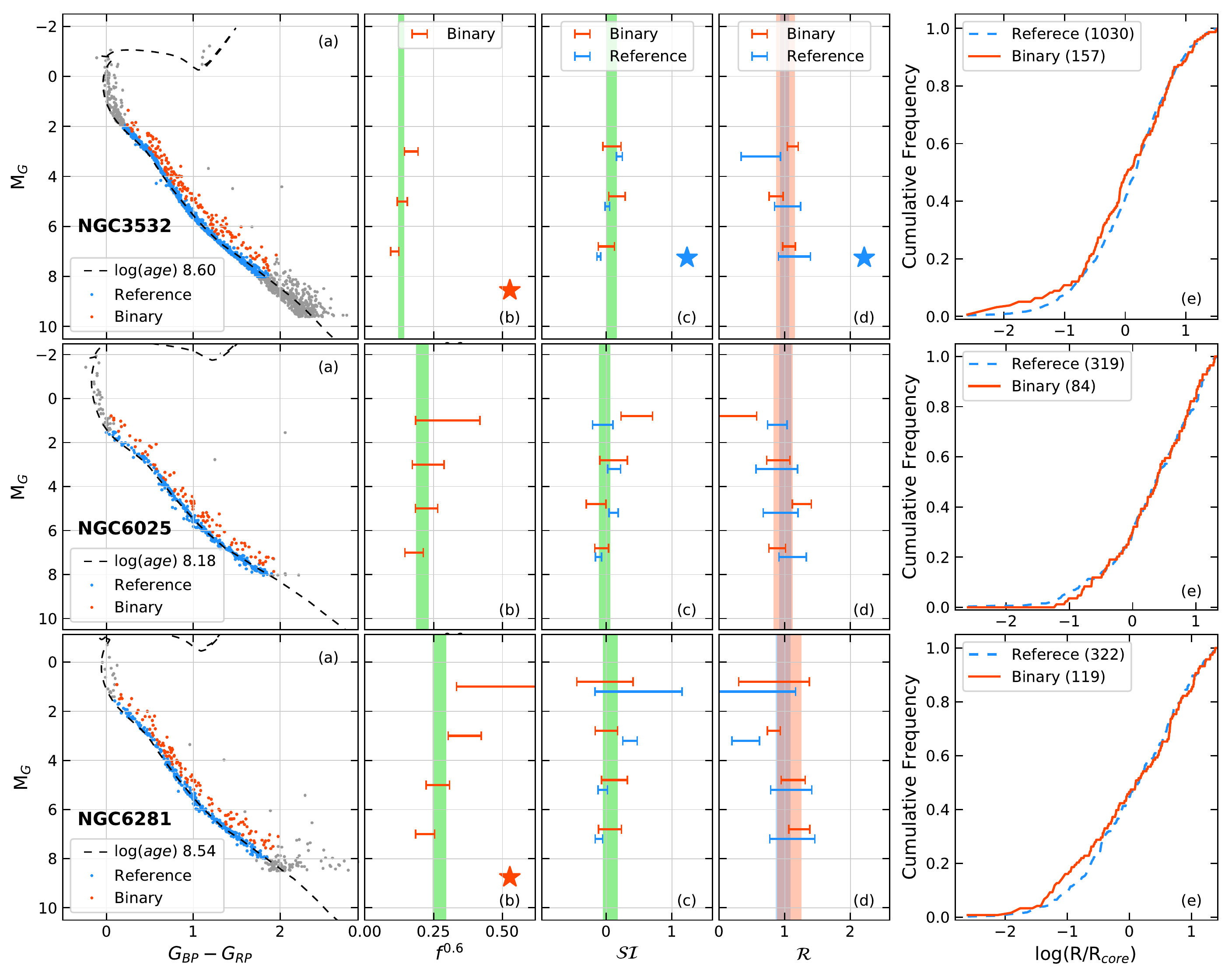}
    \caption{Plots of NGC 3532, NGC 6025 and 6281. All subplots are similar to Fig.~\ref{fig:my_label_1}.}
    \label{fig:my_label_6}
\end{figure*}

\subsection{NGC 2547} \label{sec:cluster_13}
NGC 2547 is a very young OC with an age of about 40 Myr \citep{2019MNRAS.490.1383R}.
This is the youngest cluster in our sample.
\citet{2004MNRAS.351.1401J} found out that $f^{0.5}$ is 20--35\% for M dwarfs. 
This cluster has an average $f^{0.6}$ value of $0.22\pm0.04$ and we see nearly consistent trend along the MS, though with large error bars due to fewer stars (Fig. \ref{fig:my_label_5}).
The small number of stars and young age lead to a negative value of \si\ ($= -0.23\pm0.21$). The massive reference population is slightly centrally concentrated.

\subsection{NGC 2548} \label{sec:cluster_14}
NGC 2548 is an intermediate-age OC, with an age of 420 Myr \citep{Sun_2020}.
\citet{Sun_2020} used spectroscopic data along with \textit{Gaia} data to determine a minimum BF of 11--21\%. 
The CMD shows a few red giant stars and a gap in binary sequence near 5--7 mag. 
Fitting the isochrone to giants and turnoff simultaneously was challenging, hence the estimated age of the cluster is an upper limit (Fig. \ref{fig:my_label_5}).
The average $f^{0.6}$ value of the cluster is ($0.17\pm0.02$) and shows a nearly consistent $f_{binned}^{0.6}$.
The \si\ value for this cluster ($0.19\pm0.09$) indicates binary segregation. However, the low number of binary candidates (56) and uneven binary sequence means that the bin-wise analysis could be unreliable. Overall, there is an indication that the HMR binaries are segregated in this cluster, but there is no segregation in the binned reference population.

\begin{figure*}
    \centering
    \includegraphics[width=0.97\textwidth]{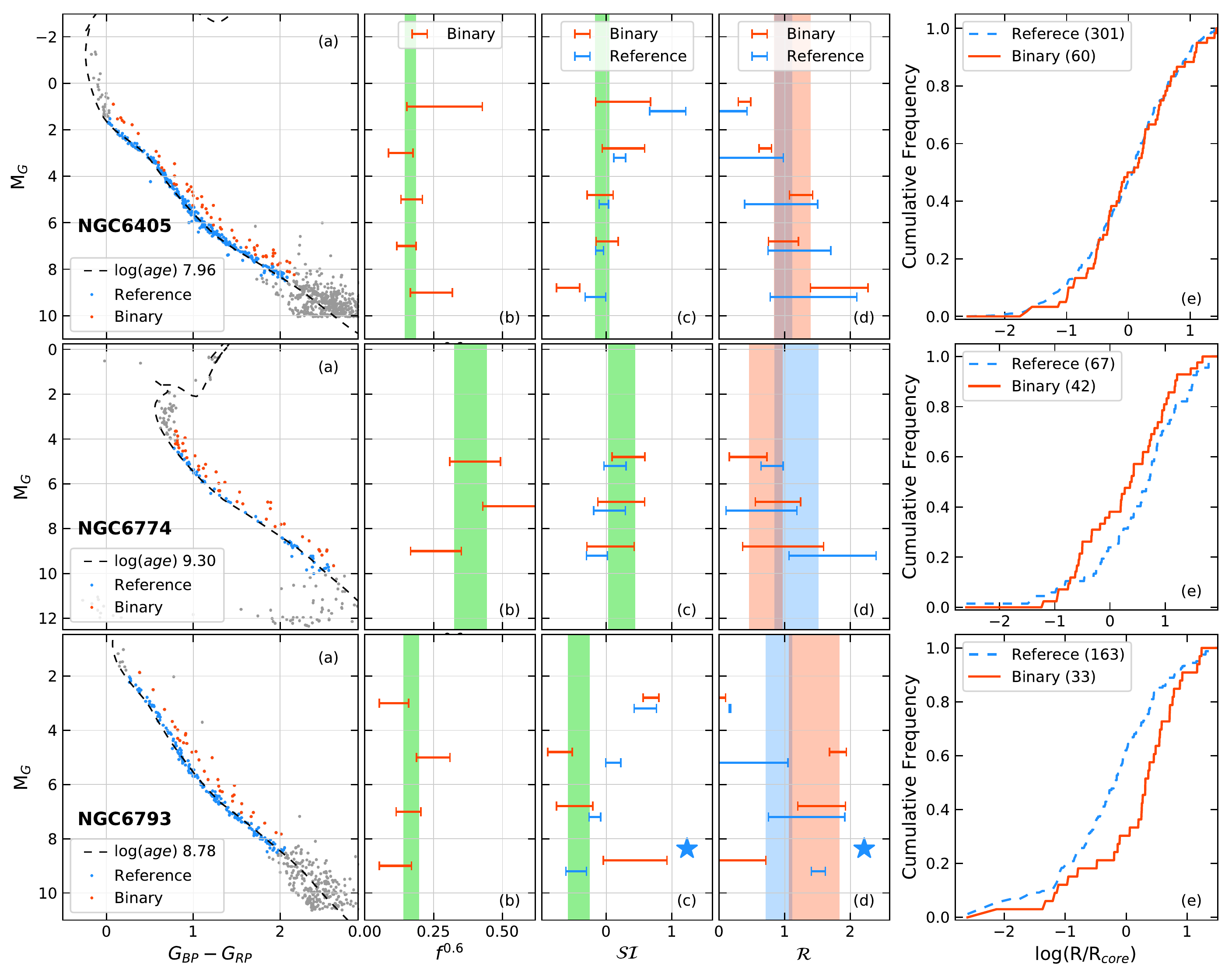}
    \caption{Plots of NGC 6405, NGC 6774 and NGC 6793. All subplots are similar to Fig.~\ref{fig:my_label_1}.}
    \label{fig:my_label_7}
\end{figure*}

\subsection{NGC 2682} \label{sec:cluster_15}
NGC 2682 (Messier 67) is an old OC with age of 4 Gyr and solar metallicity \citep{Montgomery1993}. The cluster has been studied in all wavelengths \citep{Mathieu1986, Belloni1998, Sindhu2018} due to its closeness (900 pc, \citealt{Stello2016}) and richness. The cluster contains MS stars, RGs, BSSs and has a well-defined binary sequence. \citet{Geller2015} studied the cluster and found 23\% stars are spectroscopic binaries (in their magnitude limited sample). {\citet{Geller2021AJ....161..190G} analyzed the sample of 0.7--0.13 \Mnom\ stars and found overall BF of 34$\pm3$\% which increases to 70$\pm17$\% in central region. They found the mass-ratio distribution to be uniform.}

The OC has a prominent binary sequence and HMR BF is consistent ($f^{0.6} = 0.22\pm0.02$) across the MS (Fig. \ref{fig:my_label_5}).
We found an above average segregation of binaries for NGC 2682, with \si\ $= 0.25\pm0.08$. The \si\ value increases sharply with increase in stellar mass. Due to the large number of stars in each bin, this is a statistically significant trend. The higher mass binary systems $(M_{primary} > 1.2\ \Mnom)$ are more centrally segregated than the lower mass systems. The HMR binaries are significantly segregated when compared to the reference population. Also, the high-mass reference population is centrally concentrated.

\subsection{NGC 3532} \label{sec:cluster_16}
NGC 3532 is a young age OC, with an age of about 300 Myr \citep{Dobbie_2009}.
\citet{Clem_2011} used deep wide-field CCD photometry to obtain various cluster parameters, estimating a lower limit on the BF ($27\% \pm 5\%$). \citet{Gonz_lez_2002} studied the spectroscopic binaries in this cluster. The cluster has an eMSTO \citep{Cordoni2018ApJ...869..139C}, which affects the isochrone fitting and estimation of binary stars in the brightest bin. 
\citet{2020arXiv200804684L} recently modeled the unresolved binary population and obtained a $f^{0.2} = 0.27$.
It has the {one of the} lowest $f^{0.6}$ ($0.13\pm0.01$) out of all clusters, consistent with the previous study. 
{Minimal} binary segregation can be observed from the radial profile of this cluster (Fig. \ref{fig:my_label_6}), with a \si\ value of $0.08\pm0.08$. The reference population shows an increasing \si$_{binned}^{ref}$ with mass.

\begin{figure*}
    \centering
    \includegraphics[width=0.97\textwidth]{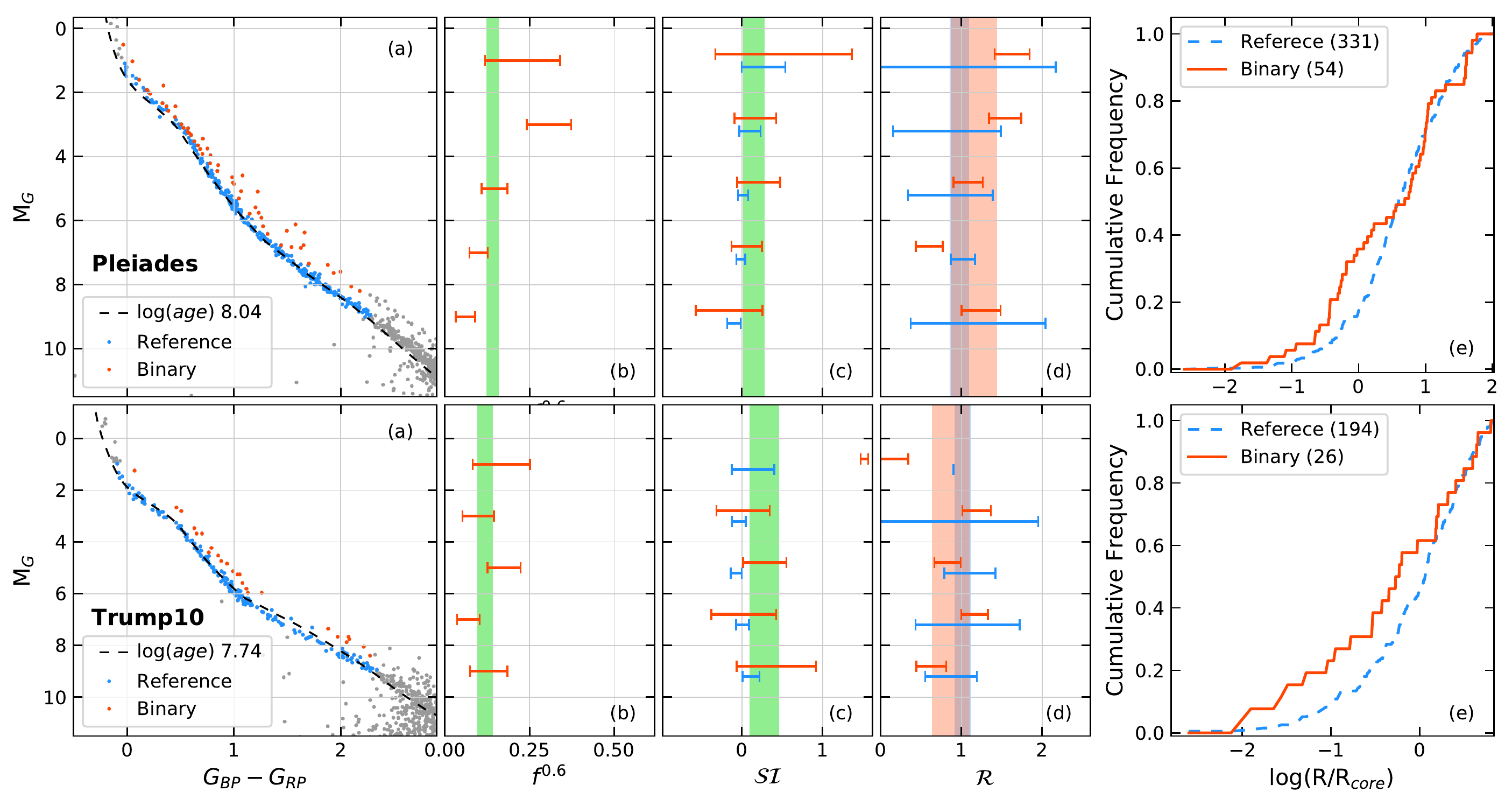}
    \caption{Plots of Pleiades and Trump 10. All subplots are similar to Fig.~\ref{fig:my_label_1}.}
    \label{fig:my_label_8}
\end{figure*}

\subsection{NGC 6025} \label{sec:cluster_17}
NGC 6025 is a young OC with an age of 151 Myr \citep{2019MNRAS.490.1383R}.
\citet{2006RMxAC..26Q.171G} calculated BF of 40--50\% for spectroscopic binaries. The average value of $f^{0.6}$ is $0.21\pm0.02$. The binaries are distributed unevenly with gaps (Fig. \ref{fig:my_label_6}).
The value of \si\ for this cluster is {near zero ($-0.03\pm0.09$)}, though \si$_{2-4}$ has a positive value. The radial profile analysis suggests absence of radial binary segregation due to younger age of the cluster.

\subsection{NGC 6281} \label{sec:cluster_18}
NGC 6281 is a young OC with an age of 302 Myr \citep{2019MNRAS.490.1383R}.
The cluster has two giant stars and a significant number of binary stars with an average $f^{0.6}$ value of $0.27\pm0.02$ (Fig. \ref{fig:my_label_6}). 
\si\ is slightly positive ($0.06\pm0.11$).
The high $f^{0.6}_{2-4}$ in the brightest bin is likely due to the narrow binary sequence in the bin. This 300 Myr old cluster shows segregation in the high-mass reference population.

\subsection{NGC 6405} \label{sec:cluster_19}
NGC 6405 (Messier 6) is a young OC, with an age of 71 Myr \citep{Paunzen_2006}.
We find the average $f^{0.6}$ value to be $0.17\pm0.02$ and nearly consistent between 2 to 8 magnitude (Fig. \ref{fig:my_label_7}). $f_{binned}^{0.6}$ increases for brightest and faintest bins, but mean value stays within errors.
The cluster shows no HMR binary segregation with \si\ $=-0.6\pm0.11$.
Low stellar counts cause the larger error bars in top and bottom bins.
This lack of binary segregation is supported by the young age of the cluster.
{Both binary and reference segregation increase with stellar mass, though the trends are only monotonous with p-value of 0.03 (more than the cutoff of 0.003 elsewhere).}

\subsection{NGC 6774} \label{sec:cluster_20}
NGC 6774 (Ruprecht 147) is an intermediate-age OC with an age of 2 Gyr \citep{2019MNRAS.490.1383R}.
\citet{Torres_2018} studied eclipsing binaries in this cluster. 
\citet{Yeh_2019} demonstrated that NGC 6774 is losing stars at a fast pace and dissolving into the general Galactic disk.
NGC 6774 has one of the highest $f^{0.6}$ value of $0.38\pm0.06$ among our sample (Fig. \ref{fig:my_label_7}).  
Due to small number of stars, $f_{binned}^{0.6}$ and \si$_{binned}$ values are not significant (Fig. \ref{fig:my_label_7}). However, the large BF and binary segregation (\si $=0.24\pm0.21$) indicates that single stars have been preferentially removed. The cluster therefore contains a majority of binaries and a very interesting target for dynamical studies. 

\subsection{NGC 6793} \label{sec:cluster_21}
NGC 6793 is an intermediate-age OC with an age of $\sim$ 600 Myr \citep{2019MNRAS.490.1383R}.
It has an average $f^{0.6}$ value of $0.17\pm0.03$ and is consistent throughout the MS. The cluster {has the lowest value of \si\ ($=-0.42\pm0.17$) among the set of clusters}.
This is a poor cluster with fewer members, and this leads to large errors in $f^{0.6}$ (Fig. \ref{fig:my_label_7}). {The reference population shows increasing segregation with stellar mass,} whereas the HMR binary population appears to have a radially outward distribution.

\subsection{Pleiades} \label{sec:cluster_22}
Pleiades (Messier 45) is a young OC with an age of 136--150 Myr \citep{Abramson_2018, Mazzei1989}.
\citet{Mermilliod1992} studied the spectroscopic binaries in the cluster. 
The \textit{Gaia} CMD of this cluster shows the presence of a large population of low luminosity stars (Fig. \ref{fig:my_label_8}). The average value of $f^{0.6}$ is $0.14\pm0.02$, with $f_{binned}^{0.6}$ showing an {increasing trend with stellar mass}.
The cluster radial profiles indicate an average binary segregation (\si\ $= 0.14\pm0.13$). Looking at the bin-wise analysis, the value of the \si$_{binned}$ parameter remains roughly constant across the magnitude bins. The reference population becomes more segregated with increasing stellar mass {, although with a p-value of only 0.04 in the Spearman test}.

\subsection{Trumpler 10} \label{sec:cluster_23}
Trumpler 10 is a young OC with an age of 60 Myr \citep{2019MNRAS.490.1383R}.
The value of $f^{0.6}$ ($0.12\pm0.02$) is one of the lowest values in our sample.
The average $f_{binned}^{0.6}$ is consistent along the magnitude bins (Fig. \ref{fig:my_label_8}).
The \si\ value for this cluster is above the average ($= 0.28\pm0.18$).
This small and very young cluster shows one of the highest binary segregation. One has to be care-full while interpreting the results, as this is a poor cluster with less number of stars.


\section{Discussion} \label{sec:discussion}

\begin{figure}
    \centering
    \includegraphics[width=0.48\textwidth]{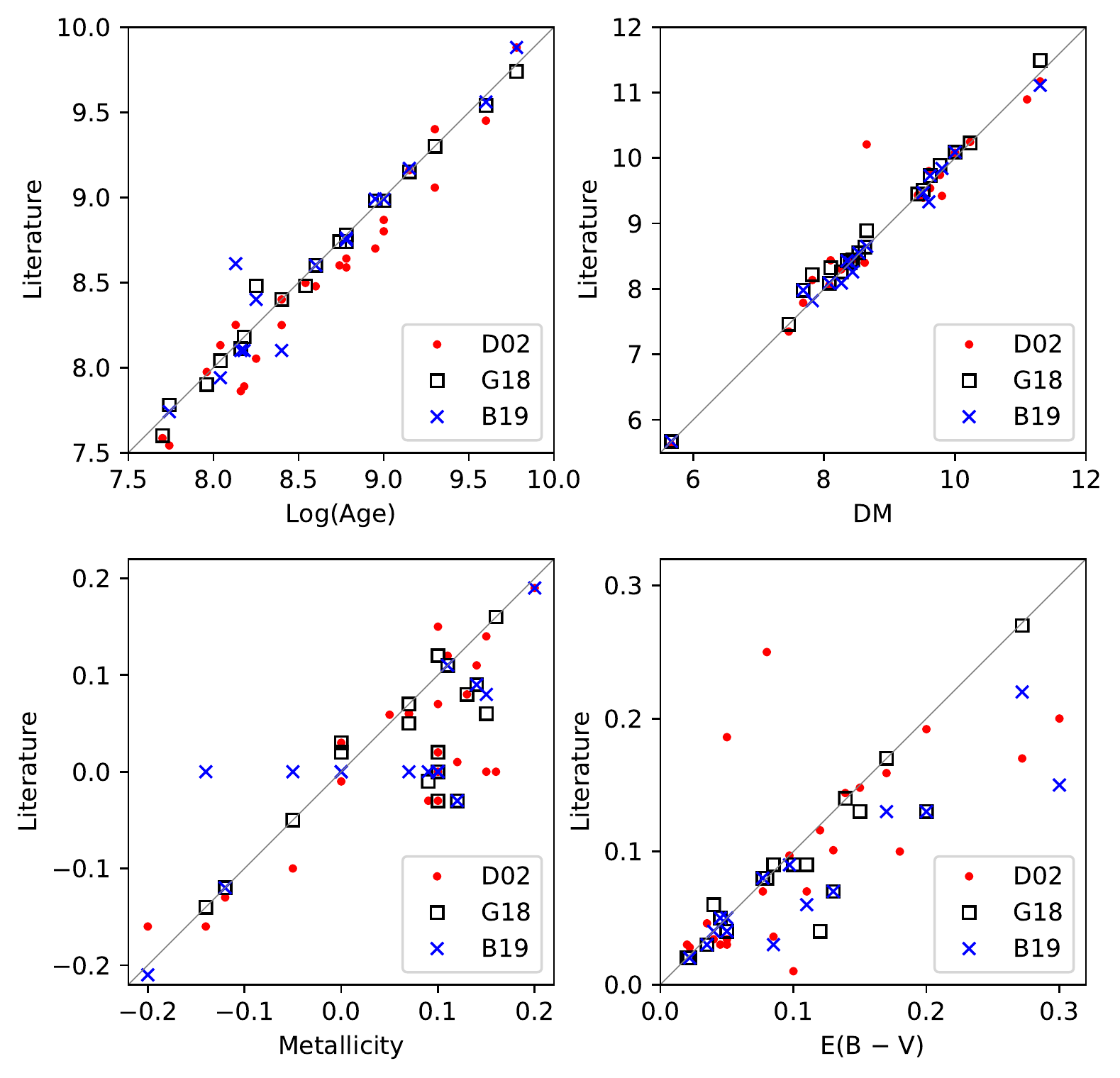}
    \caption{The panels show the comparison of isochrone fitting parameters with literature. D02: \citet{2002A&A...389..871D}; G18: \citet{Gaia2018}; B19: \citet{Bossini_2019A&A...623A.108B}}
    \label{fig:cluster_properties}
\end{figure}

\begin{figure*}
    \centering
    \includegraphics[width = 0.98\textwidth]{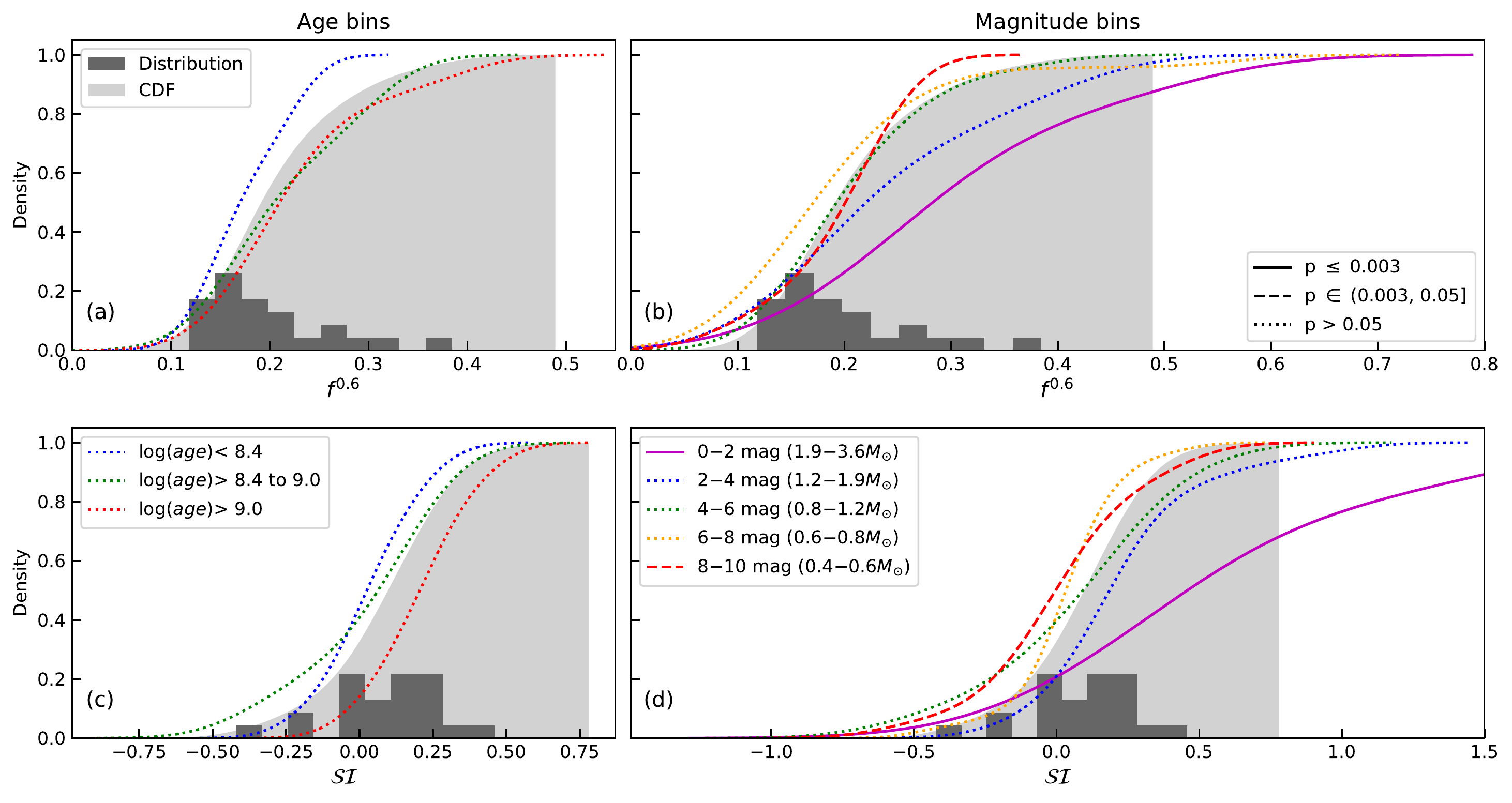}
    \caption{{The cumulative distribution functions of $f^{0.6}$ and \si\ parameters for the cluster sample. 
    The left panel has three age ranges: older than a Gyr, between 250 Myr to 1 Gyr and younger than 250 Myr.
    The right panels show the distribution of magnitude binned $f_{binned}^{0.6}$ and \si$_{binned}$ for 0 to 2 mag, 2 to 4 mag, 4 to 6 mag, 6 to 8 mag and 8 to 10 mag regions of the absolute CMD. The CDFs and scaled-down distributions of $f^{0.6}$ and \si\ are under-plotted in gray color for comparison. The curves {are shown as solid lines (p-value $<0.003$), dashed lines (p-value $\in(0.003,0.05)$) and dotted lines (p-value $>$ 0.05) according to their statistical difference from the overall distribution using KS test.}}}
    \label{fig:hist_BFBplus}
\end{figure*}

\begin{figure*}
    \centering
    \includegraphics[scale = 0.56]{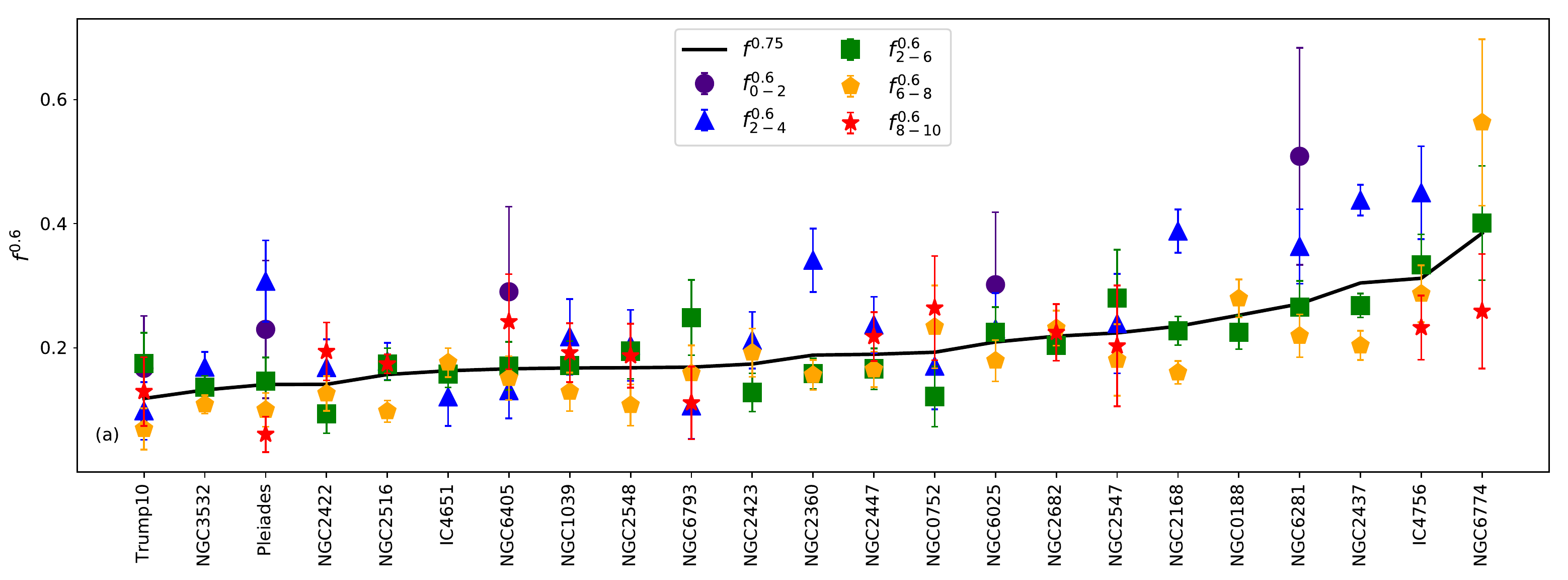}
    \includegraphics[scale = 0.56]{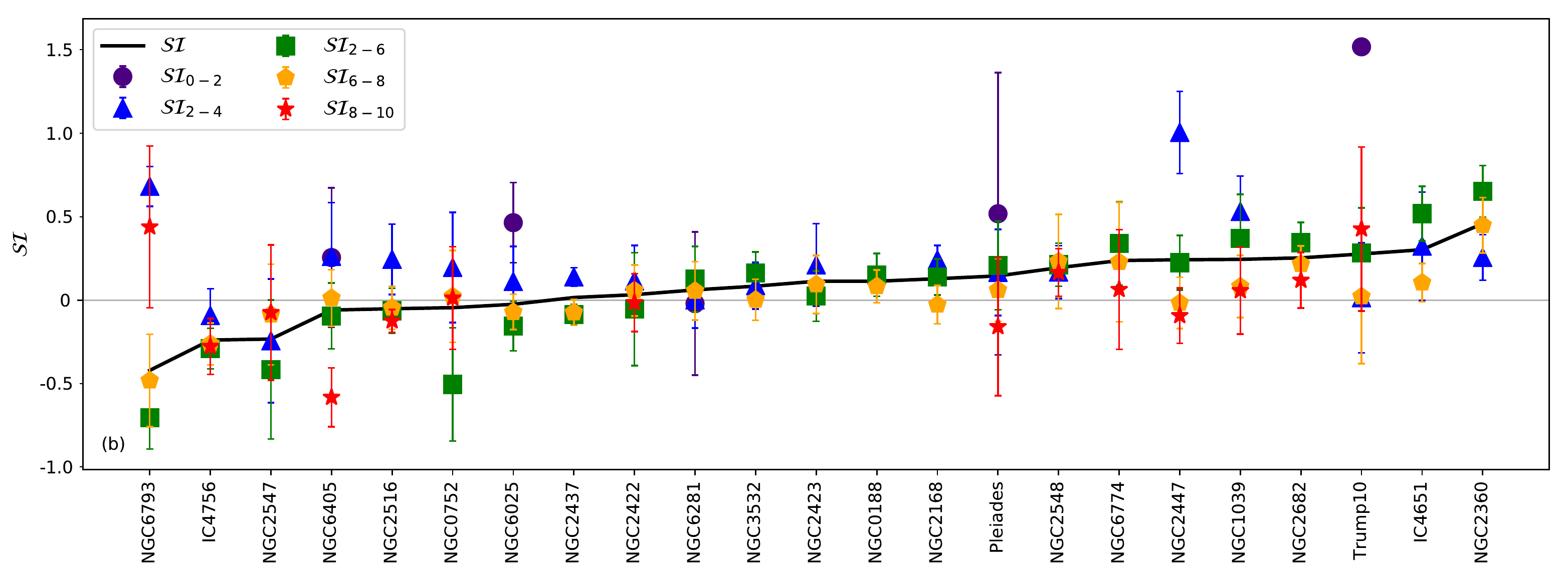}
    \caption{(a) Distribution of $f^{0.6}$ and $f_{binned}^{0.6}$ for all clusters. (b) Distribution of \si\ and \si$_{binned}$ for all clusters.
    The clusters are arranged in increasing order of $f^{0.6}$ and \si\ in (a) and (b) respectively.}  
    \label{fig:BF_comparison}
\end{figure*}

Table~\ref{tab:master_table} lists the isochrone fitting parameters (log($age$), DM, metallicity and E(B$-$V)), core radii, {number of relaxation periods passed}, total number of members, $f^{0.6}$ and \si\ for all 23 OCs. Fig.~\ref{fig:cluster_properties} show the comparison of parameters derived from our isochrone fits with the literature. 
It is to be noted that fitting the isochrones to clusters with extended MSTO (eMSTO) may not be accurate due to undefined MSTO, leading to uncertainties in log($age$) and metallicity. In such cases, we focused on fitting the MS, which was useful for selecting the binary stars. The bin-wise estimations of $f^{0.6}$, \si\ and the effective Radii are tabulated in Table \ref{tab:binned_BF}. 
The {cumulative} distributions of the average and bin-wise estimations of $f^{0.6}$ and \si\ for the sample studied are shown in Fig.~\ref{fig:hist_BFBplus}. Fig.~\ref{fig:BF_comparison} shows the average as well as bin-wise estimations of $f^{0.6}$ and \si\ along with the cluster names. 

\subsection{Improved cluster parameters} \label{sec:discussion_1}
Although the aim of this work is not to improve the cluster parameters, refitting of the isochrones was necessary to identify the binary stars, resulting in the estimations tabulated in Table~\ref{tab:master_table}.
We have compared the fitting parameters with \citet{2002A&A...389..871D}, \citet{Gaia2018} and \citet{Bossini_2019A&A...623A.108B} (see Fig.~\ref{fig:cluster_properties}).
Age and distance estimates are quite similar to the literature, while the metallicity and E(B$-$V) estimates of \citet{2002A&A...389..871D} show a larger deviation. Inspection of the individual CMDs overlaid with isochrones suggests that the fits are satisfactory, providing confidence in our estimation of these parameters.

The isochrone fits were used to calculate the absolute magnitudes of the stars. The distance modulus (DM) correction does not affect the overall $f^{0.6}$ and \si\ values. However, some stars can change their respective magnitude bin. It will induce small errors in the actual spectral class of the stars. Nevertheless, the general results for the magnitude bins remain valid.

\subsection{HMR Binary fraction} \label{sec:discussion_2}
\textit{Distribution of $f^{0.6}$:}
The distribution of the mean value of $f^{0.6}$ across clusters is shown in Fig.~\ref{fig:hist_BFBplus} (a) as dark gray histogram. The distribution has a prominent peak in the 0.12--0.20 bins, with a median value of $0.19\pm0.06$. The typical values of $f^{0.6}$ are between 0.10--0.30, with one cluster at $0.38\pm0.06$. 56\% of the clusters studied have $<$20\% HMR binaries.

Fig.~\ref{fig:hist_BFBplus} (a) also shows the cumulative distribution function (CDF) of $f^{0.6}$ values for all clusters as filled gray region. We compared the CDFs of three age groups (log($age$) $<8.4$, 8.4 to 9 and $>9$). In general, there is a tendency to have relatively more HMR binaries in older clusters compared to the younger clusters. However, KS test results show that these differences are not statistically significant. Increasing the cluster sample can confirm if the trend is significant.

\textit{Variation of $f_{binned}^{0.6}$ with magnitude/mass:}
Table \ref{tab:binned_BF} lists the bin-wise values for the cluster sample. 
The 0--2, 2--4, 4--6, 6--8 and 8--10 magnitude ranges sample 21\%, 87\%, 100\%, 100\% and 61\% clusters respectively. Fig.~\ref{fig:hist_BFBplus} (b) shows the CDFs of the BF in five bins. Overall, brighter bins seem to have larger BFs compared to fainter bins. {But only 0--2 mag bin is significantly different (3$\sigma$) while 8--10 mag bin differs from the overall distribution with 2$\sigma$ difference.}
We note that the HMR BF over the three most sampled bins show a progressive reduction between 1.9--0.6 \Mnom. 
IC 4756, NGC 2168, NGC 2360, NGC 2437, NGC 3532 and NGC 6281 show statistically significant increase in BF with stellar mass. On the other hand, IC 4651 (and to some extent NGC 188) shows a reverse trend.
This result can also be seen from Fig.~\ref{fig:BF_comparison} (a), where the blue triangles are found to be higher than the average.

\begin{table}
    \centering
    \begin{tabular}{ccc ccc}
    \toprule
Mag	&	Binary	&	Ref.	&	$f^{0.6}_{binned}$	&	$\delta\,f^{0.6}_{binned}$	&	$f^{total}_{binned}$			\\
bin	&	stars	&	stars	&		&		&				\\ \hline
0--2 & 18 & 47 & 0.28 & 0.07 & 0.41 -- 0.88 \\ 
2--4 & 677 & 1668 & 0.29 & 0.01 & 0.55 -- 0.75 \\ 
4--6 & 940 & 3664 & 0.20 & 0.01 & 0.39 -- 0.53 \\ 
6--8 & 828 & 3984 & 0.17 & 0.01 & 0.33 -- 0.45 \\ 
8--10 & 297 & 1301 & 0.19 & 0.01 & 0.26 -- 0.49 \\ \hline
Total & 2760 & 10664 & 0.206 & 0.004 & 0.30 -- 0.53 \\  
\bottomrule
    \end{tabular}
    \caption{{The average BF from all clusters for different magnitude bins. The errors in the fifth column are Poisson errors. The last column shows the expected range of total BF assuming a multiplicative factor of 1.5 to 2.5 (i.e. $1.5(x-\delta x)$ to $2.5(x+\delta x)$).}}
    \label{tab:average_BF}
\end{table}

Based on the reference and binary stars in all clusters, we calculated the average $f^{0.6}$ for different magnitude bins and the same are shown in Table~\ref{tab:average_BF}. As mentioned above, not all magnitude bins are sampled for all clusters, but all bins (except 0--2 mag bin) have a significant number of stars. Due to the selection criteria, the 8--10 mag bin values are not complete. More precise and deeper data is required to get accurate BFs in the faintest bin.

\subsection{Total Binary fraction and comparison with literature}
\label{sec:discussion_4}

We estimated the mean $f^{0.6}$ value as 0.20$\pm$0.06.
The total BF ($f^{total}$) can be estimated provided one knows the $q$ distribution ($f(q)$). If we assume a flat distribution, then the total BF becomes $2.5\times f^{0.6}$. If this is the case in all clusters, then the $f^{total}$ range becomes 0.3 to 0.96. The total BF is unlikely to be 96\%, therefore it is quite likely that the $f(q)$ is not a flat distribution, {at least in some clusters}.
\citet{Fisher2005} estimated an $f(q)$ {of solar neighborhood binaries} with a peak near $q$ = 0.9--1.0.
In their Table~3, the total binaries are 369, while binaries with $q > 0.6$ are 190. Hence, total BF can be calculated as $f^{total} \approx 2 \times f^{0.6}$.
However, such total BF strongly depends on the empirical $f(q)$. Moreover, $f(q)$ may not be the same for all clusters. Hence, we only used $f^{0.6}$ in this work for analysis.

\citet{Thompson2021} used SED fitting techniques to identify the binary stars in eight OCs, and three clusters are common between our sample. The BF obtained by \citeauthor{Thompson2021} is approximately {1.7--2.6} $\times$ $f^{0.6}$, which is close to the multiple obtained from \citet{Fisher2005} {and a possible flat distribution}. 
Comparison of \citet{2020arXiv200804684L} and our results for NGC 3532 shows that $f^{0.2} \sim 2.1 \times f^{0.6}$. 
Recently, \citet{Sun2021} derived $f^{total}$ of IC 4756 to be 0.48 using stars in absolute magnitude range of 2.8--4.8 mag. The value is similar to $f_{2-4}^{0.6}\ (=0.45\pm0.07)$ or {1.5}$\times f^{0.6}$. As we see that this cluster has a very strange distribution of binary stars, the factor estimated here may not be similar to that found in other clusters.
If we assume a multiplicative factor of {1.5--2.5} to obtain the total BF, then the {mean BF} ($f^{total}$) will be 30\%--50\%, which is not very different from the BF known from the literature.

{We compared the BF values of past literature to our study:}
\begin{itemize}
    \item \citet{Cohen2020AJ....159...11C} found $f^{0.5}$ to be 42$\pm$4\% using multi-band CMDs of NGC 188, whereas $f^{0.6}$ is $25\pm2$\%. Contrary to \citet{Fisher2005}, \citeauthor{Cohen2020AJ....159...11C} found that the $f(q)$ distribution decreases from $q=0.5$ to 1.0.
    \item \citet{Sung_1999} {and \citet{Leiner2015AJ....150...10L}} found a minimum BF of $35\pm5\%$ {and $24\pm3\%$ respectively} in NGC 2168, whereas we estimate $f^{0.6}$ of $0.23\pm0.01$ and total BF of 34--57\%. 
    \item \citet{Jeffries_2001} suggested a BF of 65--85\% for NGC 2516, and we estimate {a total BF of 24--39\%}, which is on the lower side.
    \item \citet{2004MNRAS.351.1401J} found $f^{0.5}$ of 20--35\% for M dwarfs in NGC 2547. On the other hand, we estimate an HMR BF for M stars in this cluster as $0.20\pm0.09$, which amounts to a total BF of 30--50\%. 
    \item \citet{Sun_2020} used spectroscopic data along with \textit{Gaia} data to determine a minimum BF of 11--21\% in NGC 2548, and we estimate an HMR BF of $0.17\pm0.02$.
    \item \citet{Geller2015} {and \citet{Geller2021AJ....161..190G} found 23--34\%} of the stars are spectroscopic binaries in NGC 2682, where we find a comparable HMR BF to be $0.22\pm0.02$, considering the fact that HMR binaries are potential candidates to be detected as spectroscopic binaries.
\end{itemize}
{The values of total BF depend on the exact $f(q)$ distribution for photometric studies and on the orbits and timeline of followups in case of spectroscopic studies. Most of the above literature studies have accounted for these biases, and we have quoted the total BF values wherever possible. Overall,}
our estimations are found to be more or less matching with the estimations in the literature, wherever they are available.

Based on average BF across all clusters, there is reduction in the BF from B {(0.28$\pm$0.07)} to K-type {(0.17$\pm$0.01)} and a slight increase in the early M-type {(0.19$\pm$0.01)}. It is known that BF in the Galactic field decreases as a function of mass. Over 70\% of massive B and A-type stars are observed in binary or multiple systems \citep{Kouwenhoven2007A&A...474...77K, Peter2012A&A...538A..74P}. The BF decreases to 50--60\% for Solar-type stars \citep{Duquennoy1991A&A...248..485D, Raghavan2010ApJS..190....1R} and around 30--40\% of the less massive M-type stars are in multiple systems \citep{Fischer1992ApJ...396..178F, Delfosse2004ASPC..318..166D, Janson2012ApJ...754...44J}. A similar trend is observed among the young systems (\citealt{Lee2020SSRv..216...70L} and references therein).  
The reducing trend seen in the Galactic field is replicated in the clusters.
{Unlike the field, we do not see the BF reducing from K to M-type stars. A possible reason could be reducing reference population due to evaporation of low mass stars, but more sample size and deeper data is needed to confirm whether the effect is universal and its cause.}

\subsection{Binary formation mechanisms}
The binary formation mechanism must be able to explain the range of observed separations and the observed mass ratio distribution among binary systems. \citet{Lee2020SSRv..216...70L} presents a discussion of theoretical models. Suggested mechanisms are turbulent fragmentation of a core,  the gravitational fragmentation of an unstable accretion disk, or gravitational capture during dynamical interactions. Hydro-dynamical simulations suggest that all of the above three mechanisms operate during the formation of star clusters \citep{Lee2020SSRv..216...70L}. The dominant mode will depend on the parent cloud's physical processes (such as feedback mechanisms from forming stars, turbulence, and magnetic field). Studies show that increased radiative feedback reduces disk fragmentation \citep{Bate2009MNRAS.392..590B,Bate2012MNRAS.419.3115B}, while magnetized clouds could produce more binaries \citep{Cunningham2018MNRAS.476..771C,Lee2019ApJ...887..232L}.

In principle, observed estimates of BFs could be used to constrain numerical calculations by introducing limits on initial conditions. However, due to insufficient statistics and significant errors in observations, such realistic comparisons have not been possible. Our BF estimates for a reasonably large number of clusters can be a valuable input to the numerical simulations. Several clusters have a statistically good number of members and, therefore, will help in constraining the initial conditions of their formation and identifying the major mechanisms responsible for regulating binary formation.

\subsection{Segregation of HMR binaries} \label{sec:discussion_3}
\textit{Distribution of \si:}
Fig.~\ref{fig:hist_BFBplus} (c) shows the distribution of the mean value of \si\ across clusters in dark gray. It shows a peak between 0.0--0.2, though the estimated range is from $-$0.42 to 0.46. A positive \si\ value suggests a moderate amount of mass segregation of the HMR binaries with respect to the reference population. 7 OCs show negative values of the \si\ parameter, however 4 of the values are within 1$\sigma$ of 0 (\ref{fig:stats} a) indicating statistically insignificant differences in radial profiles.

The age binned CDFs in Fig.~\ref{fig:hist_BFBplus} (c) show that older clusters typically have positive values with peak higher than the average value. The intermediate and the younger clusters show a CDF similar to the average. In the case of intermediate-age clusters, there are more clusters with a positive value when compared to those with a negative value. Nevertheless, there is no statistically significant monotonous correlation between age and mass of the cluster, and \si. Fig \ref{fig:BF_comparison} (b) shows that there are slightly more number of clusters with positive values of \si. Among clusters with a significant number of members, NGC 2360 has the highest \si\ value, followed by IC 4651 and NGC 2682, suggesting that the HMR binaries in these clusters show a radial segregation, with respect to the reference population. Our estimates show that all clusters older than 300 Myr are relaxed, and therefore expect segregation of more massive binaries/stars. On the other hand, IC 4756 shows a negative index, even though it is expected to be relaxed ($N_{relax} = 7.48$). We note that the lower mass bins show similar segregation for both the HMR and the reference population, whereas the more massive bins show a preferential segregation of single stars. This is an unexpected result. We suggest a couple of possibilities. It may be that a good fraction of stars falling in the HMR binary region for the massive bins are peculiar single stars (such as fast rotators, chemically peculiar stars) and not binaries. Another possibility is that a {good} fraction of the radially segregated massive single stars are the products of mergers in binaries, that are already segregated. A detailed study is required to understand this reverse segregation.

\textit{Variation of $\mathbf{\mathcal{SI}}$ with magnitude:}
Fig.~\ref{fig:hist_BFBplus} (d) shows the bin-wise CDFs of \si. The CDFs of brighter stars are rightwards, indicating more segregation for the massive stars. While CDFs of 6--8 and 8--10 mag bins are leftwards of the mean \si, indicating less segregation for the lower mass stars. Overall, {0--2 (3$\sigma$), and 8--10 mag (2$\sigma$) bins} show significant difference from the total \si\ distribution according to KS test. 

The second columns in Fig.~\ref{fig:binwise_trends} and Fig.~\ref{fig:binwise_trends_relax} shows the variation of \si$_{binned}$ $-$ \si\ with mass and magnitude. From these panels, it becomes clear that overall the massive HMR binaries are more segregated compared to the less massive ones. However, the monotonous trend is absent in the most relaxed/oldest clusters. However, the reference population (third columns in Fig.~\ref{fig:binwise_trends} and Fig.~\ref{fig:binwise_trends_relax}) shows the monotonous trends in older/relaxed clusters, but not in young/unrelaxed clusters. This suggests that in unrelaxed clusters, HMR binary segregation depends on primary mass. In relaxed clusters, the reference population segregation increases with stellar mass, but the HMR binary segregation is not impacted by the primary mass. We note that older clusters have comparatively larger segregation (Fig.~\ref{fig:hist_BFBplus} c), just that there is no internal stellar mass dependence.

In summary, the more massive HMR binaries ($>$ 1.0 \Mnom) are segregated in general, across clusters of all ages, though there are a few exceptions. The HMR binaries (B and A/F types) in some clusters are radially segregated even when the clusters are not relaxed. However, in relaxed clusters, the segregation of HMR binaries does not depend on the primary mass.

\textit{Normalized effective radii:}
The normalized effective radii ($\mathcal{R}$) for all the clusters are tabulated in Table~\ref{tab:binned_BF}. We see that 4 clusters show a statistically significant deviation between the HMR binaries' values and the reference population. These are IC 4756, NGC 2360 ,NGC 2548 and NGC 2682. Among these 4, we find that the HMR binaries in NGC 2360, NGC 2548 and NGC 2682 have smaller radii than the reference population, suggesting that they are preferentially located well inside the cluster. On the other hand, the HMR binaries in IC 4756 have larger radii with respect to the reference population, suggesting that they are located relatively outward in the cluster.

The bin-wise distribution of the $\mathcal{R}$ suggests that the massive HMR binaries have a relatively smaller $\mathcal{R}$, with respect to the less massive ones, similar to that observed for \si\ values. This is shown in the fourth columns of Fig.~\ref{fig:binwise_trends} {and Fig.~\ref{fig:binwise_trends_relax}.} The intermediate age clusters shown in the third row suggests a scatter for the high-mass end of the HMR binaries and an outward distribution in the low mass end. This trend continues to the older clusters. Therefore, it is interesting to note that in general, the low mass HMR binaries have a relatively extended distribution for older clusters.

\textit{Segregation of reference population:}
The segregation index for reference population (\si$_{binned}^{ref}$) is a valuable tool to detect the segregation within different magnitude bins of the reference population. Furthermore, {9 out of 23} clusters showed a {significant} monotonous increase in \si$_{binned}^{ref}$ with stellar mass. The third column of Fig.~\ref{fig:binwise_trends_relax} {shows that relaxed clusters have a monotonous increase in segregation with primary mass.}
The effective radii of the reference population also decrease with increasing mass for these clusters (fifth column of Fig.~\ref{fig:binwise_trends}). Therefore, we find evidence for an increase in segregation with stellar mass, for the reference population for {relaxed clusters and} clusters older than 250 Myr.

We find an interesting trend for the {intermediately relaxed } clusters {(third row in Fig.~\ref{fig:binwise_trends_relax})}. Here, when the HMR binaries ($>$ 1 \Mnom) do not show clear segregation, the reference population in the same mass range shows clear segregation. We note that this group has several {suspected} eMSTO clusters {and hence the binary selection in brighter bins can be affected due to the degeneracy in unresolved binaries, rotational velocities and differential metallicity/extinction}.

\textit{Nature vs. nurture:} The mass segregation found in star clusters could be either dynamical or primordial.
{We detect radial segregation of bright HMR binaries in unrelaxed clusters but not for bright reference population (second row Fig.~\ref{fig:binwise_trends_relax}).}
These clusters are not relaxed, and therefore the mass segregation is likely to be primordial.  

Many studies find mass segregation in young clusters that are not relaxed \citep{Meylan2000ASPC..211..215M}. The primordial mass segregation would demand a preferential central formation of massive stars \citep{Murray1996ApJ...467..728M}. Recently, studies of cloud fragments found evidence of mass segregation \citep{Plunkett2018A&A...615A...9P,Dib2019A&A...629A.135D}. A recent study by \citet{Nony2021A&A...645A..94N} finds that the most massive clumps (M = 9.3--53 \Mnom) are located in the central proto-clusters. They proposed that the observed mass segregation could be inherited from that of clumps, originating from the mass assembly phase of molecular clouds, in agreement with the scenario of the global hierarchical collapse of molecular clouds \citep{Vazquez2019MNRAS.490.3061V}. \citet{Konyves2020A&A...635A..34K} found that the most massive proto-stellar cores of their sample (2--10 \Mnom) in Orion B were found to be spatially segregated. They also suggested that intermediate-mass and high-mass proto-stars may form preferentially in the dense inner parts of clouds, leading to primordial mass segregation.
Not many studies have been carried out to trace the segregation of binaries in very young star clusters. The mass range of clumps reported for the cases mentioned above is massive enough to produce the mass range of the early type HMR binaries studied here. We suggest that primordial segregation of HMR binaries in young clusters can arise in the massive clumps that are spatially segregated in the collapse phase of the molecular cloud (for which there is observational evidence) as HMR binaries require massive clumps for their formation.

\section{Summary} \label{sec:conclusions}
\begin{enumerate}
    \item We have improved the estimates of cluster parameters for 23 clusters. We identified high mass-ratio ($q>0.6$) binaries and studied their fraction ($f^{0.6}$) in clusters across a range of age and stellar mass. We also defined a new parameter (\si) to quantitatively trace radial segregation of HMR binaries with respect to the rest of the population.
    
    \item The estimates of HMR BF range between 0.12 to 0.38, with the peak between 0.12--0.20 (56\% clusters). NGC 6774 (0.38$\pm$0.06) and IC 4756 (0.31$\pm$0.03) have the highest BF, while Trumpler 10 (0.12$\pm$0.02) and NGC 3532 (0.13$\pm$0.01) have the lowest.
    \item We find that clusters older than 1 Gyr have relatively higher $f^{0.6}$, whereas the younger clusters have relatively lower $f^{0.6}$.
    
    \item Magnitude bin-wise analysis of HMR BF shows a decreasing BF from late B-type to {K-type}. The average $f^{0.6}$ in the 0--2 magnitude bin (late B-type stars) is 28$\pm$7\%, 2--4 bin (A\&F-type) is 29$\pm$1\%, 4--6 bin (G-type) is 20$\pm$1\%, 6--8 bin (K-type) is 17$\pm$1\% and 8--10 bin (early M-type) is 19$\pm$1\%.
    
    \item The comparison to literature and field binary population suggests a multiplicative factor of 1.5--2.5 to get total BF from HMR binaries ($f^{total} \approx 1.5\ to\ 2.5 \times f^{0.6}$). We estimate the total BF as: late B-type: 41--88\%; A\&F-type: 55--75\%; G-type: 39--53\%; K-type: 33--45\%; early M-type: 26--49\%. The observed decrease in BF for B-K type stars is in agreement with Galactic field stars.
    
    \item In unrelaxed clusters, the brighter HMR binaries are more segregated with respect to low-mass HMR binaries, whereas the reference population do not show mass-dependent segregation. This segregation of B and A/F HMR binaries could be of primordial origin.
    B and A/F type reference population shows mass dependent segregation for clusters older than 250 Myr, consistent with their relaxation time. Furthermore, the HMR binaries and reference stars are similarly segregated with respect to their mass in relaxed clusters.
    
    \item The effective radii of the HMR binaries show a more or less similar trend as \si. NGC 2360 and {NGC 6774 have the smallest} effective radii, whereas IC 4756, {NGC 2547, and NGC 6793 have the highest} effective radii for HMR binaries when compared to the reference population. While the massive HMR binaries have smaller effective radii, the lower mass HMR binaries show a relatively extended distribution.
    
    \item IC 4756 stands out as a cluster with large HMR BF. However, it shows a relatively extended distribution of the HMR binaries with respect to the reference population, even though the cluster is expected to be dynamically relaxed.
\end{enumerate}

\acknowledgments
We would like to thank the anonymous referee for constructive comments and suggestions which helped in improving the manuscript.
NJ thanks the Indian Science Academies for the support as part of the SRFP-20. 
This work has made use of data from the European Space Agency (ESA) mission {\it Gaia} (\url{https://www.cosmos.esa.int/gaia}), processed by the {\it Gaia} Data Processing and Analysis Consortium (DPAC, \url{https://www.cosmos.esa.int/web/gaia/dpac/consortium}). Funding for the DPAC has been provided by national institutions, in particular the institutions participating in the {\it Gaia} Multilateral Agreement.

\vspace{5mm}
\facilities{\textit{Gaia}}

\vspace{5mm}
\textit{Software:} {{\sc topcat} \citep{Taylor2005}, matplotlib \citep{Hunter2007}, numpy \citep{oliphant2006guide}, pandas \citep{mckinney-proc-scipy-2010}, scipy \citep{2020SciPy-NMeth}}
          
\appendix
\section{Appendix information}
\subsection{Selection of binary stars} \label{sec:binary_selection}

We selected the sample of high mass-ratio (HMR) binaries as follows:
\begin{enumerate}
    \item Fitted an isochrone to the single stellar population of the MS. Created additional isochrone for $q=1$ and verified that most of the stars lie within $q=0$ and $q=1$ isochrones. 
    
    \item Identified the usable region of the MS by removing fainter stars and stars near and above the MSTO. {The magnitude cut-offs were limited to multiples of 0.25 mag. The faint limit was chosen as the magnitude at which the width of the MS becomes $6\times bp\_rp_{error}$. This ensures that the single and binary sequences are at least 3$\sigma$ apart in color axis. The faint limit was increased for NGC 2422, 2682 and 3532 due to unsatisfactory isochrone fits to the lower MS.} {The upper cut-off was selected by checking the goodness of the isochrone fit and the $q=1$ and $q=0$ isochrones do not overlap.}
    For all further analytical purposes, we use only this portion of the stars.
    
    \item {We selected stars bluer than $q=0.6$ isochrone (up to 0.2 mag bluer than the $q=0$ isochrone) as reference stars. Stars redder than $q=0.6$ isochrone (and up to 0.2 mag redder than the $q=1$ isochrone) are classified as HMR binaries.} 
     
\end{enumerate}

\subsection{Error estimation} \label{sec:error}
To estimate the errors in $f^{0.6}$, \si and $\mathcal{R}$, we utilized the bootstrap method. This entails taking random samples of the data and estimating parameters of the sub-population. Here, we measured $f^{0.6}$ and \si\ for 1000 random samples of each cluster. The mean and standard deviations of the 1000 measurements are quoted in Table~\ref{tab:master_table}. 
$f_{binned}^{0.6}$, \si$_{binned}$, \si$^{ref}_{binned}$, $\mathcal{R}$, $\mathcal{R}_{binned}$, $\mathcal{R}^{ref}$ and $\mathcal{R}^{ref}_{binned}$ were also calculated in each sub-population, whose mean values and standard deviations are quoted in Table~\ref{tab:binned_BF} for the 23 clusters studied here.

\begin{figure}
    \centering
    \includegraphics[width=0.98\textwidth]{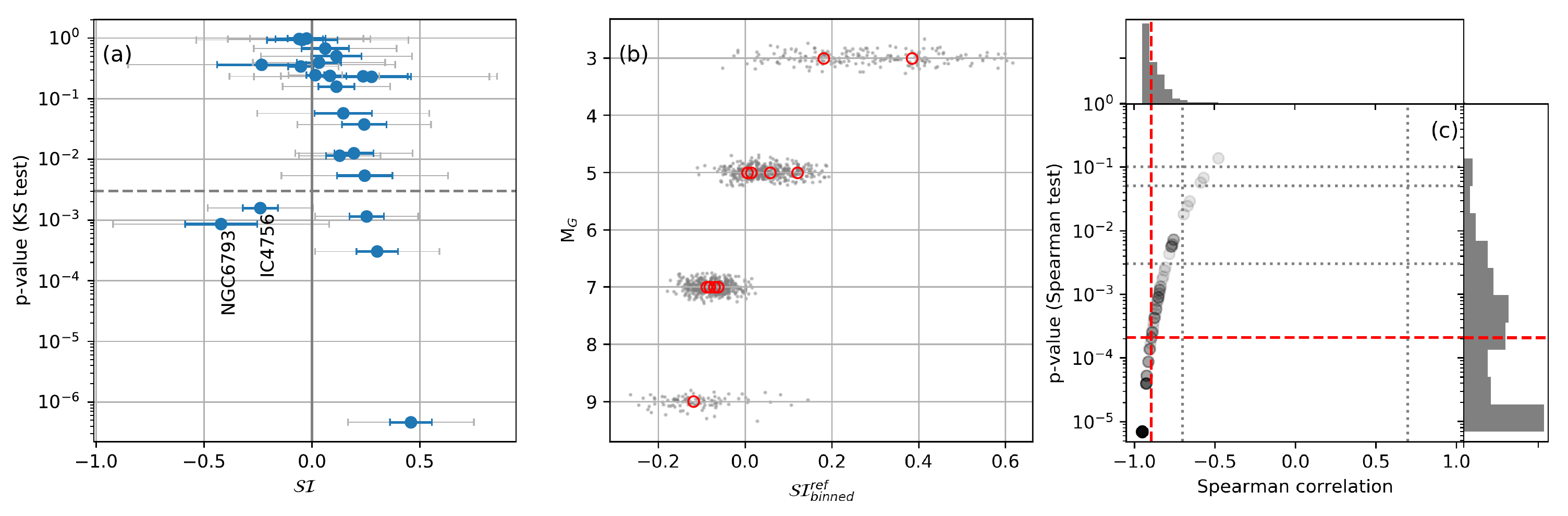}
    \caption{(a) Comparison of \si\ and p-values of KS test for radial distribution of HMR binary stars. {The blue and gray error-bars represent 1$\sigma$ and 3$\sigma$ errors.} (b) Variation of \si$^{ref}_{binned}$ with magnitudes for old clusters. The red circles are actual values and the gray dots are values after adding a Gaussian noise. (c) Distribution of Spearman correlation coefficient and p-values for 100 noisy distributions in (b). The red dashed lines show median values of correlation coefficient and p-value.}
    \label{fig:stats}
\end{figure} 

\subsection{Statistical significance} \label{sec:stats}

{We used Kolmogorov--Smirnov (KS) test to compare whether any two populations are statistically significantly different. 
For example, we compared the binned BF values with total $f^{0.6}$ distribution in Fig.~\ref{fig:hist_BFBplus}. The solid lines in the plot indicate the distributions that are different with {99.7\% confidence (p-value $<$ 0.003; $\sim 3 \sigma$ significance), dashed lines have 95\% confidence (p-value $<$ 0.05; $\sim 2 \sigma$ significance) and dotted lines have less than 95\% confidence.}
Similarly, we compared the \si\ values with p-values of KS test ran on the radial distributions of reference and HMR binary stars. The p-values are small for clusters with $|\mathcal{SI}|>\mathcal{SI}_{error}$ (Fig.~\ref{fig:stats} a). This reflects the fact that the two populations are significantly different if \si\ is {roughly 3$\sigma$} away from 0.}

For analyzing increasing and decreasing parameters, we used Spearman rank-order correlation coefficient. As our parameters have uncertainties, we used the Monte Carlo approach by adding Gaussian noise to the data and calculating Spearman coefficient distribution for 100 iterations. For example,  the red circles in Fig.~\ref{fig:stats} (b) indicate that \si$^{ref}_{binned}$, for clusters older than Gyr, is decreasing with magnitude. The gray dots show the values of parameters after adding the Gaussian noise. Fig.~\ref{fig:stats} (c) shows the distribution of Spearman coefficients and p-values for the 100 iterations. If the median of p-value is {less than 0.003 and the Spearman correlation coefficient is more than 0.7 (or less than -0.7)}, then we deemed the trend to be significantly monotonous. In Fig.~\ref{fig:stats} (c), the distribution peaks in the bottom left corner, illustrating a strong monotonous trend.

\subsection{Data limitations} \label{sec:data_limitation}
\textit{Field contamination:} We have assumed the membership catalog contains no field stars, however this is not completely accurate. We used a method similar to \citet[\S 4.1.1]{2012A&A...540A..16M} for estimating field star contamination from vector point diagrams of up to {5\%}. As this method does not account for parallax, we can consider the {5\%} as the upper bound on the field contamination, that will affect  both the single and binary population. 
{The field contamination increases with apparent magnitude and depending on the closeness of members and field in the proper motion space. However, this should impact the binary and reference population uniformly. Fortunately, the effect is most prominent below 18 G-mag, and this section was typically removed due to selection criteria mentioned in \S~\ref{sec:binary_selection} point 2.}
The exact field contamination analysis needs to be done during the membership determination process, and is beyond the scope of this paper.

{\textit{Selection biases in Gaia data:} The quality checks applied in \citet{Gaia2018} may lead to removal of around 12\% member stars from the both main sequence and binary region \citep[\S 5.3]{Jadhav2021MNRAS.503..236J}. It has been seen that binary stars have higher statistical errors in \textit{Gaia} data \citep{Gaia2018, Riello2021A&A...649A...3R}, which can lead to slight increase in the BF values. Additionally, approximately 1\% of stars do not have proper motions in \textit{Gaia} data due to various reasons, which does not affect the overall results presented here.}

\begin{figure}
    \centering
    \includegraphics[width=0.94\textwidth]{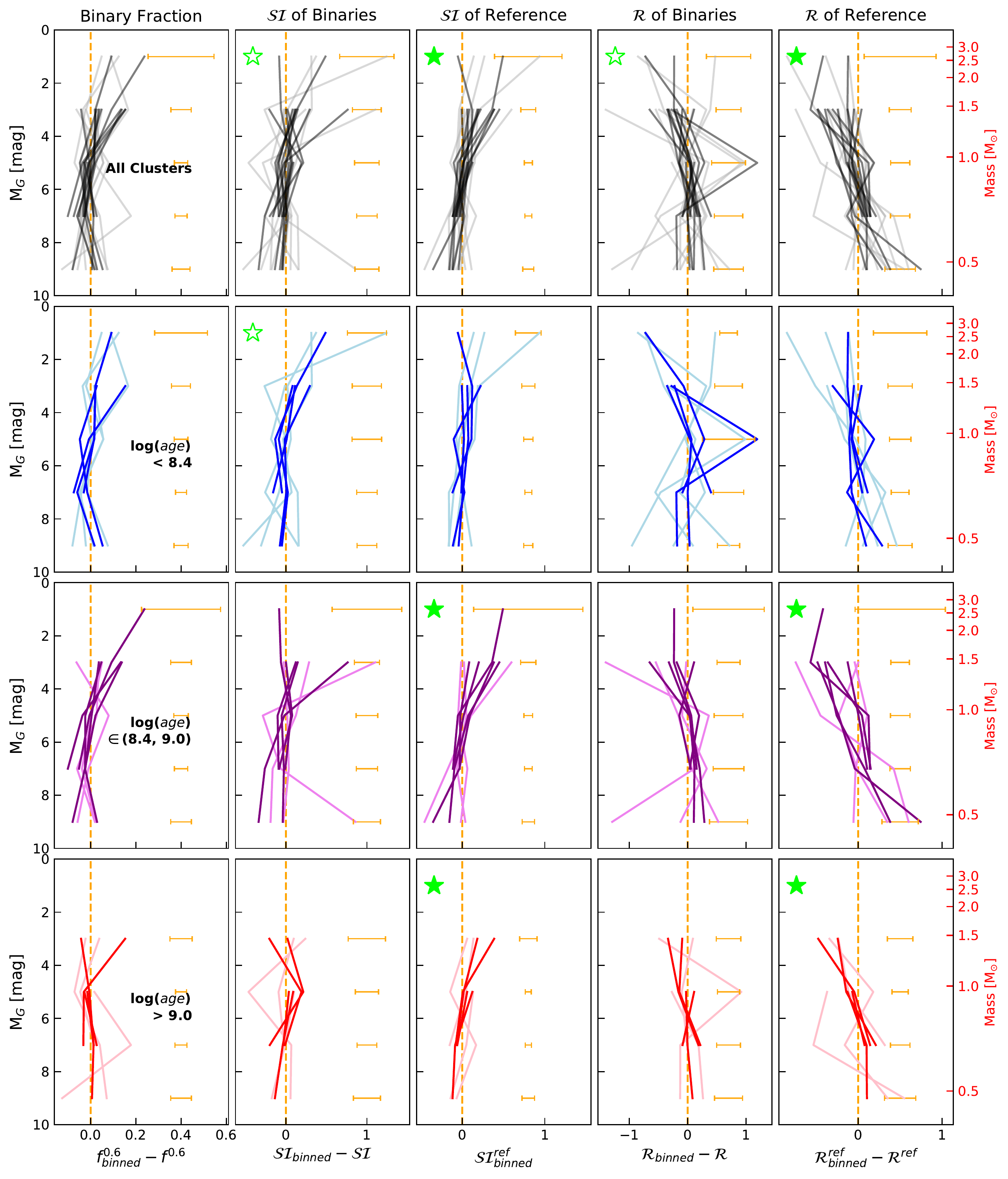}
    \caption{The trends of different binned parameters are shown here. The columns show the trends in $f_{binned}^{0.6}$, \si$_{binned}$, \si$_{binned}^{ref}$, $\mathcal{R}_{binned}$ and $\mathcal{R}_{binned}^{ref}$ with magnitude and mass. The different rows show the different sets of cluster based on age (first: all clusters; second: young clusters; third: medium age clusters; fourth: old clusters).
    The clusters with more than 400 stars ($N_A + N_B$) are shown with darker colors in each subplot. The rightmost subplots show the approximate mass of the primary stars. 
    {Orange error-bars on the right side of the panels indicate the mean errors in the parameters. {A filled green star in the top left corner indicates a monotonic trend with p-value $<$ 0.003, and a hollow green star represents a monotonic trend with p-value $\in (0.003,0.05)$.}}}
    \label{fig:binwise_trends}
\end{figure} 

\begin{figure}
    \centering
    \includegraphics[width=0.94\textwidth]{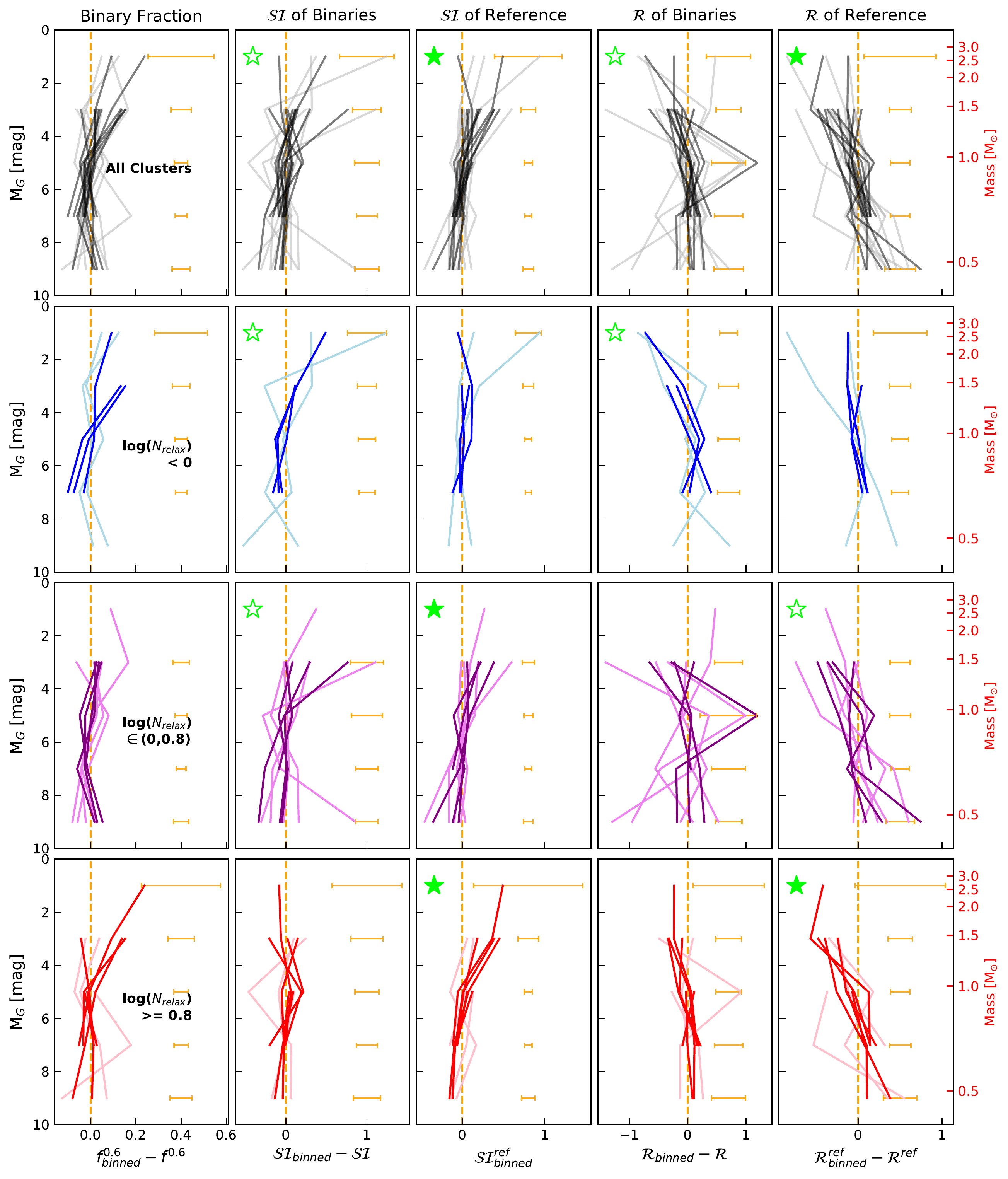}
    \caption{The trends of different binned parameters are shown here. The columns show the trends in $f_{binned}^{0.6}$, \si$_{binned}$, \si$_{binned}^{ref}$, $\mathcal{R}_{binned}$ and $\mathcal{R}_{binned}^{ref}$ with magnitude and mass. The different rows show the different sets of cluster based on age (first: all clusters; second: unrelaxed clusters; third: moderately relaxed clusters; fourth: relaxed clusters).
    The clusters with more than 400 stars ($N_A + N_B$) are shown with darker colors in each subplot. The rightmost subplots show the approximate mass of the primary stars. Orange error-bars on the right side of the panels indicate the mean errors in the parameters. {A filled green star in the top left corner indicates a monotonic trend with p-value $<$ 0.003, and a hollow green star represents a monotonic trend with p-value $\in (0.003,0.05)$.}}
    \label{fig:binwise_trends_relax}
\end{figure} 

\begin{table*}
\begin{rotatetable*}
\begin{center}
\caption{The table contain following parameters: The number of reference and HMR binary stars in each cluster; the bin-wise BF ($f^{0.6}_{binned}$); bin-wise segregation index for HMR binaries ({\si$_{binned}$}) and reference stars ({\si$_{binned}^{ref}$}); \textit{Continued...}}
\label{tab:binned_BF}
\tiny
\begin{tabular}{ccc|ccccc|ccccc|ccc}
\toprule
{Name}	&	{$N_A$}	&	{$N_B$}	&	{$f^{0.6}_{0-2}$}	&	{$f^{0.6}_{2-4}$}	&	{$f^{0.6}_{4-6}$}	&	{$f^{0.6}_{6-8}$}	&	{$f^{0.6}_{8-10}$}	&	{\si$_{0-2}$}	&	{\si$_{2-4}$}	&	{\si$_{4-6}$}	&	{\si$_{6-8}$}	&	{\si$_{8-10}$}	&	{\si$^{ref}_{0-2}$}	&	{\si$^{ref}_{2-4}$}	&	{\si$^{ref}_{4-6}$}	\\ \toprule
IC4651	&	598	&	117	&		&	0.12	&	0.16	&	0.18	&		&		&	0.32	&	0.52	&	0.1	&		&		&	0.39	&	0.01	\\ 
	&		&		&		&	$\pm$0.05	&	$\pm$0.02	&	$\pm$0.02	&		&		&	$\pm$0.32	&	$\pm$0.16	&	$\pm$0.11	&		&		&	$\pm$0.12	&	$\pm$0.03	\\ 
\hline IC4756	&	313	&	142	&		&	0.45	&	0.33	&	0.29	&	0.23	&		&	-0.09	&	-0.29	&	-0.26	&	-0.28	&		&	0.45	&	0.07	\\ 
	&		&		&		&	$\pm$0.07	&	$\pm$0.05	&	$\pm$0.05	&	$\pm$0.05	&		&	$\pm$0.16	&	$\pm$0.12	&	$\pm$0.13	&	$\pm$0.17	&		&	$\pm$0.17	&	$\pm$0.07	\\ 
\hline NGC0188	&	440	&	148	&		&		&	0.22	&	0.28	&		&		&		&	0.15	&	0.08	&		&		&		&	0.06	\\ 
	&		&		&		&		&	$\pm$0.03	&	$\pm$0.03	&		&		&		&	$\pm$0.13	&	$\pm$0.1	&		&		&		&	$\pm$0.03	\\ 
\hline NGC0752	&	139	&	33	&		&	0.17	&	0.12	&	0.23	&	0.26	&		&	0.2	&	-0.51	&	0.02	&	0.01	&		&	0.06	&	-0.15	\\ 
	&		&		&		&	$\pm$0.07	&	$\pm$0.05	&	$\pm$0.07	&	$\pm$0.08	&		&	$\pm$0.33	&	$\pm$0.34	&	$\pm$0.27	&	$\pm$0.31	&		&	$\pm$0.12	&	$\pm$0.1	\\ 
\hline NGC1039	&	320	&	65	&		&	0.22	&	0.17	&	0.13	&	0.19	&		&	0.53	&	0.37	&	0.08	&	0.06	&		&	0.02	&	0.05	\\ 
	&		&		&		&	$\pm$0.06	&	$\pm$0.04	&	$\pm$0.03	&	$\pm$0.05	&		&	$\pm$0.21	&	$\pm$0.27	&	$\pm$0.19	&	$\pm$0.26	&		&	$\pm$0.12	&	$\pm$0.07	\\ 
\hline NGC2168	&	895	&	274	&		&	0.39	&	0.23	&	0.16	&		&		&	0.24	&	0.14	&	-0.03	&		&		&	-0.01	&	0.02	\\ 
	&		&		&		&	$\pm$0.03	&	$\pm$0.02	&	$\pm$0.02	&		&		&	$\pm$0.09	&	$\pm$0.11	&	$\pm$0.11	&		&		&	$\pm$0.06	&	$\pm$0.04	\\ 
\hline NGC2360	&	534	&	124	&		&	0.34	&	0.16	&	0.16	&		&		&	0.26	&	0.65	&	0.45	&		&		&	0.18	&	0.01	\\ 
	&		&		&		&	$\pm$0.05	&	$\pm$0.02	&	$\pm$0.02	&		&		&	$\pm$0.14	&	$\pm$0.15	&	$\pm$0.17	&		&		&	$\pm$0.09	&	$\pm$0.04	\\ 
\hline NGC2422	&	367	&	60	&		&	0.17	&	0.09	&	0.13	&	0.19	&		&	0.12	&	-0.05	&	0.06	&	-0.02	&		&	0.22	&	-0.11	\\ 
	&		&		&		&	$\pm$0.05	&	$\pm$0.03	&	$\pm$0.03	&	$\pm$0.05	&		&	$\pm$0.21	&	$\pm$0.34	&	$\pm$0.15	&	$\pm$0.17	&		&	$\pm$0.1	&	$\pm$0.07	\\ 
\hline NGC2423	&	293	&	62	&		&	0.21	&	0.13	&	0.19	&		&		&	0.21	&	0.02	&	0.09	&		&		&	0.13	&	0.04	\\ 
	&		&		&		&	$\pm$0.05	&	$\pm$0.03	&	$\pm$0.04	&		&		&	$\pm$0.25	&	$\pm$0.15	&	$\pm$0.17	&		&		&	$\pm$0.08	&	$\pm$0.06	\\ 
\hline NGC2437	&	1241	&	544	&		&	0.44	&	0.27	&	0.2	&		&		&	0.14	&	-0.09	&	-0.07	&		&		&	0.08	&	-0.02	\\ 
	&		&		&		&	$\pm$0.02	&	$\pm$0.02	&	$\pm$0.02	&		&		&	$\pm$0.06	&	$\pm$0.06	&	$\pm$0.08	&		&		&	$\pm$0.04	&	$\pm$0.03	\\ 
\hline NGC2447	&	508	&	119	&		&	0.24	&	0.17	&	0.16	&	0.22	&		&	1	&	0.22	&	-0.02	&	-0.09	&		&	0.38	&	0.09	\\ 
	&		&		&		&	$\pm$0.05	&	$\pm$0.03	&	$\pm$0.03	&	$\pm$0.04	&		&	$\pm$0.25	&	$\pm$0.16	&	$\pm$0.15	&	$\pm$0.17	&		&	$\pm$0.09	&	$\pm$0.06	\\ 
\hline NGC2516	&	1312	&	244	&		&	0.18	&	0.17	&	0.1	&	0.17	&		&	0.24	&	-0.06	&	-0.05	&	-0.12	&		&	0.06	&	0.07	\\ 
	&		&		&		&	$\pm$0.03	&	$\pm$0.03	&	$\pm$0.02	&	$\pm$0.02	&		&	$\pm$0.21	&	$\pm$0.13	&	$\pm$0.13	&	$\pm$0.07	&		&	$\pm$0.07	&	$\pm$0.05	\\ 
\hline NGC2547	&	123	&	35	&		&	0.24	&	0.28	&	0.18	&	0.2	&		&	-0.24	&	-0.42	&	-0.09	&	-0.07	&		&	0.19	&	0.15	\\ 
	&		&		&		&	$\pm$0.08	&	$\pm$0.08	&	$\pm$0.06	&	$\pm$0.1	&		&	$\pm$0.37	&	$\pm$0.42	&	$\pm$0.3	&	$\pm$0.4	&		&	$\pm$0.17	&	$\pm$0.16	\\ 
\hline NGC2548	&	277	&	56	&		&	0.2	&	0.19	&	0.11	&	0.19	&		&	0.17	&	0.21	&	0.23	&	0.17	&		&	-0.01	&	-0.05	\\ 
	&		&		&		&	$\pm$0.06	&	$\pm$0.04	&	$\pm$0.03	&	$\pm$0.05	&		&	$\pm$0.16	&	$\pm$0.13	&	$\pm$0.28	&	$\pm$0.14	&		&	$\pm$0.08	&	$\pm$0.06	\\ 
\hline NGC2682	&	577	&	162	&		&		&	0.2	&	0.23	&	0.22	&		&		&	0.34	&	0.22	&	0.12	&		&		&	0.12	\\ 
	&		&		&		&		&	$\pm$0.02	&	$\pm$0.03	&	$\pm$0.05	&		&		&	$\pm$0.12	&	$\pm$0.11	&	$\pm$0.17	&		&		&	$\pm$0.04	\\ 
\hline NGC3532	&	1030	&	157	&		&	0.17	&	0.14	&	0.11	&		&		&	0.09	&	0.16	&	0	&		&		&	0.2	&	0.02	\\ 
	&		&		&		&	$\pm$0.02	&	$\pm$0.02	&	$\pm$0.01	&		&		&	$\pm$0.14	&	$\pm$0.13	&	$\pm$0.12	&		&		&	$\pm$0.04	&	$\pm$0.04	\\ 
\hline NGC6025	&	319	&	84	&	0.3	&	0.23	&	0.22	&	0.18	&		&	0.46	&	0.11	&	-0.16	&	-0.07	&		&	-0.05	&	0.12	&	0.11	\\ 
	&		&		&	$\pm$0.12	&	$\pm$0.06	&	$\pm$0.04	&	$\pm$0.03	&		&	$\pm$0.24	&	$\pm$0.21	&	$\pm$0.15	&	$\pm$0.11	&		&	$\pm$0.16	&	$\pm$0.1	&	$\pm$0.07	\\ 
\hline NGC6281	&	322	&	119	&	0.51	&	0.36	&	0.27	&	0.22	&		&	-0.02	&	0	&	0.12	&	0.06	&		&	0.49	&	0.36	&	-0.05	\\ 
	&		&		&	$\pm$0.17	&	$\pm$0.06	&	$\pm$0.04	&	$\pm$0.03	&		&	$\pm$0.43	&	$\pm$0.17	&	$\pm$0.2	&	$\pm$0.18	&		&	$\pm$0.66	&	$\pm$0.11	&	$\pm$0.07	\\ 
\hline NGC6405	&	301	&	60	&	0.29	&	0.13	&	0.17	&	0.15	&	0.24	&	0.25	&	0.26	&	-0.09	&	0.01	&	-0.58	&	0.94	&	0.2	&	-0.04	\\ 
	&		&		&	$\pm$0.14	&	$\pm$0.04	&	$\pm$0.04	&	$\pm$0.04	&	$\pm$0.08	&	$\pm$0.42	&	$\pm$0.32	&	$\pm$0.2	&	$\pm$0.17	&	$\pm$0.18	&	$\pm$0.27	&	$\pm$0.09	&	$\pm$0.07	\\ 
\hline NGC6774	&	67	&	42	&		&		&	0.4	&	0.56	&	0.26	&		&		&	0.34	&	0.23	&	0.06	&		&		&	0.13	\\ 
	&		&		&		&		&	$\pm$0.09	&	$\pm$0.13	&	$\pm$0.09	&		&		&	$\pm$0.25	&	$\pm$0.36	&	$\pm$0.36	&		&		&	$\pm$0.17	\\ 
\hline NGC6793	&	163	&	33	&		&	0.11	&	0.25	&	0.16	&	0.11	&		&	0.68	&	-0.71	&	-0.48	&	0.44	&		&	0.59	&	0.11	\\ 
	&		&		&		&	$\pm$0.05	&	$\pm$0.06	&	$\pm$0.04	&	$\pm$0.06	&		&	$\pm$0.12	&	$\pm$0.19	&	$\pm$0.28	&	$\pm$0.48	&		&	$\pm$0.17	&	$\pm$0.12	\\ 
\hline Pleiades	&	331	&	54	&	0.23	&	0.31	&	0.15	&	0.1	&	0.06	&	0.52	&	0.17	&	0.21	&	0.06	&	-0.16	&	0.27	&	0.1	&	0.01	\\ 
	&		&		&	$\pm$0.11	&	$\pm$0.07	&	$\pm$0.04	&	$\pm$0.03	&	$\pm$0.03	&	$\pm$0.85	&	$\pm$0.26	&	$\pm$0.27	&	$\pm$0.19	&	$\pm$0.41	&	$\pm$0.27	&	$\pm$0.13	&	$\pm$0.06	\\ 
\hline Trump10	&	194	&	26	&	0.17	&	0.1	&	0.17	&	0.07	&	0.13	&	1.52	&	0.01	&	0.28	&	0.02	&	0.43	&	0.14	&	-0.04	&	-0.07	\\ 
	&		&		&	$\pm$0.08	&	$\pm$0.05	&	$\pm$0.05	&	$\pm$0.03	&	$\pm$0.06	&	$\pm$0.05	&	$\pm$0.33	&	$\pm$0.27	&	$\pm$0.4	&	$\pm$0.49	&	$\pm$0.26	&	$\pm$0.08	&	$\pm$0.07	\\

 \bottomrule

\end{tabular}
\end{center}
\end{rotatetable*}
\end{table*}

\addtocounter{table}{-1}

\begin{table*}
\begin{rotatetable*}
\begin{center}
\caption{\textit{Continued...} effective radii of the clusters (r$_{eff}(cluster)$), normalized effective radii for HMR binaries ($\mathcal{R}$) and reference stars ($\mathcal{R}^{ref}$); bin-wise normalized effective radii of HMR binaries ($\mathcal{R}_{binned}$) and reference stars ($\mathcal{R}_{binned}^{ref}$).}
\tiny
\begin{tabular}{ccc|ccc|ccccc|ccccc}
\toprule
{Name}	&	{\si$^{ref}_{6-8}$}	&	{\si$^{ref}_{8-10}$}	&	{r$_{eff}(cluster)$ [deg]}	&	{$\mathcal{R}$}	&	{$\mathcal{R}^{ref}$}	&	{$\mathcal{R}_{0-2}$}	&	{$\mathcal{R}_{2-4}$}	&	{$\mathcal{R}_{4-6}$}	&	{$\mathcal{R}_{6-8}$}	&	{$\mathcal{R}_{8-10}$}	&	{$\mathcal{R}_{0-2}^{ref}$}	&	{$\mathcal{R}_{2-4}^{ref}$}	&	{$\mathcal{R}_{4-6}^{ref}$}	&	{$\mathcal{R}_{6-8}^{ref}$}	&	{$\mathcal{R}_{8-10}^{ref}$}	\\ \toprule
IC4651	&	-0.08	&		&	0.118	&	1.02	&	0.89	&		&	0.55	&	0.73	&	1.07	&		&		&	0.06	&	0.12	&	0.14	&		\\ 
	&	$\pm$0.03	&		&	0.006	&	$\pm$0.07	&	$\pm$0.14	&		&	$\pm$0.31	&	$\pm$0.18	&	$\pm$0.22	&		&		&	$\pm$0.01	&	$\pm$0.01	&	$\pm$0.01	&		\\ 
\hline IC4756	&	-0.1	&	-0.16	&	0.85	&	0.89	&	1.24	&		&	0.92	&	1.28	&	1.36	&	1.35	&		&	0.42	&	0.54	&	0.82	&	1.08	\\ 
	&	$\pm$0.07	&	$\pm$0.09	&	0.057	&	$\pm$0.1	&	$\pm$0.15	&		&	$\pm$0.24	&	$\pm$0.24	&	$\pm$0.29	&	$\pm$0.34	&		&	$\pm$0.14	&	$\pm$0.09	&	$\pm$0.11	&	$\pm$0.16	\\ 
\hline NGC0188	&	-0.06	&		&	0.042	&	0.98	&	1.07	&		&		&	1.18	&	0.98	&		&		&		&	0.04	&	0.04	&		\\ 
	&	$\pm$0.04	&		&	0.003	&	$\pm$0.09	&	$\pm$0.17	&		&		&	$\pm$0.29	&	$\pm$0.19	&		&		&		&	$\pm$0.0	&	$\pm$0.0	&		\\ 
\hline NGC0752	&	0.17	&	-0.07	&	0.912	&	0.98	&	1.1	&		&	0.62	&	2.02	&	0.98	&	0.98	&		&	0.58	&	1.05	&	0.74	&	1.21	\\ 
	&	$\pm$0.13	&	$\pm$0.17	&	0.132	&	$\pm$0.21	&	$\pm$0.38	&		&	$\pm$0.63	&	$\pm$1.14	&	$\pm$0.53	&	$\pm$0.76	&		&	$\pm$0.23	&	$\pm$0.26	&	$\pm$0.26	&	$\pm$0.42	\\ 
\hline NGC1039	&	-0.07	&	0.04	&	0.395	&	1.04	&	0.8	&		&	0.26	&	0.64	&	0.87	&	1.33	&		&	0.41	&	0.32	&	0.4	&	0.55	\\ 
	&	$\pm$0.06	&	$\pm$0.11	&	0.037	&	$\pm$0.14	&	$\pm$0.23	&		&	$\pm$0.17	&	$\pm$0.3	&	$\pm$0.47	&	$\pm$0.65	&		&	$\pm$0.11	&	$\pm$0.06	&	$\pm$0.06	&	$\pm$0.12	\\ 
\hline NGC2168	&	-0.01	&		&	0.253	&	1	&	1	&		&	0.65	&	1.04	&	1.4	&		&		&	0.26	&	0.23	&	0.27	&		\\ 
	&	$\pm$0.03	&		&	0.011	&	$\pm$0.07	&	$\pm$0.11	&		&	$\pm$0.11	&	$\pm$0.17	&	$\pm$0.22	&		&		&	$\pm$0.03	&	$\pm$0.02	&	$\pm$0.02	&		\\ 
\hline NGC2360	&	-0.07	&		&	0.084	&	1.09	&	0.63	&		&	0.54	&	0.49	&	0.85	&		&		&	0.07	&	0.08	&	0.11	&		\\ 
	&	$\pm$0.04	&		&	0.005	&	$\pm$0.09	&	$\pm$0.11	&		&	$\pm$0.12	&	$\pm$0.14	&	$\pm$0.26	&		&		&	$\pm$0.01	&	$\pm$0.01	&	$\pm$0.01	&		\\ 
\hline NGC2422	&	0.03	&	-0.11	&	0.223	&	1	&	0.99	&		&	0.72	&	2.18	&	0.8	&	0.81	&		&	0.16	&	0.27	&	0.19	&	0.29	\\ 
	&	$\pm$0.05	&	$\pm$0.08	&	0.016	&	$\pm$0.11	&	$\pm$0.23	&		&	$\pm$0.3	&	$\pm$1.15	&	$\pm$0.34	&	$\pm$0.24	&		&	$\pm$0.03	&	$\pm$0.04	&	$\pm$0.02	&	$\pm$0.04	\\ 
\hline NGC2423	&	-0.15	&		&	0.17	&	1.01	&	0.93	&		&	1.02	&	0.83	&	0.92	&		&		&	0.13	&	0.15	&	0.23	&		\\ 
	&	$\pm$0.06	&		&	0.013	&	$\pm$0.11	&	$\pm$0.2	&		&	$\pm$0.37	&	$\pm$0.32	&	$\pm$0.28	&		&		&	$\pm$0.02	&	$\pm$0.02	&	$\pm$0.03	&		\\ 
\hline NGC2437	&	-0.03	&		&	0.094	&	1.02	&	0.95	&		&	0.77	&	1.15	&	0.99	&		&		&	0.08	&	0.1	&	0.11	&		\\ 
	&	$\pm$0.04	&		&	0.003	&	$\pm$0.05	&	$\pm$0.06	&		&	$\pm$0.08	&	$\pm$0.11	&	$\pm$0.13	&		&		&	$\pm$0.01	&	$\pm$0.01	&	$\pm$0.01	&		\\ 
\hline NGC2447	&	-0.04	&	-0.35	&	0.147	&	1.01	&	0.96	&		&	0.31	&	0.99	&	1.16	&	1.24	&		&	0.08	&	0.11	&	0.14	&	0.26	\\ 
	&	$\pm$0.05	&	$\pm$0.06	&	0.009	&	$\pm$0.09	&	$\pm$0.15	&		&	$\pm$0.15	&	$\pm$0.37	&	$\pm$0.3	&	$\pm$0.31	&		&	$\pm$0.02	&	$\pm$0.01	&	$\pm$0.02	&	$\pm$0.03	\\ 
\hline NGC2516	&	0.01	&	-0.04	&	0.882	&	0.99	&	1.07	&		&	0.84	&	1.13	&	1.07	&	1.1	&		&	0.83	&	0.78	&	0.8	&	0.95	\\ 
	&	$\pm$0.04	&	$\pm$0.03	&	0.033	&	$\pm$0.05	&	$\pm$0.1	&		&	$\pm$0.22	&	$\pm$0.22	&	$\pm$0.27	&	$\pm$0.14	&		&	$\pm$0.1	&	$\pm$0.07	&	$\pm$0.07	&	$\pm$0.06	\\ 
\hline NGC2547	&	-0.16	&	-0.16	&	0.978	&	0.91	&	1.32	&		&	0.99	&	2.31	&	0.86	&	0.37	&		&	0.52	&	0.73	&	1.2	&	0.94	\\ 
	&	$\pm$0.13	&	$\pm$0.22	&	0.157	&	$\pm$0.22	&	$\pm$0.44	&		&	$\pm$0.53	&	$\pm$0.95	&	$\pm$0.68	&	$\pm$0.26	&		&	$\pm$0.25	&	$\pm$0.28	&	$\pm$0.32	&	$\pm$0.48	\\ 
\hline NGC2548	&	0.06	&	-0.02	&	0.105	&	1.05	&	0.73	&		&	0.71	&	0.63	&	1.06	&	0.61	&		&	0.11	&	0.12	&	0.11	&	0.1	\\ 
	&	$\pm$0.06	&	$\pm$0.07	&	0.005	&	$\pm$0.07	&	$\pm$0.12	&		&	$\pm$0.23	&	$\pm$0.17	&	$\pm$0.37	&	$\pm$0.14	&		&	$\pm$0.01	&	$\pm$0.01	&	$\pm$0.01	&	$\pm$0.01	\\ 
\hline NGC2682	&	-0.09	&	-0.12	&	0.157	&	1.07	&	0.75	&		&		&	0.72	&	0.74	&	0.83	&		&		&	0.15	&	0.18	&	0.18	\\ 
	&	$\pm$0.04	&	$\pm$0.08	&	0.008	&	$\pm$0.08	&	$\pm$0.1	&		&		&	$\pm$0.15	&	$\pm$0.15	&	$\pm$0.24	&		&		&	$\pm$0.01	&	$\pm$0.02	&	$\pm$0.03	\\ 
\hline NGC3532	&	-0.11	&		&	0.535	&	1	&	1.01	&		&	1.12	&	0.87	&	1.06	&		&		&	0.34	&	0.56	&	0.61	&		\\ 
	&	$\pm$0.03	&		&	0.025	&	$\pm$0.07	&	$\pm$0.15	&		&	$\pm$0.3	&	$\pm$0.2	&	$\pm$0.24	&		&		&	$\pm$0.04	&	$\pm$0.05	&	$\pm$0.04	&		\\ 
\hline NGC6025	&	-0.12	&		&	0.189	&	1.01	&	0.97	&	0.25	&	0.9	&	1.26	&	0.88	&		&	0.17	&	0.17	&	0.18	&	0.21	&		\\ 
	&	$\pm$0.05	&		&	0.012	&	$\pm$0.1	&	$\pm$0.15	&	$\pm$0.15	&	$\pm$0.32	&	$\pm$0.26	&	$\pm$0.21	&		&	$\pm$0.06	&	$\pm$0.03	&	$\pm$0.02	&	$\pm$0.02	&		\\ 
\hline NGC6281	&	-0.11	&		&	0.216	&	0.98	&	1.07	&	0.84	&	0.83	&	1.13	&	1.22	&		&	0.12	&	0.09	&	0.24	&	0.24	&		\\ 
	&	$\pm$0.06	&		&	0.017	&	$\pm$0.12	&	$\pm$0.19	&	$\pm$0.61	&	$\pm$0.21	&	$\pm$0.31	&	$\pm$0.34	&		&	$\pm$0.12	&	$\pm$0.02	&	$\pm$0.03	&	$\pm$0.03	&		\\ 
\hline NGC6405	&	-0.1	&	-0.17	&	0.228	&	0.98	&	1.11	&	0.39	&	0.7	&	1.25	&	0.98	&	1.83	&	0.03	&	0.11	&	0.22	&	0.28	&	0.33	\\ 
	&	$\pm$0.06	&	$\pm$0.16	&	0.021	&	$\pm$0.13	&	$\pm$0.28	&	$\pm$0.3	&	$\pm$0.51	&	$\pm$0.56	&	$\pm$0.48	&	$\pm$0.66	&	$\pm$0.02	&	$\pm$0.02	&	$\pm$0.03	&	$\pm$0.05	&	$\pm$0.1	\\ 
\hline NGC6774	&	0.05	&	-0.14	&	1.371	&	1.18	&	0.71	&		&		&	0.44	&	0.9	&	0.97	&		&		&	1.11	&	0.88	&	2.37	\\ 
	&	$\pm$0.24	&	$\pm$0.16	&	0.249	&	$\pm$0.34	&	$\pm$0.26	&		&		&	$\pm$0.17	&	$\pm$0.54	&	$\pm$0.66	&		&		&	$\pm$0.34	&	$\pm$0.44	&	$\pm$0.73	\\ 
\hline NGC6793	&	-0.18	&	-0.46	&	0.172	&	0.91	&	1.45	&		&	0.05	&	1.81	&	1.56	&	0.16	&		&	0.03	&	0.08	&	0.23	&	0.26	\\ 
	&	$\pm$0.09	&	$\pm$0.16	&	0.025	&	$\pm$0.2	&	$\pm$0.39	&		&	$\pm$0.01	&	$\pm$0.59	&	$\pm$0.58	&	$\pm$0.11	&		&	$\pm$0.01	&	$\pm$0.02	&	$\pm$0.05	&	$\pm$0.09	\\ 
\hline Pleiades	&	-0.01	&	-0.1	&	4.424	&	0.97	&	1.15	&	1.63	&	1.54	&	1.08	&	0.6	&	1.24	&	2.6	&	3.62	&	3.81	&	4.5	&	5.34	\\ 
	&	$\pm$0.06	&	$\pm$0.08	&	0.376	&	$\pm$0.12	&	$\pm$0.29	&	$\pm$1.58	&	$\pm$0.67	&	$\pm$0.52	&	$\pm$0.15	&	$\pm$0.83	&	$\pm$0.93	&	$\pm$0.83	&	$\pm$0.73	&	$\pm$0.63	&	$\pm$0.96	\\ 
\hline Trump10	&	0.01	&	0.11	&	0.661	&	1.02	&	0.87	&	0.02	&	1.19	&	0.83	&	1.16	&	0.63	&	0.6	&	0.64	&	0.73	&	0.71	&	0.57	\\ 
	&	$\pm$0.08	&	$\pm$0.11	&	0.047	&	$\pm$0.1	&	$\pm$0.23	&	$\pm$0.0	&	$\pm$0.99	&	$\pm$0.32	&	$\pm$0.64	&	$\pm$0.32	&	$\pm$0.21	&	$\pm$0.11	&	$\pm$0.09	&	$\pm$0.1	&	$\pm$0.11	\\ 

\bottomrule

\end{tabular}
\end{center}
\end{rotatetable*}
\end{table*}

\bibliography{references}{}

\begin{thebibliography}{}
\expandafter\ifx\csname natexlab\endcsname\relax\def\natexlab#1{#1}\fi
\providecommand{\url}[1]{\href{#1}{#1}}
\providecommand{\dodoi}[1]{doi:~\href{http://doi.org/#1}{\nolinkurl{#1}}}
\providecommand{\doeprint}[1]{\href{http://ascl.net/#1}{\nolinkurl{http://ascl.net/#1}}}
\providecommand{\doarXiv}[1]{\href{https://arxiv.org/abs/#1}{\nolinkurl{https://arxiv.org/abs/#1}}}

\bibitem[{Abramson(2018)}]{Abramson_2018}
Abramson, G. 2018, Research Notes of the {AAS}, 2, 150,
  \dodoi{10.3847/2515-5172/aada8b}

\bibitem[{{Abt} \& {Levy}(1972)}]{Abt_1972}
{Abt}, H.~A., \& {Levy}, S.~G. 1972, \apj, 172, 355, \dodoi{10.1086/151353}

\bibitem[{Alessandrini {et~al.}(2016)Alessandrini, Lanzoni, Ferraro, Miocchi,
  \& Vesperini}]{Alessandrini_2016}
Alessandrini, E., Lanzoni, B., Ferraro, F.~R., Miocchi, P., \& Vesperini, E.
  2016, The Astrophysical Journal, 833, 252,
  \dodoi{10.3847/1538-4357/833/2/252}

\bibitem[{Bailyn(1995)}]{bailyn1995blue}
Bailyn, C.~D. 1995, Annual Review of Astronomy and Astrophysics, 33, 133

\bibitem[{{Bate}(2009)}]{Bate2009MNRAS.392..590B}
{Bate}, M.~R. 2009, \mnras, 392, 590, \dodoi{10.1111/j.1365-2966.2008.14106.x}

\bibitem[{{Bate}(2012)}]{Bate2012MNRAS.419.3115B}
---. 2012, \mnras, 419, 3115, \dodoi{10.1111/j.1365-2966.2011.19955.x}

\bibitem[{{Belloni} {et~al.}(1998){Belloni}, {Verbunt}, \&
  {Mathieu}}]{Belloni1998}
{Belloni}, T., {Verbunt}, F., \& {Mathieu}, R.~D. 1998, \aap, 339, 431.
\newblock \doarXiv{astro-ph/9808329}

\bibitem[{{Bergfors} {et~al.}(2010){Bergfors}, {Brandner}, {Janson}, {Daemgen},
  {Geissler}, {Henning}, {Hippler}, {Hormuth}, {Joergens}, \&
  {K{\"o}hler}}]{Bergfors2010A&A...520A..54B}
{Bergfors}, C., {Brandner}, W., {Janson}, M., {et~al.} 2010, \aap, 520, A54,
  \dodoi{10.1051/0004-6361/201014114}

\bibitem[{Bonatto {et~al.}(2008)Bonatto, Bica, \& Santos}]{Bonatto_2008}
Bonatto, C., Bica, E., \& Santos, J. F.~C., J. 2008, Monthly Notices of the
  Royal Astronomical Society, 386, 324,
  \dodoi{10.1111/j.1365-2966.2008.13042.x}

\bibitem[{{Bossini} {et~al.}(2019){Bossini}, {Vallenari}, {Bragaglia},
  {Cantat-Gaudin}, {Sordo}, {Balaguer-N{\'u}{\~n}ez}, {Jordi}, {Moitinho},
  {Soubiran}, {Casamiquela}, {Carrera}, \&
  {Heiter}}]{Bossini_2019A&A...623A.108B}
{Bossini}, D., {Vallenari}, A., {Bragaglia}, A., {et~al.} 2019, \aap, 623,
  A108, \dodoi{10.1051/0004-6361/201834693}

\bibitem[{{Bressan} {et~al.}(2012){Bressan}, {Marigo}, {Girardi}, {Salasnich},
  {Dal Cero}, {Rubele}, \& {Nanni}}]{2012MNRAS.427..127B}
{Bressan}, A., {Marigo}, P., {Girardi}, L., {et~al.} 2012, \mnras, 427, 127,
  \dodoi{10.1111/j.1365-2966.2012.21948.x}

\bibitem[{Clem {et~al.}(2011)Clem, Landolt, Hoard, \& Wachter}]{Clem_2011}
Clem, J.~L., Landolt, A.~U., Hoard, D.~W., \& Wachter, S. 2011, The
  Astronomical Journal, 141, 115, \dodoi{10.1088/0004-6256/141/4/115}

\bibitem[{{Cohen} {et~al.}(2020){Cohen}, {Geller}, \& {von
  Hippel}}]{Cohen2020AJ....159...11C}
{Cohen}, R.~E., {Geller}, A.~M., \& {von Hippel}, T. 2020, \aj, 159, 11,
  \dodoi{10.3847/1538-3881/ab59d7}

\bibitem[{{Cordoni} {et~al.}(2018){Cordoni}, {Milone}, {Marino}, {Di
  Criscienzo}, {D'Antona}, {Dotter}, {Lagioia}, \&
  {Tailo}}]{Cordoni2018ApJ...869..139C}
{Cordoni}, G., {Milone}, A.~P., {Marino}, A.~F., {et~al.} 2018, \apj, 869, 139,
  \dodoi{10.3847/1538-4357/aaedc1}

\bibitem[{{Cunningham} {et~al.}(2018){Cunningham}, {Krumholz}, {McKee}, \&
  {Klein}}]{Cunningham2018MNRAS.476..771C}
{Cunningham}, A.~J., {Krumholz}, M.~R., {McKee}, C.~F., \& {Klein}, R.~I. 2018,
  \mnras, 476, 771, \dodoi{10.1093/mnras/sty154}

\bibitem[{Davidge(2013)}]{Davidge_2013}
Davidge, T.~J. 2013, Publications of the Astronomical Society of the Pacific,
  125, 115, \dodoi{10.1086/669823}

\bibitem[{{De Rosa} {et~al.}(2014){De Rosa}, {Patience}, {Wilson}, {Schneider},
  {Wiktorowicz}, {Vigan}, {Marois}, {Song}, {Macintosh}, {Graham}, {Doyon},
  {Bessell}, {Thomas}, \& {Lai}}]{DeRosa2014MNRAS.437.1216D}
{De Rosa}, R.~J., {Patience}, J., {Wilson}, P.~A., {et~al.} 2014, \mnras, 437,
  1216, \dodoi{10.1093/mnras/stt1932}

\bibitem[{{Delfosse} {et~al.}(2004){Delfosse}, {Beuzit}, {Marchal}, {Bonfils},
  {Perrier}, {S{\'e}gransan}, {Udry}, {Mayor}, \&
  {Forveille}}]{Delfosse2004ASPC..318..166D}
{Delfosse}, X., {Beuzit}, J.~L., {Marchal}, L., {et~al.} 2004, in Astronomical
  Society of the Pacific Conference Series, Vol. 318, Spectroscopically and
  Spatially Resolving the Components of the Close Binary Stars, ed. R.~W.
  {Hilditch}, H.~{Hensberge}, \& K.~{Pavlovski}, 166--174

\bibitem[{{Dias} {et~al.}(2002){Dias}, {Alessi}, {Moitinho}, \&
  {L{\'e}pine}}]{2002A&A...389..871D}
{Dias}, W.~S., {Alessi}, B.~S., {Moitinho}, A., \& {L{\'e}pine}, J.~R.~D. 2002,
  \aap, 389, 871, \dodoi{10.1051/0004-6361:20020668}

\bibitem[{{Dib} \& {Henning}(2019)}]{Dib2019A&A...629A.135D}
{Dib}, S., \& {Henning}, T. 2019, \aap, 629, A135,
  \dodoi{10.1051/0004-6361/201834080}

\bibitem[{{Dinescu} {et~al.}(1995){Dinescu}, {Demarque}, {Guenther}, \&
  {Pinsonneault}}]{Dinescu_1995}
{Dinescu}, D.~I., {Demarque}, P., {Guenther}, D.~B., \& {Pinsonneault}, M.~H.
  1995, \aj, 109, 2090, \dodoi{10.1086/117434}

\bibitem[{Dobbie {et~al.}(2009)Dobbie, Napiwotzki, Burleigh, Williams, Sharp,
  Barstow, Casewell, \& Hubeny}]{Dobbie_2009}
Dobbie, P.~D., Napiwotzki, R., Burleigh, M.~R., {et~al.} 2009, Monthly Notices
  of the Royal Astronomical Society, 395, 2248,
  \dodoi{10.1111/j.1365-2966.2009.14688.x}

\bibitem[{Duquennoy \& Mayor(1991)}]{duquennoy1991multiplicity}
Duquennoy, A., \& Mayor, M. 1991, Astronomy and Astrophysics, 248, 485

\bibitem[{{Duquennoy} \& {Mayor}(1991)}]{Duquennoy1991A&A...248..485D}
{Duquennoy}, A., \& {Mayor}, M. 1991, \aap, 500, 337

\bibitem[{{Dworetsky}(1975)}]{1975AJ.....80..131D}
{Dworetsky}, M.~M. 1975, \aj, 80, 131, \dodoi{10.1086/111723}

\bibitem[{{Eyer} {et~al.}(2010){Eyer}, {Eggenberger}, {Greco}, {Saesen},
  {Anderson}, \& {Mowlavi}}]{Eyer_2010}
{Eyer}, L., {Eggenberger}, P., {Greco}, C., {et~al.} 2010, in JENAM 2010, Joint
  European and National Astronomy Meeting, 212.
\newblock \doarXiv{1011.4952}

\bibitem[{{Fischer} \& {Marcy}(1992)}]{Fischer1992ApJ...396..178F}
{Fischer}, D.~A., \& {Marcy}, G.~W. 1992, \apj, 396, 178,
  \dodoi{10.1086/171708}

\bibitem[{{Fisher} {et~al.}(2005){Fisher}, {Schr{\"o}der}, \&
  {Smith}}]{Fisher2005}
{Fisher}, J., {Schr{\"o}der}, K.-P., \& {Smith}, R.~C. 2005, \mnras, 361, 495,
  \dodoi{10.1111/j.1365-2966.2005.09193.x}

\bibitem[{{Fregeau} {et~al.}(2009){Fregeau}, {Ivanova}, \&
  {Rasio}}]{Fregeau2009}
{Fregeau}, J.~M., {Ivanova}, N., \& {Rasio}, F.~A. 2009, \apj, 707, 1533,
  \dodoi{10.1088/0004-637X/707/2/1533}

\bibitem[{{Gaia Collaboration} {et~al.}(2016){Gaia Collaboration}, {Prusti},
  {de Bruijne}, {Brown}, {Vallenari}, {Babusiaux}, {Bailer-Jones}, {Bastian},
  {Biermann}, {Evans}, {Eyer}, {Jansen}, {Jordi}, {Klioner}, {Lammers},
  {Lindegren}, {Luri}, {Mignard}, {Milligan}, {Panem}, {Poinsignon},
  {Pourbaix}, {Randich}, {Sarri}, {Sartoretti}, {Siddiqui}, {Soubiran},
  {Valette}, {van Leeuwen}, {Walton}, {Aerts}, {Arenou}, {Cropper}, {Drimmel},
  {H{\o}g}, {Katz}, {Lattanzi}, {O'Mullane}, {Grebel}, {Holland}, {Huc},
  {Passot}, {Bramante}, {Cacciari}, {Casta{\~n}eda}, {Chaoul}, {Cheek}, {De
  Angeli}, {Fabricius}, {Guerra}, {Hern{\'a}ndez}, {Jean-Antoine-Piccolo},
  {Masana}, {Messineo}, {Mowlavi}, {Nienartowicz}, {Ord{\'o}{\~n}ez-Blanco},
  {Panuzzo}, {Portell}, {Richards}, {Riello}, {Seabroke}, {Tanga},
  {Th{\'e}venin}, {Torra}, {Els}, {Gracia-Abril}, {Comoretto},
  {Garcia-Reinaldos}, {Lock}, {Mercier}, {Altmann}, {Andrae}, {Astraatmadja},
  {Bellas-Velidis}, {Benson}, {Berthier}, {Blomme}, {Busso}, {Carry},
  {Cellino}, {Clementini}, {Cowell}, {Creevey}, {Cuypers}, {Davidson}, {De
  Ridder}, {de Torres}, {Delchambre}, {Dell'Oro}, {Ducourant}, {Fr{\'e}mat},
  {Garc{\'\i}a-Torres}, {Gosset}, {Halbwachs}, {Hambly}, {Harrison}, {Hauser},
  {Hestroffer}, {Hodgkin}, {Huckle}, {Hutton}, {Jasniewicz}, {Jordan},
  {Kontizas}, {Korn}, {Lanzafame}, {Manteiga}, {Moitinho}, {Muinonen},
  {Osinde}, {Pancino}, {Pauwels}, {Petit}, {Recio-Blanco}, {Robin}, {Sarro},
  {Siopis}, {Smith}, {Smith}, {Sozzetti}, {Thuillot}, {van Reeven}, {Viala},
  {Abbas}, {Abreu Aramburu}, {Accart}, {Aguado}, {Allan}, {Allasia},
  {Altavilla}, {{\'A}lvarez}, {Alves}, {Anderson}, {Andrei}, {Anglada Varela},
  {Antiche}, {Antoja}, {Ant{\'o}n}, {Arcay}, {Atzei}, {Ayache}, {Bach},
  {Baker}, {Balaguer-N{\'u}{\~n}ez}, {Barache}, {Barata}, {Barbier}, {Barblan},
  {Baroni}, {Barrado y Navascu{\'e}s}, {Barros}, {Barstow}, {Becciani},
  {Bellazzini}, {Bellei}, {Bello Garc{\'\i}a}, {Belokurov}, {Bendjoya},
  {Berihuete}, {Bianchi}, {Bienaym{\'e}}, {Billebaud}, {Blagorodnova},
  {Blanco-Cuaresma}, {Boch}, {Bombrun}, {Borrachero}, {Bouquillon}, {Bourda},
  {Bouy}, {Bragaglia}, {Breddels}, {Brouillet}, {Br{\"u}semeister},
  {Bucciarelli}, {Budnik}, {Burgess}, {Burgon}, {Burlacu}, {Busonero}, {Buzzi},
  {Caffau}, {Cambras}, {Campbell}, {Cancelliere}, {Cantat-Gaudin}, {Carlucci},
  {Carrasco}, {Castellani}, {Charlot}, {Charnas}, {Charvet}, {Chassat},
  {Chiavassa}, {Clotet}, {Cocozza}, {Collins}, {Collins}, {Costigan}, {Crifo},
  {Cross}, {Crosta}, {Crowley}, {Dafonte}, {Damerdji}, {Dapergolas}, {David},
  {David}, {De Cat}, {de Felice}, {de Laverny}, {De Luise}, {De March}, {de
  Martino}, {de Souza}, {Debosscher}, {del Pozo}, {Delbo}, {Delgado},
  {Delgado}, {di Marco}, {Di Matteo}, {Diakite}, {Distefano}, {Dolding}, {Dos
  Anjos}, {Drazinos}, {Dur{\'a}n}, {Dzigan}, {Ecale}, {Edvardsson}, {Enke},
  {Erdmann}, {Escolar}, {Espina}, {Evans}, {Eynard Bontemps}, {Fabre},
  {Fabrizio}, {Faigler}, {Falc{\~a}o}, {Farr{\`a}s Casas}, {Faye}, {Federici},
  {Fedorets}, {Fern{\'a}ndez-Hern{\'a}ndez}, {Fernique}, {Fienga}, {Figueras},
  {Filippi}, {Findeisen}, {Fonti}, {Fouesneau}, {Fraile}, {Fraser}, {Fuchs},
  {Furnell}, {Gai}, {Galleti}, {Galluccio}, {Garabato}, {Garc{\'\i}a-Sedano},
  {Gar{\'e}}, {Garofalo}, {Garralda}, {Gavras}, {Gerssen}, {Geyer}, {Gilmore},
  {Girona}, {Giuffrida}, {Gomes}, {Gonz{\'a}lez-Marcos},
  {Gonz{\'a}lez-N{\'u}{\~n}ez}, {Gonz{\'a}lez-Vidal}, {Granvik}, {Guerrier},
  {Guillout}, {Guiraud}, {G{\'u}rpide}, {Guti{\'e}rrez-S{\'a}nchez}, {Guy},
  {Haigron}, {Hatzidimitriou}, {Haywood}, {Heiter}, {Helmi}, {Hobbs},
  {Hofmann}, {Holl}, {Holland}, {Hunt}, {Hypki}, {Icardi}, {Irwin}, {Jevardat
  de Fombelle}, {Jofr{\'e}}, {Jonker}, {Jorissen}, {Julbe}, {Karampelas},
  {Kochoska}, {Kohley}, {Kolenberg}, {Kontizas}, {Koposov}, {Kordopatis},
  {Koubsky}, {Kowalczyk}, {Krone-Martins}, {Kudryashova}, {Kull}, {Bachchan},
  {Lacoste-Seris}, {Lanza}, {Lavigne}, {Le Poncin-Lafitte}, {Lebreton},
  {Lebzelter}, {Leccia}, {Leclerc}, {Lecoeur-Taibi}, {Lemaitre}, {Lenhardt},
  {Leroux}, {Liao}, {Licata}, {Lindstr{\o}m}, {Lister}, {Livanou}, {Lobel},
  {L{\"o}ffler}, {L{\'o}pez}, {Lopez-Lozano}, {Lorenz}, {Loureiro},
  {MacDonald}, {Magalh{\~a}es Fernandes}, {Managau}, {Mann}, {Mantelet},
  {Marchal}, {Marchant}, {Marconi}, {Marie}, {Marinoni}, {Marrese},
  {Marschalk{\'o}}, {Marshall}, {Mart{\'\i}n-Fleitas}, {Martino}, {Mary},
  {Matijevi{\v{c}}}, {Mazeh}, {McMillan}, {Messina}, {Mestre}, {Michalik},
  {Millar}, {Miranda}, {Molina}, {Molinaro}, {Molinaro}, {Moln{\'a}r},
  {Moniez}, {Montegriffo}, {Monteiro}, {Mor}, {Mora}, {Morbidelli}, {Morel},
  {Morgenthaler}, {Morley}, {Morris}, {Mulone}, {Muraveva}, {Musella},
  {Narbonne}, {Nelemans}, {Nicastro}, {Noval}, {Ord{\'e}novic},
  {Ordieres-Mer{\'e}}, {Osborne}, {Pagani}, {Pagano}, {Pailler}, {Palacin},
  {Palaversa}, {Parsons}, {Paulsen}, {Pecoraro}, {Pedrosa}, {Pentik{\"a}inen},
  {Pereira}, {Pichon}, {Piersimoni}, {Pineau}, {Plachy}, {Plum}, {Poujoulet},
  {Pr{\v{s}}a}, {Pulone}, {Ragaini}, {Rago}, {Rambaux}, {Ramos-Lerate},
  {Ranalli}, {Rauw}, {Read}, {Regibo}, {Renk}, {Reyl{\'e}}, {Ribeiro},
  {Rimoldini}, {Ripepi}, {Riva}, {Rixon}, {Roelens}, {Romero-G{\'o}mez},
  {Rowell}, {Royer}, {Rudolph}, {Ruiz-Dern}, {Sadowski}, {Sagrist{\`a}
  Sell{\'e}s}, {Sahlmann}, {Salgado}, {Salguero}, {Sarasso}, {Savietto},
  {Schnorhk}, {Schultheis}, {Sciacca}, {Segol}, {Segovia}, {Segransan},
  {Serpell}, {Shih}, {Smareglia}, {Smart}, {Smith}, {Solano}, {Solitro},
  {Sordo}, {Soria Nieto}, {Souchay}, {Spagna}, {Spoto}, {Stampa}, {Steele},
  {Steidelm{\"u}ller}, {Stephenson}, {Stoev}, {Suess}, {S{\"u}veges}, {Surdej},
  {Szabados}, {Szegedi-Elek}, {Tapiador}, {Taris}, {Tauran}, {Taylor},
  {Teixeira}, {Terrett}, {Tingley}, {Trager}, {Turon}, {Ulla}, {Utrilla},
  {Valentini}, {van Elteren}, {Van Hemelryck}, {van Leeuwen}, {Varadi},
  {Vecchiato}, {Veljanoski}, {Via}, {Vicente}, {Vogt}, {Voss}, {Votruba},
  {Voutsinas}, {Walmsley}, {Weiler}, {Weingrill}, {Werner}, {Wevers},
  {Whitehead}, {Wyrzykowski}, {Yoldas}, {{\v{Z}}erjal}, {Zucker}, {Zurbach},
  {Zwitter}, {Alecu}, {Allen}, {Allende Prieto}, {Amorim},
  {Anglada-Escud{\'e}}, {Arsenijevic}, {Azaz}, {Balm}, {Beck}, {Bernstein},
  {Bigot}, {Bijaoui}, {Blasco}, {Bonfigli}, {Bono}, {Boudreault}, {Bressan},
  {Brown}, {Brunet}, {Bunclark}, {Buonanno}, {Butkevich}, {Carret}, {Carrion},
  {Chemin}, {Ch{\'e}reau}, {Corcione}, {Darmigny}, {de Boer}, {de Teodoro}, {de
  Zeeuw}, {Delle Luche}, {Domingues}, {Dubath}, {Fodor}, {Fr{\'e}zouls},
  {Fries}, {Fustes}, {Fyfe}, {Gallardo}, {Gallegos}, {Gardiol}, {Gebran},
  {Gomboc}, {G{\'o}mez}, {Grux}, {Gueguen}, {Heyrovsky}, {Hoar}, {Iannicola},
  {Isasi Parache}, {Janotto}, {Joliet}, {Jonckheere}, {Keil}, {Kim},
  {Klagyivik}, {Klar}, {Knude}, {Kochukhov}, {Kolka}, {Kos}, {Kutka}, {Lainey},
  {LeBouquin}, {Liu}, {Loreggia}, {Makarov}, {Marseille}, {Martayan},
  {Martinez-Rubi}, {Massart}, {Meynadier}, {Mignot}, {Munari}, {Nguyen},
  {Nordlander}, {Ocvirk}, {O'Flaherty}, {Olias Sanz}, {Ortiz}, {Osorio},
  {Oszkiewicz}, {Ouzounis}, {Palmer}, {Park}, {Pasquato}, {Peltzer}, {Peralta},
  {P{\'e}turaud}, {Pieniluoma}, {Pigozzi}, {Poels}, {Prat}, {Prod'homme},
  {Raison}, {Rebordao}, {Risquez}, {Rocca-Volmerange}, {Rosen}, {Ruiz-Fuertes},
  {Russo}, {Sembay}, {Serraller Vizcaino}, {Short}, {Siebert}, {Silva},
  {Sinachopoulos}, {Slezak}, {Soffel}, {Sosnowska}, {Strai{\v{z}}ys}, {ter
  Linden}, {Terrell}, {Theil}, {Tiede}, {Troisi}, {Tsalmantza}, {Tur},
  {Vaccari}, {Vachier}, {Valles}, {Van Hamme}, {Veltz}, {Virtanen}, {Wallut},
  {Wichmann}, {Wilkinson}, {Ziaeepour}, \& {Zschocke}}]{2016A&A...595A...1G}
{Gaia Collaboration}, {Prusti}, T., {de Bruijne}, J.~H.~J., {et~al.} 2016,
  \aap, 595, A1, \dodoi{10.1051/0004-6361/201629272}

\bibitem[{{Gaia Collaboration} {et~al.}(2018{\natexlab{a}}){Gaia
  Collaboration}, {Brown}, {Vallenari}, {Prusti}, {de Bruijne}, {Babusiaux},
  {Bailer-Jones}, {Biermann}, {Evans}, {Eyer}, {Jansen}, {Jordi}, {Klioner},
  {Lammers}, {Lindegren}, {Luri}, {Mignard}, {Panem}, {Pourbaix}, {Randich},
  {Sartoretti}, {Siddiqui}, {Soubiran}, {van Leeuwen}, {Walton}, {Arenou},
  {Bastian}, {Cropper}, {Drimmel}, {Katz}, {Lattanzi}, {Bakker}, {Cacciari},
  {Casta{\~n}eda}, {Chaoul}, {Cheek}, {De Angeli}, {Fabricius}, {Guerra},
  {Holl}, {Masana}, {Messineo}, {Mowlavi}, {Nienartowicz}, {Panuzzo},
  {Portell}, {Riello}, {Seabroke}, {Tanga}, {Th{\'e}venin}, {Gracia-Abril},
  {Comoretto}, {Garcia-Reinaldos}, {Teyssier}, {Altmann}, {Andrae}, {Audard},
  {Bellas-Velidis}, {Benson}, {Berthier}, {Blomme}, {Burgess}, {Busso},
  {Carry}, {Cellino}, {Clementini}, {Clotet}, {Creevey}, {Davidson}, {De
  Ridder}, {Delchambre}, {Dell'Oro}, {Ducourant},
  {Fern{\'a}ndez-Hern{\'a}ndez}, {Fouesneau}, {Fr{\'e}mat}, {Galluccio},
  {Garc{\'\i}a-Torres}, {Gonz{\'a}lez-N{\'u}{\~n}ez}, {Gonz{\'a}lez-Vidal},
  {Gosset}, {Guy}, {Halbwachs}, {Hambly}, {Harrison}, {Hern{\'a}ndez},
  {Hestroffer}, {Hodgkin}, {Hutton}, {Jasniewicz}, {Jean-Antoine-Piccolo},
  {Jordan}, {Korn}, {Krone-Martins}, {Lanzafame}, {Lebzelter}, {L{\"o}ffler},
  {Manteiga}, {Marrese}, {Mart{\'\i}n-Fleitas}, {Moitinho}, {Mora}, {Muinonen},
  {Osinde}, {Pancino}, {Pauwels}, {Petit}, {Recio-Blanco}, {Richards},
  {Rimoldini}, {Robin}, {Sarro}, {Siopis}, {Smith}, {Sozzetti}, {S{\"u}veges},
  {Torra}, {van Reeven}, {Abbas}, {Abreu Aramburu}, {Accart}, {Aerts},
  {Altavilla}, {{\'A}lvarez}, {Alvarez}, {Alves}, {Anderson}, {Andrei},
  {Anglada Varela}, {Antiche}, {Antoja}, {Arcay}, {Astraatmadja}, {Bach},
  {Baker}, {Balaguer-N{\'u}{\~n}ez}, {Balm}, {Barache}, {Barata}, {Barbato},
  {Barblan}, {Barklem}, {Barrado}, {Barros}, {Barstow}, {Bartholom{\'e}
  Mu{\~n}oz}, {Bassilana}, {Becciani}, {Bellazzini}, {Berihuete}, {Bertone},
  {Bianchi}, {Bienaym{\'e}}, {Blanco-Cuaresma}, {Boch}, {Boeche}, {Bombrun},
  {Borrachero}, {Bossini}, {Bouquillon}, {Bourda}, {Bragaglia}, {Bramante},
  {Breddels}, {Bressan}, {Brouillet}, {Br{\"u}semeister}, {Brugaletta},
  {Bucciarelli}, {Burlacu}, {Busonero}, {Butkevich}, {Buzzi}, {Caffau},
  {Cancelliere}, {Cannizzaro}, {Cantat-Gaudin}, {Carballo}, {Carlucci},
  {Carrasco}, {Casamiquela}, {Castellani}, {Castro-Ginard}, {Charlot},
  {Chemin}, {Chiavassa}, {Cocozza}, {Costigan}, {Cowell}, {Crifo}, {Crosta},
  {Crowley}, {Cuypers}, {Dafonte}, {Damerdji}, {Dapergolas}, {David}, {David},
  {de Laverny}, {De Luise}, {De March}, {de Martino}, {de Souza}, {de Torres},
  {Debosscher}, {del Pozo}, {Delbo}, {Delgado}, {Delgado}, {Di Matteo},
  {Diakite}, {Diener}, {Distefano}, {Dolding}, {Drazinos}, {Dur{\'a}n},
  {Edvardsson}, {Enke}, {Eriksson}, {Esquej}, {Eynard Bontemps}, {Fabre},
  {Fabrizio}, {Faigler}, {Falc{\~a}o}, {Farr{\`a}s Casas}, {Federici},
  {Fedorets}, {Fernique}, {Figueras}, {Filippi}, {Findeisen}, {Fonti},
  {Fraile}, {Fraser}, {Fr{\'e}zouls}, {Gai}, {Galleti}, {Garabato},
  {Garc{\'\i}a-Sedano}, {Garofalo}, {Garralda}, {Gavel}, {Gavras}, {Gerssen},
  {Geyer}, {Giacobbe}, {Gilmore}, {Girona}, {Giuffrida}, {Glass}, {Gomes},
  {Granvik}, {Gueguen}, {Guerrier}, {Guiraud}, {Guti{\'e}rrez-S{\'a}nchez},
  {Haigron}, {Hatzidimitriou}, {Hauser}, {Haywood}, {Heiter}, {Helmi}, {Heu},
  {Hilger}, {Hobbs}, {Hofmann}, {Holland}, {Huckle}, {Hypki}, {Icardi},
  {Jan{\ss}en}, {Jevardat de Fombelle}, {Jonker}, {Juh{\'a}sz}, {Julbe},
  {Karampelas}, {Kewley}, {Klar}, {Kochoska}, {Kohley}, {Kolenberg},
  {Kontizas}, {Kontizas}, {Koposov}, {Kordopatis}, {Kostrzewa-Rutkowska},
  {Koubsky}, {Lambert}, {Lanza}, {Lasne}, {Lavigne}, {Le Fustec}, {Le
  Poncin-Lafitte}, {Lebreton}, {Leccia}, {Leclerc}, {Lecoeur-Taibi},
  {Lenhardt}, {Leroux}, {Liao}, {Licata}, {Lindstr{\o}m}, {Lister}, {Livanou},
  {Lobel}, {L{\'o}pez}, {Managau}, {Mann}, {Mantelet}, {Marchal}, {Marchant},
  {Marconi}, {Marinoni}, {Marschalk{\'o}}, {Marshall}, {Martino}, {Marton},
  {Mary}, {Massari}, {Matijevi{\v{c}}}, {Mazeh}, {McMillan}, {Messina},
  {Michalik}, {Millar}, {Molina}, {Molinaro}, {Moln{\'a}r}, {Montegriffo},
  {Mor}, {Morbidelli}, {Morel}, {Morris}, {Mulone}, {Muraveva}, {Musella},
  {Nelemans}, {Nicastro}, {Noval}, {O'Mullane}, {Ord{\'e}novic},
  {Ord{\'o}{\~n}ez-Blanco}, {Osborne}, {Pagani}, {Pagano}, {Pailler},
  {Palacin}, {Palaversa}, {Panahi}, {Pawlak}, {Piersimoni}, {Pineau}, {Plachy},
  {Plum}, {Poggio}, {Poujoulet}, {Pr{\v{s}}a}, {Pulone}, {Racero}, {Ragaini},
  {Rambaux}, {Ramos-Lerate}, {Regibo}, {Reyl{\'e}}, {Riclet}, {Ripepi}, {Riva},
  {Rivard}, {Rixon}, {Roegiers}, {Roelens}, {Romero-G{\'o}mez}, {Rowell},
  {Royer}, {Ruiz-Dern}, {Sadowski}, {Sagrist{\`a} Sell{\'e}s}, {Sahlmann},
  {Salgado}, {Salguero}, {Sanna}, {Santana-Ros}, {Sarasso}, {Savietto},
  {Schultheis}, {Sciacca}, {Segol}, {Segovia}, {S{\'e}gransan}, {Shih},
  {Siltala}, {Silva}, {Smart}, {Smith}, {Solano}, {Solitro}, {Sordo}, {Soria
  Nieto}, {Souchay}, {Spagna}, {Spoto}, {Stampa}, {Steele},
  {Steidelm{\"u}ller}, {Stephenson}, {Stoev}, {Suess}, {Surdej}, {Szabados},
  {Szegedi-Elek}, {Tapiador}, {Taris}, {Tauran}, {Taylor}, {Teixeira},
  {Terrett}, {Teyssandier}, {Thuillot}, {Titarenko}, {Torra Clotet}, {Turon},
  {Ulla}, {Utrilla}, {Uzzi}, {Vaillant}, {Valentini}, {Valette}, {van Elteren},
  {Van Hemelryck}, {van Leeuwen}, {Vaschetto}, {Vecchiato}, {Veljanoski},
  {Viala}, {Vicente}, {Vogt}, {von Essen}, {Voss}, {Votruba}, {Voutsinas},
  {Walmsley}, {Weiler}, {Wertz}, {Wevers}, {Wyrzykowski}, {Yoldas},
  {{\v{Z}}erjal}, {Ziaeepour}, {Zorec}, {Zschocke}, {Zucker}, {Zurbach}, \&
  {Zwitter}}]{2018A&A...616A...1G}
{Gaia Collaboration}, {Brown}, A.~G.~A., {Vallenari}, A., {et~al.}
  2018{\natexlab{a}}, \aap, 616, A1, \dodoi{10.1051/0004-6361/201833051}

\bibitem[{{Gaia Collaboration} {et~al.}(2018{\natexlab{b}}){Gaia
  Collaboration}, {Babusiaux}, {van Leeuwen}, {Barstow}, {Jordi}, {Vallenari},
  {Bossini}, {Bressan}, {Cantat-Gaudin}, {van Leeuwen}, {Brown}, {Prusti}, {de
  Bruijne}, {Bailer-Jones}, {Biermann}, {Evans}, {Eyer}, {Jansen}, {Klioner},
  {Lammers}, {Lindegren}, {Luri}, {Mignard}, {Panem}, {Pourbaix}, {Randich},
  {Sartoretti}, {Siddiqui}, {Soubiran}, {Walton}, {Arenou}, {Bastian},
  {Cropper}, {Drimmel}, {Katz}, {Lattanzi}, {Bakker}, {Cacciari},
  {Casta{\~n}eda}, {Chaoul}, {Cheek}, {De Angeli}, {Fabricius}, {Guerra},
  {Holl}, {Masana}, {Messineo}, {Mowlavi}, {Nienartowicz}, {Panuzzo},
  {Portell}, {Riello}, {Seabroke}, {Tanga}, {Th{\'e}venin}, {Gracia-Abril},
  {Comoretto}, {Garcia-Reinaldos}, {Teyssier}, {Altmann}, {Andrae}, {Audard},
  {Bellas-Velidis}, {Benson}, {Berthier}, {Blomme}, {Burgess}, {Busso},
  {Carry}, {Cellino}, {Clementini}, {Clotet}, {Creevey}, {Davidson}, {De
  Ridder}, {Delchambre}, {Dell'Oro}, {Ducourant},
  {Fern{\'a}ndez-Hern{\'a}ndez}, {Fouesneau}, {Fr{\'e}mat}, {Galluccio},
  {Garc{\'\i}a-Torres}, {Gonz{\'a}lez-N{\'u}{\~n}ez}, {Gonz{\'a}lez-Vidal},
  {Gosset}, {Guy}, {Halbwachs}, {Hambly}, {Harrison}, {Hern{\'a}ndez},
  {Hestroffer}, {Hodgkin}, {Hutton}, {Jasniewicz}, {Jean-Antoine-Piccolo},
  {Jordan}, {Korn}, {Krone-Martins}, {Lanzafame}, {Lebzelter}, {L{\"o}ffler},
  {Manteiga}, {Marrese}, {Mart{\'\i}n-Fleitas}, {Moitinho}, {Mora}, {Muinonen},
  {Osinde}, {Pancino}, {Pauwels}, {Petit}, {Recio-Blanco}, {Richards},
  {Rimoldini}, {Robin}, {Sarro}, {Siopis}, {Smith}, {Sozzetti}, {S{\"u}veges},
  {Torra}, {van Reeven}, {Abbas}, {Abreu Aramburu}, {Accart}, {Aerts},
  {Altavilla}, {{\'A}lvarez}, {Alvarez}, {Alves}, {Anderson}, {Andrei},
  {Anglada Varela}, {Antiche}, {Antoja}, {Arcay}, {Astraatmadja}, {Bach},
  {Baker}, {Balaguer-N{\'u}{\~n}ez}, {Balm}, {Barache}, {Barata}, {Barbato},
  {Barblan}, {Barklem}, {Barrado}, {Barros}, {Bartholom{\'e} Mu{\~n}oz},
  {Bassilana}, {Becciani}, {Bellazzini}, {Berihuete}, {Bertone}, {Bianchi},
  {Bienaym{\'e}}, {Blanco-Cuaresma}, {Boch}, {Boeche}, {Bombrun}, {Borrachero},
  {Bouquillon}, {Bourda}, {Bragaglia}, {Bramante}, {Breddels}, {Brouillet},
  {Br{\"u}semeister}, {Brugaletta}, {Bucciarelli}, {Burlacu}, {Busonero},
  {Butkevich}, {Buzzi}, {Caffau}, {Cancelliere}, {Cannizzaro}, {Carballo},
  {Carlucci}, {Carrasco}, {Casamiquela}, {Castellani}, {Castro-Ginard},
  {Charlot}, {Chemin}, {Chiavassa}, {Cocozza}, {Costigan}, {Cowell}, {Crifo},
  {Crosta}, {Crowley}, {Cuypers}, {Dafonte}, {Damerdji}, {Dapergolas}, {David},
  {David}, {de Laverny}, {De Luise}, {De March}, {de Martino}, {de Souza}, {de
  Torres}, {Debosscher}, {del Pozo}, {Delbo}, {Delgado}, {Delgado}, {Diakite},
  {Diener}, {Distefano}, {Dolding}, {Drazinos}, {Dur{\'a}n}, {Edvardsson},
  {Enke}, {Eriksson}, {Esquej}, {Eynard Bontemps}, {Fabre}, {Fabrizio},
  {Faigler}, {Falc{\~a}o}, {Farr{\`a}s Casas}, {Federici}, {Fedorets},
  {Fernique}, {Figueras}, {Filippi}, {Findeisen}, {Fonti}, {Fraile}, {Fraser},
  {Fr{\'e}zouls}, {Gai}, {Galleti}, {Garabato}, {Garc{\'\i}a-Sedano},
  {Garofalo}, {Garralda}, {Gavel}, {Gavras}, {Gerssen}, {Geyer}, {Giacobbe},
  {Gilmore}, {Girona}, {Giuffrida}, {Glass}, {Gomes}, {Granvik}, {Gueguen},
  {Guerrier}, {Guiraud}, {Guti{\'e}}, {Haigron}, {Hatzidimitriou}, {Hauser},
  {Haywood}, {Heiter}, {Helmi}, {Heu}, {Hilger}, {Hobbs}, {Hofmann}, {Holland},
  {Huckle}, {Hypki}, {Icardi}, {Jan{\ss}en}, {Jevardat de Fombelle}, {Jonker},
  {Juh{\'a}sz}, {Julbe}, {Karampelas}, {Kewley}, {Klar}, {Kochoska}, {Kohley},
  {Kolenberg}, {Kontizas}, {Kontizas}, {Koposov}, {Kordopatis},
  {Kostrzewa-Rutkowska}, {Koubsky}, {Lambert}, {Lanza}, {Lasne}, {Lavigne}, {Le
  Fustec}, {Le Poncin-Lafitte}, {Lebreton}, {Leccia}, {Leclerc},
  {Lecoeur-Taibi}, {Lenhardt}, {Leroux}, {Liao}, {Licata}, {Lindstr{\o}m},
  {Lister}, {Livanou}, {Lobel}, {L{\'o}pez}, {Managau}, {Mann}, {Mantelet},
  {Marchal}, {Marchant}, {Marconi}, {Marinoni}, {Marschalk{\'o}}, {Marshall},
  {Martino}, {Marton}, {Mary}, {Massari}, {Matijevi{\v{c}}}, {Mazeh},
  {McMillan}, {Messina}, {Michalik}, {Millar}, {Molina}, {Molinaro},
  {Moln{\'a}r}, {Montegriffo}, {Mor}, {Morbidelli}, {Morel}, {Morris},
  {Mulone}, {Muraveva}, {Musella}, {Nelemans}, {Nicastro}, {Noval},
  {O'Mullane}, {Ord{\'e}novic}, {Ord{\'o}{\~n}ez-Blanco}, {Osborne}, {Pagani},
  {Pagano}, {Pailler}, {Palacin}, {Palaversa}, {Panahi}, {Pawlak},
  {Piersimoni}, {Pineau}, {Plachy}, {Plum}, {Poggio}, {Poujoulet},
  {Pr{\v{s}}a}, {Pulone}, {Racero}, {Ragaini}, {Rambaux}, {Ramos-Lerate},
  {Regibo}, {Reyl{\'e}}, {Riclet}, {Ripepi}, {Riva}, {Rivard}, {Rixon},
  {Roegiers}, {Roelens}, {Romero-G{\'o}mez}, {Rowell}, {Royer}, {Ruiz-Dern},
  {Sadowski}, {Sagrist{\`a} Sell{\'e}s}, {Sahlmann}, {Salgado}, {Salguero},
  {Sanna}, {Santana-Ros}, {Sarasso}, {Savietto}, {Schultheis}, {Sciacca},
  {Segol}, {Segovia}, {S{\'e}gransan}, {Shih}, {Siltala}, {Silva}, {Smart},
  {Smith}, {Solano}, {Solitro}, {Sordo}, {Soria Nieto}, {Souchay}, {Spagna},
  {Spoto}, {Stampa}, {Steele}, {Steidelm{\"u}ller}, {Stephenson}, {Stoev},
  {Suess}, {Surdej}, {Szabados}, {Szegedi-Elek}, {Tapiador}, {Taris}, {Tauran},
  {Taylor}, {Teixeira}, {Terrett}, {Teyssand ier}, {Thuillot}, {Titarenko},
  {Torra Clotet}, {Turon}, {Ulla}, {Utrilla}, {Uzzi}, {Vaillant}, {Valentini},
  {Valette}, {van Elteren}, {Van Hemelryck}, {Vaschetto}, {Vecchiato},
  {Veljanoski}, {Viala}, {Vicente}, {Vogt}, {von Essen}, {Voss}, {Votruba},
  {Voutsinas}, {Walmsley}, {Weiler}, {Wertz}, {Wevers}, {Wyrzykowski},
  {Yoldas}, {{\v{Z}}erjal}, {Ziaeepour}, {Zorec}, {Zschocke}, {Zucker},
  {Zurbach}, \& {Zwitter}}]{Gaia2018}
{Gaia Collaboration}, {Babusiaux}, C., {van Leeuwen}, F., {et~al.}
  2018{\natexlab{b}}, \aap, 616, A10, \dodoi{10.1051/0004-6361/201832843}

\bibitem[{{Geller} {et~al.}(2015){Geller}, {Latham}, \& {Mathieu}}]{Geller2015}
{Geller}, A.~M., {Latham}, D.~W., \& {Mathieu}, R.~D. 2015, \aj, 150, 97,
  \dodoi{10.1088/0004-6256/150/3/97}

\bibitem[{Geller \& Mathieu(2012)}]{Geller_2012}
Geller, A.~M., \& Mathieu, R.~D. 2012, The Astronomical Journal, 144, 54,
  \dodoi{10.1088/0004-6256/144/2/54}

\bibitem[{{Geller} {et~al.}(2021){Geller}, {Mathieu}, {Latham}, {Pollack},
  {Torres}, \& {Leiner}}]{Geller2021AJ....161..190G}
{Geller}, A.~M., {Mathieu}, R.~D., {Latham}, D.~W., {et~al.} 2021, \aj, 161,
  190, \dodoi{10.3847/1538-3881/abdd23}

\bibitem[{Gonz{\'{a}}lez \& Lapasset(2000)}]{Gonz_lez_2000}
Gonz{\'{a}}lez, J.~F., \& Lapasset, E. 2000, The Astronomical Journal, 119,
  2296, \dodoi{10.1086/301364}

\bibitem[{Gonz{\'{a}}lez \& Lapasset(2002)}]{Gonz_lez_2002}
---. 2002, The Astronomical Journal, 123, 3318, \dodoi{10.1086/340566}

\bibitem[{{Gonz{\'a}lez} \& {Levato}(2006)}]{2006RMxAC..26Q.171G}
{Gonz{\'a}lez}, J.~F., \& {Levato}, H. 2006, in Revista Mexicana de Astronomia
  y Astrofisica Conference Series, Vol.~26, Revista Mexicana de Astronomia y
  Astrofisica Conference Series, 171

\bibitem[{Güneş {et~al.}(2012)Güneş, Karataş, \& Bonatto}]{Gunes2012}
Güneş, O., Karataş, Y., \& Bonatto, C. 2012, New Astronomy, 17,
  \dodoi{10.1016/j.newast.2012.05.003}

\bibitem[{Hamdani {et~al.}(2000)Hamdani, North, Mowlavi, Raboud, \&
  Mermilliod}]{hamdani2000chemical}
Hamdani, S., North, P., Mowlavi, N., Raboud, D., \& Mermilliod, J.~C. 2000,
  Chemical abundances in seven red giants of NGC 2360 and NGC 2447.
\newblock \doarXiv{astro-ph/0006442}

\bibitem[{{Hunter}(2007)}]{Hunter2007}
{Hunter}, J.~D. 2007, Computing in Science and Engineering, 9, 90,
  \dodoi{10.1109/MCSE.2007.55}

\bibitem[{{Jadhav} {et~al.}(2021){Jadhav}, {Pennock}, {Subramaniam}, {Sagar},
  \& {Nayak}}]{Jadhav2021MNRAS.503..236J}
{Jadhav}, V.~V., {Pennock}, C.~M., {Subramaniam}, A., {Sagar}, R., \& {Nayak},
  P.~K. 2021, \mnras, 503, 236, \dodoi{10.1093/mnras/stab213}

\bibitem[{{Jadhav} \& {Subramaniam}(2021)}]{Jadhav2021MNRAS.tmp.2079J}
{Jadhav}, V.~V., \& {Subramaniam}, A. 2021, \mnras,
  \dodoi{10.1093/mnras/stab2264}

\bibitem[{{Janson} {et~al.}(2012){Janson}, {Hormuth}, {Bergfors}, {Brandner},
  {Hippler}, {Daemgen}, {Kudryavtseva}, {Schmalzl}, {Schnupp}, \&
  {Henning}}]{Janson2012ApJ...754...44J}
{Janson}, M., {Hormuth}, F., {Bergfors}, C., {et~al.} 2012, \apj, 754, 44,
  \dodoi{10.1088/0004-637X/754/1/44}

\bibitem[{{Jeffries} {et~al.}(2004){Jeffries}, {Naylor}, {Devey}, \&
  {Totten}}]{2004MNRAS.351.1401J}
{Jeffries}, R.~D., {Naylor}, T., {Devey}, C.~R., \& {Totten}, E.~J. 2004,
  \mnras, 351, 1401, \dodoi{10.1111/j.1365-2966.2004.07886.x}

\bibitem[{{Jeffries} {et~al.}(2001){Jeffries}, {Thurston}, \&
  {Hambly}}]{Jeffries_2001}
{Jeffries}, R.~D., {Thurston}, M.~R., \& {Hambly}, N.~C. 2001, \aap, 375, 863,
  \dodoi{10.1051/0004-6361:20010918}

\bibitem[{{Jha} {et~al.}(2019){Jha}, {Maguire}, \&
  {Sullivan}}]{Jha2019NatAs...3..706J}
{Jha}, S.~W., {Maguire}, K., \& {Sullivan}, M. 2019, Nature Astronomy, 3, 706,
  \dodoi{10.1038/s41550-019-0858-0}

\bibitem[{Kalirai {et~al.}(2003)Kalirai, Fahlman, Richer, \&
  Ventura}]{Kalirai_2003}
Kalirai, J.~S., Fahlman, G.~G., Richer, H.~B., \& Ventura, P. 2003, The
  Astronomical Journal, 126, 1402, \dodoi{10.1086/377320}

\bibitem[{{King}(1962)}]{King_1962}
{King}, I. 1962, \aj, 67, 471, \dodoi{10.1086/108756}

\bibitem[{{K{\"o}nyves} {et~al.}(2020){K{\"o}nyves}, {Andr{\'e}},
  {Arzoumanian}, {Schneider}, {Men'shchikov}, {Bontemps}, {Ladjelate},
  {Didelon}, {Pezzuto}, {Benedettini}, {Bracco}, {Di Francesco}, {Goodwin},
  {Rygl}, {Shimajiri}, {Spinoglio}, {Ward-Thompson}, \&
  {White}}]{Konyves2020A&A...635A..34K}
{K{\"o}nyves}, V., {Andr{\'e}}, P., {Arzoumanian}, D., {et~al.} 2020, \aap,
  635, A34, \dodoi{10.1051/0004-6361/201834753}

\bibitem[{{Kouwenhoven} {et~al.}(2007){Kouwenhoven}, {Brown}, {Portegies
  Zwart}, \& {Kaper}}]{Kouwenhoven2007A&A...474...77K}
{Kouwenhoven}, M.~B.~N., {Brown}, A.~G.~A., {Portegies Zwart}, S.~F., \&
  {Kaper}, L. 2007, \aap, 474, 77, \dodoi{10.1051/0004-6361:20077719}

\bibitem[{{Lee} {et~al.}(2019){Lee}, {Offner}, {Kratter}, {Smullen}, \&
  {Li}}]{Lee2019ApJ...887..232L}
{Lee}, A.~T., {Offner}, S. S.~R., {Kratter}, K.~M., {Smullen}, R.~A., \& {Li},
  P.~S. 2019, \apj, 887, 232, \dodoi{10.3847/1538-4357/ab584b}

\bibitem[{{Lee} {et~al.}(2020){Lee}, {Offner}, {Hennebelle}, {Andr{\'e}},
  {Zinnecker}, {Ballesteros-Paredes}, {Inutsuka}, \&
  {Kruijssen}}]{Lee2020SSRv..216...70L}
{Lee}, Y.-N., {Offner}, S. S.~R., {Hennebelle}, P., {et~al.} 2020, \ssr, 216,
  70, \dodoi{10.1007/s11214-020-00699-2}

\bibitem[{{Leiner} {et~al.}(2015){Leiner}, {Mathieu}, {Gosnell}, \&
  {Geller}}]{Leiner2015AJ....150...10L}
{Leiner}, E.~M., {Mathieu}, R.~D., {Gosnell}, N.~M., \& {Geller}, A.~M. 2015,
  \aj, 150, 10, \dodoi{10.1088/0004-6256/150/1/10}

\bibitem[{{Li} {et~al.}(2020){Li}, {Shao}, {Li}, {Yu}, {Zhong}, \&
  {Chen}}]{2020arXiv200804684L}
{Li}, L., {Shao}, Z., {Li}, Z.-Z., {et~al.} 2020, arXiv e-prints,
  arXiv:2008.04684.
\newblock \doarXiv{2008.04684}

\bibitem[{{Lovis, C.} \& {Mayor, M.}(2007)}]{Lovis_2007}
{Lovis, C.}, \& {Mayor, M.} 2007, A\&A, 472, 657,
  \dodoi{10.1051/0004-6361:20077375}

\bibitem[{Mathieu \& Geller(2009)}]{mathieu2009binary}
Mathieu, R.~D., \& Geller, A.~M. 2009, Nature, 462, 1032

\bibitem[{{Mathieu} \& {Latham}(1986)}]{Mathieu1986}
{Mathieu}, R.~D., \& {Latham}, D.~W. 1986, \aj, 92, 1364,
  \dodoi{10.1086/114269}

\bibitem[{{Mazzei} \& {Pigatto}(1989)}]{Mazzei1989}
{Mazzei}, P., \& {Pigatto}, L. 1989, \aap, 213, L1

\bibitem[{{Mermilliod} {et~al.}(1989){Mermilliod}, {Mayor}, {Andersen},
  {Nordstrom}, {Lindgren}, \& {Duquennoy}}]{Mermilliod1989}
{Mermilliod}, J.~C., {Mayor}, M., {Andersen}, J., {et~al.} 1989, \aaps, 79, 11

\bibitem[{{Mermilliod} {et~al.}(1992){Mermilliod}, {Rosvick}, {Duquennoy}, \&
  {Mayor}}]{Mermilliod1992}
{Mermilliod}, J.~C., {Rosvick}, J.~M., {Duquennoy}, A., \& {Mayor}, M. 1992,
  \aap, 265, 513

\bibitem[{{Meylan}(2000)}]{Meylan2000ASPC..211..215M}
{Meylan}, G. 2000, in Astronomical Society of the Pacific Conference Series,
  Vol. 211, Massive Stellar Clusters, ed. A.~{Lan{\c{c}}on} \& C.~M. {Boily},
  215.
\newblock \doarXiv{astro-ph/0003390}

\bibitem[{{Milone} {et~al.}(2012){Milone}, {Piotto}, {Bedin}, {Aparicio},
  {Anderson}, {Sarajedini}, {Marino}, {Moretti}, {Davies}, {Chaboyer},
  {Dotter}, {Hempel}, {Mar{\'\i}n-Franch}, {Majewski}, {Paust}, {Reid},
  {Rosenberg}, \& {Siegel}}]{2012A&A...540A..16M}
{Milone}, A.~P., {Piotto}, G., {Bedin}, L.~R., {et~al.} 2012, \aap, 540, A16,
  \dodoi{10.1051/0004-6361/201016384}

\bibitem[{{Moe} \& {Di Stefano}(2017)}]{Moe2017ApJS..230...15M}
{Moe}, M., \& {Di Stefano}, R. 2017, \apjs, 230, 15,
  \dodoi{10.3847/1538-4365/aa6fb6}

\bibitem[{{Montgomery} {et~al.}(1993){Montgomery}, {Marschall}, \&
  {Janes}}]{Montgomery1993}
{Montgomery}, K.~A., {Marschall}, L.~A., \& {Janes}, K.~A. 1993, \aj, 106, 181,
  \dodoi{10.1086/116628}

\bibitem[{{Murray} \& {Lin}(1996)}]{Murray1996ApJ...467..728M}
{Murray}, S.~D., \& {Lin}, D. N.~C. 1996, \apj, 467, 728,
  \dodoi{10.1086/177648}

\bibitem[{{Netopil, M.} {et~al.}(2016){Netopil, M.}, {Paunzen, E.}, {Heiter,
  U.}, \& {Soubiran, C.}}]{Netopil2016}
{Netopil, M.}, {Paunzen, E.}, {Heiter, U.}, \& {Soubiran, C.} 2016, A\&A, 585,
  A150, \dodoi{10.1051/0004-6361/201526370}

\bibitem[{{Nony} {et~al.}(2021){Nony}, {Robitaille}, {Motte}, {Gonzalez},
  {Joncour}, {Moraux}, {Men'shchikov}, {Didelon}, {Louvet}, {Buckner},
  {Schneider}, {Lumsden}, {Bontemps}, {Pouteau}, {Cunningham}, {Fiorellino},
  {Oudmaijer}, {Andr{\'e}}, \& {Thomasson}}]{Nony2021A&A...645A..94N}
{Nony}, T., {Robitaille}, J.~F., {Motte}, F., {et~al.} 2021, \aap, 645, A94,
  \dodoi{10.1051/0004-6361/202039353}

\bibitem[{Oliphant(2015)}]{oliphant2006guide}
Oliphant, T.~E. 2015, Guide to NumPy, 2nd edn. (North Charleston, SC, USA:
  CreateSpace Independent Publishing Platform)

\bibitem[{{Parker} \& {Meyer}(2014)}]{Parker2014MNRAS.442.3722P}
{Parker}, R.~J., \& {Meyer}, M.~R. 2014, \mnras, 442, 3722,
  \dodoi{10.1093/mnras/stu1101}

\bibitem[{Paunzen \& Netopil(2006)}]{Paunzen_2006}
Paunzen, E., \& Netopil, M. 2006, Monthly Notices of the Royal Astronomical
  Society, 371, 1641, \dodoi{10.1111/j.1365-2966.2006.10783.x}

\bibitem[{{Peter} {et~al.}(2012){Peter}, {Feldt}, {Henning}, \&
  {Hormuth}}]{Peter2012A&A...538A..74P}
{Peter}, D., {Feldt}, M., {Henning}, T., \& {Hormuth}, F. 2012, \aap, 538, A74,
  \dodoi{10.1051/0004-6361/201015027}

\bibitem[{{Plunkett} {et~al.}(2018){Plunkett}, {Fern{\'a}ndez-L{\'o}pez},
  {Arce}, {Busquet}, {Mardones}, \& {Dunham}}]{Plunkett2018A&A...615A...9P}
{Plunkett}, A.~L., {Fern{\'a}ndez-L{\'o}pez}, M., {Arce}, H.~G., {et~al.} 2018,
  \aap, 615, A9, \dodoi{10.1051/0004-6361/201732372}

\bibitem[{{Raghavan} {et~al.}(2010){Raghavan}, {McAlister}, {Henry}, {Latham},
  {Marcy}, {Mason}, {Gies}, {White}, \& {ten
  Brummelaar}}]{Raghavan2010ApJS..190....1R}
{Raghavan}, D., {McAlister}, H.~A., {Henry}, T.~J., {et~al.} 2010, \apjs, 190,
  1, \dodoi{10.1088/0067-0049/190/1/1}

\bibitem[{{Rangwal} {et~al.}(2019){Rangwal}, {Yadav}, {Durgapal}, {Bisht}, \&
  {Nardiello}}]{2019MNRAS.490.1383R}
{Rangwal}, G., {Yadav}, R.~K.~S., {Durgapal}, A., {Bisht}, D., \& {Nardiello},
  D. 2019, \mnras, 490, 1383, \dodoi{10.1093/mnras/stz2642}

\bibitem[{{Riello} {et~al.}(2021){Riello}, {De Angeli}, {Evans}, {Montegriffo},
  {Carrasco}, {Busso}, {Palaversa}, {Burgess}, {Diener}, {Davidson}, {Rowell},
  {Fabricius}, {Jordi}, {Bellazzini}, {Pancino}, {Harrison}, {Cacciari}, {van
  Leeuwen}, {Hambly}, {Hodgkin}, {Osborne}, {Altavilla}, {Barstow}, {Brown},
  {Castellani}, {Cowell}, {De Luise}, {Gilmore}, {Giuffrida}, {Hidalgo},
  {Holland}, {Marinoni}, {Pagani}, {Piersimoni}, {Pulone}, {Ragaini}, {Rainer},
  {Richards}, {Sanna}, {Walton}, {Weiler}, \&
  {Yoldas}}]{Riello2021A&A...649A...3R}
{Riello}, M., {De Angeli}, F., {Evans}, D.~W., {et~al.} 2021, \aap, 649, A3,
  \dodoi{10.1051/0004-6361/202039587}

\bibitem[{Silva {et~al.}(2014)Silva, Su{\'{a}}rez, Santrich, Pereira, Drake, \&
  Roig}]{Sales_Silva_2014}
Silva, J. V.~S., Su{\'{a}}rez, V. J.~P., Santrich, O. J.~K., {et~al.} 2014, The
  Astronomical Journal, 148, 83, \dodoi{10.1088/0004-6256/148/5/83}

\bibitem[{{Sindhu} {et~al.}(2018){Sindhu}, {Subramaniam}, \&
  {Radha}}]{Sindhu2018}
{Sindhu}, N., {Subramaniam}, A., \& {Radha}, C.~A. 2018, \mnras, 481, 226,
  \dodoi{10.1093/mnras/sty2283}

\bibitem[{{Spitzer} \& {Hart}(1971)}]{Spitzer1971ApJ...164..399S}
{Spitzer}, Lyman, J., \& {Hart}, M.~H. 1971, \apj, 164, 399,
  \dodoi{10.1086/150855}

\bibitem[{{Stello} {et~al.}(2016){Stello}, {Vanderburg}, {Casagrande},
  {Gilliland}, {Silva Aguirre}, {Sandquist}, {Leiner}, {Mathieu}, \&
  {Soderblom}}]{Stello2016}
{Stello}, D., {Vanderburg}, A., {Casagrande}, L., {et~al.} 2016, \apj, 832,
  133, \dodoi{10.3847/0004-637X/832/2/133}

\bibitem[{{Strassmeier} {et~al.}(2015){Strassmeier}, {Weingrill}, {Granzer},
  {Bihain}, {Weber}, \& {Barnes}}]{2015A&A...580A..66S}
{Strassmeier}, K.~G., {Weingrill}, J., {Granzer}, T., {et~al.} 2015, \aap, 580,
  A66, \dodoi{10.1051/0004-6361/201525756}

\bibitem[{Sun {et~al.}(2020)Sun, Deliyannis, Steinhauer, Twarog, \&
  Anthony-Twarog}]{Sun_2020}
Sun, Q., Deliyannis, C.~P., Steinhauer, A., Twarog, B.~A., \& Anthony-Twarog,
  B.~J. 2020, The Astronomical Journal, 159, 220,
  \dodoi{10.3847/1538-3881/ab83ef}

\bibitem[{{Sun} {et~al.}(2021){Sun}, {de Grijs}, {Deng}, \& {Albrow}}]{Sun2021}
{Sun}, W., {de Grijs}, R., {Deng}, L., \& {Albrow}, M.~D. 2021, arXiv e-prints,
  arXiv:2102.02352.
\newblock \doarXiv{2102.02352}

\bibitem[{Sung \& Bessell(1999)}]{Sung_1999}
Sung, H., \& Bessell, M.~S. 1999, Monthly Notices of the Royal Astronomical
  Society, 306, 361, \dodoi{10.1046/j.1365-8711.1999.02522.x}

\bibitem[{{Taylor}(2005)}]{Taylor2005}
{Taylor}, M.~B. 2005, Astronomical Society of the Pacific Conference Series,
  Vol. 347, {TOPCAT \&amp; STIL: Starlink Table/VOTable Processing Software},
  ed. P.~{Shopbell}, M.~{Britton}, \& R.~{Ebert}, 29

\bibitem[{{Thompson} {et~al.}(2021){Thompson}, {Frinchaboy}, {Spoo}, \&
  {Donor}}]{Thompson2021}
{Thompson}, B., {Frinchaboy}, P.~M., {Spoo}, T., \& {Donor}, J. 2021, arXiv
  e-prints, arXiv:2101.07857.
\newblock \doarXiv{2101.07857}

\bibitem[{Torres {et~al.}(2018)Torres, Curtis, Vanderburg, Kraus, \&
  Rizzuto}]{Torres_2018}
Torres, G., Curtis, J.~L., Vanderburg, A., Kraus, A.~L., \& Rizzuto, A. 2018,
  The Astrophysical Journal, 866, 67, \dodoi{10.3847/1538-4357/aadca8}

\bibitem[{{Twarog}(1983)}]{Twarog_1983}
{Twarog}, B.~A. 1983, \apj, 267, 207, \dodoi{10.1086/160860}

\bibitem[{{V{\'a}zquez-Semadeni} {et~al.}(2019){V{\'a}zquez-Semadeni}, {Palau},
  {Ballesteros-Paredes}, {G{\'o}mez}, \&
  {Zamora-Avil{\'e}s}}]{Vazquez2019MNRAS.490.3061V}
{V{\'a}zquez-Semadeni}, E., {Palau}, A., {Ballesteros-Paredes}, J.,
  {G{\'o}mez}, G.~C., \& {Zamora-Avil{\'e}s}, M. 2019, \mnras, 490, 3061,
  \dodoi{10.1093/mnras/stz2736}

\bibitem[{{Virtanen} {et~al.}(2020){Virtanen}, {Gommers}, {Oliphant},
  {Haberland}, {Reddy}, {Cournapeau}, {Burovski}, {Peterson}, {Weckesser},
  {Bright}, {van der Walt}, {Brett}, {Wilson}, {Jarrod Millman}, {Mayorov},
  {Nelson}, {Jones}, {Kern}, {Larson}, {Carey}, {Polat}, {Feng}, {Moore}, {Vand
  erPlas}, {Laxalde}, {Perktold}, {Cimrman}, {Henriksen}, {Quintero}, {Harris},
  {Archibald}, {Ribeiro}, {Pedregosa}, {van Mulbregt}, \&
  {Contributors}}]{2020SciPy-NMeth}
{Virtanen}, P., {Gommers}, R., {Oliphant}, T.~E., {et~al.} 2020, Nature
  Methods, 17, 261, \dodoi{https://doi.org/10.1038/s41592-019-0686-2}

\bibitem[{{W}es {M}c{K}inney(2010)}]{mckinney-proc-scipy-2010}
{W}es {M}c{K}inney. 2010, in {P}roceedings of the 9th {P}ython in {S}cience
  {C}onference, ed. {S}t\'efan van~der {W}alt \& {J}arrod {M}illman, 56 -- 61,
  \dodoi{10.25080/Majora-92bf1922-00a}

\bibitem[{Yeh {et~al.}(2019)Yeh, Carraro, Montalto, \& Seleznev}]{Yeh_2019}
Yeh, F.~C., Carraro, G., Montalto, M., \& Seleznev, A.~F. 2019, The
  Astronomical Journal, 157, 115, \dodoi{10.3847/1538-3881/aaff6c}

\end{thebibliography}
\bibliographystyle{aasjournal}



\end{document}